\documentclass[twocolumn,prl,superscriptaddress]{revtex4-2}
\usepackage{bm}
\usepackage{graphicx}
\usepackage{color}
\usepackage{braket}
\usepackage{amsmath,amssymb,amsfonts,amsthm,mathtools}
\usepackage{enumerate}
\usepackage[colorlinks=true,linkcolor=blue,anchorcolor=red,citecolor=blue,urlcolor=blue]{hyperref}
\usepackage{titlesec}
\usepackage{tikz-cd} 
\usepackage{float}
\usepackage{multirow}
\usepackage{mathrsfs}
\usepackage{lipsum} 
\usepackage{titletoc}

\def \L {{\mathcal{L}}}
\def \t {{\bm{t}}}

\makeatletter
\renewcommand{\bibsection}{%
	\@ifx{\@empty\refname}{}{%
		\par\vspace{6\p@ plus 6\p@}%
	}%
}
\makeatother

\begin{document}
	
	\title{Flux-Mediated Correspondence Between Real- and Momentum-Space Nonsymmorphicity}     
	\author{Z. Y. Chen}
	\affiliation{Department of Physics and HK Institute of Quantum Science \& Technology, The University of Hong Kong, Pokfulam Road, Hong Kong, China}
	
	\author{Y. X. Zhao}
	\email[]{yuxinphy@hku.hk}
	\affiliation{Department of Physics and HK Institute of Quantum Science \& Technology, The University of Hong Kong, Pokfulam Road, Hong Kong, China}
	
	\begin{abstract}
		Momentum-space nonsymmorphic symmetries have recently attracted significant interest in both artificial and condensed-matter crystals, whereas real-space nonsymmorphic symmetries have long played an important role in the study of crystalline topological phases. Here, we establish a general theory of momentum-space crystallographic groups that emerge from projective representations of real-space crystallographic groups in the presence of gauge flux, applicable in particular to real-space nonsymmorphic groups. A central result is a flux-mediated ``bi-nonsymmorphicity'' relation that reveals a structural correspondence between real-space and momentum-space nonsymmorphicity mediated by gauge flux. This relation implies that, under a symmetric gauge flux, real-space nonsymmorphicity can enforce momentum-space nonsymmorphicity, and that in some cases a symmetric gauge flux requires nonsymmorphicity in both real and momentum space. Our work not only identifies a fundamental structure in projective crystal symmetries, but also provides guiding principles for designing artificial crystals and condensed-matter platforms that exhibit both real-space and momentum-space nonsymmorphic symmetries.
	\end{abstract}
	\maketitle
	
	\textit{Introduction} Recently, momentum-space nonsymmorphic symmetries have emerged as a research focus in both artificial crystals of metamaterials~\cite{pu2023acoustic,liu2024topological,jiang2024symmetry,lai2024real,tao2024higher,hu2024higher,ZhuYangWu2024,fonseca2024weyl,qiu2025octupole,hu2025acoustic,li2026higher} and condensed-matter crystals~\cite{li2023klein,xiao2024spin,cualuguaru2025moire,Chen_Zhang}. According to the general theory of projective crystal symmetries, momentum-space nonsymmorphic symmetries occur when the conjugation of lattice translations by point-group symmetries leads to nontrivial phase factors~\cite{Zhiyi_NC_2022,Zhang2023General}. Such projective crystal symmetries can be realized by engineerable gauge-flux patterns in artificial crystals~\cite{Haoran_PRL_2022,li2022acoustic,chen2023classification} and by more intrinsic mechanisms in various condensed-matter systems~\cite{Fan_Zhang_PRL,yoon2026majorana}, including flux phases of spin liquids~\cite{wen2002quantum}, moir\'e systems~\cite{cualuguaru2025moire}, and magnetic materials with certain spin-space groups~\cite{xiao2024spin}.
	
	On the other hand, real-space nonsymmorphic symmetries have long been studied and were revived in topological insulators and semimetals because they can lead to symmetry-enforced band crossings~\cite{zhao2016nonsymmorphic}. Typical examples include nonsymmorphic Dirac semimetals~\cite{young2012dirac,yang2014classification,young2015dirac}, M\"obius topological insulators~\cite{shiozaki2015z}, and hourglass topological insulators~\cite{wang2016hourglass,shiozaki_cls}.
	
	A natural question therefore arises: what happens when real-space and momentum-space nonsymmorphic symmetries coexist? The present work is devoted to addressing this fundamental question within the general framework of projective crystal symmetries, particularly in the presence of gauge flux.
	
	In this work, an element $\gamma$ of a crystallographic group $\Gamma$ in arithmetic class $c$ is denoted by $\gamma=(\bm{t},g)$. Here, $\bm{t}$ is a lattice translation in the lattice $L$, $g$ is an element of the point group $G$, and the action of $G$ on $L$ is specified by $c$~\cite{bradley_book}. The multiplication law is
	$
	\gamma_2\gamma_1=(\bm{t}_2+g_2\bm{t}_1+\bm{\omega}(g_2,g_1), g_2g_1)
	$.
	Here, $\bm{\omega}(g_2,g_1)$ is an $L$-valued 2-cocycle characterizing the real-space nonsymmorphicity.
	
	
	Under lattice-translation symmetry, the gauge-flux configuration can be described by a skew-symmetric matrix $\Phi$ via
	\begin{equation}
		\mathcal{W}(\bm{t}_2,\bm{t}_1)=e^{2\pi i\, t_2^T \Phi  t_1}.
	\end{equation}
	Here, $t$ is the integer column vector of $\bm{t}$ under a basis $\bm{e}_i$ of $L$, i.e., $\bm{t}=\sum_{i}\bm{e}_it_i$ with $t_i\in\mathbb{Z}$. On a monoclinic lattice with only nearest-neighbor hoppings, $2\pi \Phi_{ij}$ may be interpreted as the gauge flux through the plaquette spanned by $\bm{e}_i$ and $\bm{e}_j$, and therefore $\Phi$ is referred to as the flux form. For the corresponding multiplier $\sigma$ of $L$, $\mathcal{W}(\bm{t}_2,\bm{t}_1)={\sigma(\bm{t}_2,\bm{t}_1)}/{\sigma(\bm{t}_1,\bm{t}_2)}$ is called the commutator function. The equivalence classes of multipliers for the translation group $L$ are in one-to-one correspondence with the commutator function~\cite{backhouse1970projective, Moore2022GroupTheory}. 
	
	As in the Hofstadter model, an irrational flux leads to an infinite-dimensional representation~\cite{zak1964magnetic,hofstadter1976energy}, intrinsically related to the Aubry-Andr\'e model~\cite{aubry1980analyticity}, a deep mathematical subject. Hereafter, we assume that each entry of $\Phi$ is rational, which should be sufficient for physical purposes. In particular, flux forms $\Phi$ with entries $0$ and $1/2$ preserve time-reversal invariance and therefore can be simulated readily in various artificial crystals~\cite{Keil_2016_UniversalSignControl,Xue_2020_AcousticOctupole,Haoran_PRL_2022,li2022acoustic,Zhan_Peng} and arise naturally in $\mathbb{Z}_2$ spin liquids and spin-space groups~\cite{wen2002quantum,xiao2024spin}.
	
	To construct the Brillouin zone in the presence of $\Phi$, we note that the projective algebra of $L$ has an underlying finite Heisenberg algebra~\cite{weyl1950theory,Moore2022GroupTheory}, as shown in the Supplemental Material (SM)~\cite{SM}. Its center, consisting of translation operators that commute with all translation operators, is precisely the mod-$\mathbb{Z}$ kernel $Z$ of $\Phi$, referred to as the central sublattice of $L$. Introducing the dual lattice $Z_F$ of $Z$, we find that $T^d_F=\mathbb{R}^d/Z_F$ is the space of all irreducible projective representations of $L$~\cite{mackey1989unitary}. Thus, $T^d_F$ is the Brillouin zone in the presence of $\Phi$. We shall show that, in general, a point-group element $g\in G$ acts on momentum space as $\bm{k}\mapsto g\bm{k}+\bm{\kappa}_g$, where $\bm{\kappa}_{g}$ is a fractional reciprocal-lattice vector satisfying  
	\begin{equation}\label{eq:rec_cocycle}
		\bm{\omega}_F(g_2,g_1) = g_2 \bm{\kappa}_{g_1} - \bm{\kappa}_{g_2 g_1} + \bm{\kappa}_{g_2} \in Z_F,
	\end{equation}
	i.e., $\bm{\omega}_F(g_2,g_1)$ is $Z_F$-valued for any $g_2,g_1\in G$. The momentum-space crystallographic group $\Gamma_F$ is then defined as $\Gamma_F=Z_F\rtimes_{(\tilde{c}_F,\bm{\omega}_F)} G$. Here, $\tilde{c}$ denotes the arithmetic class of $Z$ with the natural $G$ action, which is in general different from $c$, and $\tilde{c}_F$ is the Fourier dual of $\tilde{c}$ specifying the $G$ action on $Z_F$.
	
	\textit{Main results} We now have the real-space lattice $L$ and its central sublattice $Z$, together with their dual lattices $L_F$ and $Z_F$. Notably, in momentum space, $L_F$ is a sublattice of $Z_F$ with $Z_F/L_F\cong L/Z\cong \mathbb{Z}_p\times \mathbb{Z}_p$ for some integer $p$, as we will show below. Group cohomologies can be formulated with coefficients in these lattices and quotients (see the SM~\cite{SM}). We are now ready to state one of our main results, termed the flux-mediated bi-nonsymmorphicity relation, or simply the bi-nonsymmorphicity relation:
	\begin{equation}\label{eq:duality}
		[\overline{\bm{\omega}}_F]
		= [\overline{\bm{\Phi}\cdot \bm{\omega}}]
		+ [\overline{\bm{\Lambda}_{\Phi}}],
	\end{equation}
	which relates real-space and momentum-space nonsymmorphicity through the flux form $\Phi$. Here, $\bm{\omega}_F$ is a cocycle valued in $Z_F$, and $\overline{\bm{\omega}}_F$ denotes $\bm{\omega}_F \!\!\mod L_F$, corresponding to a cohomology class $[\overline{\bm{\omega}}_F] \in H^{2,\tilde{c}_F}(G,Z_F/L_F)$. Due to the $G$-invariance of $\Phi$, $\bm{\Phi}\cdot \bm{\omega}= \sum_{i,j} \bm{G}_i \,\Phi_{ij}\,\bm{G}_j \cdot \bm{\omega}$ is naturally a cocycle modulo $L_F$, defining a cohomology class $[\overline{\bm{\Phi}\cdot \bm{\omega}}] \in H^{2,\tilde{c}_F}(G,Z_F/L_F)$ (see the SM~\cite{SM}). Finally, as we will specify, $[\overline{\bm{\Lambda}_{\Phi}}]$ is a cohomology class in $H^{2,\tilde{c}_F}(G,Z_F/L_F)$ determined solely by the flux form $\Phi$, and will be referred to as the flux twist.
	
	In fact, for almost all of the 73 arithmetic classes, $[\overline{\bm{\Lambda}_{\Phi}}]$ is trivial, reducing the duality relation to 
	\begin{equation}\label{eq:s-duality}
		[\overline{\bm{\omega}}_F]
		= [\overline{\bm{\Phi}\cdot \bm{\omega}}].
	\end{equation}
The duality relation characterizes the compatibility between real-space and momentum-space nonsymmorphicity in the presence of a gauge-flux configuration. In particular, for certain real-space nonsymmorphic groups, specific flux configurations force the corresponding momentum-space crystallographic groups to be nonsymmorphic as well. Moreover, \eqref{eq:s-duality} imposes constraints on the admissible flux forms for a nonsymmorphic $\Gamma$, since for some 
G-invariant flux forms, no $\bm{\omega}_F$ satisfies \eqref{eq:s-duality}.

The three exceptional arithmetic classes are $mm2F$, $mmmF$, and $m\bar{3}F$, all with the same $G$-invariant flux form. For all of them, it is impossible to find $\bm{\omega}_F$ such that $[\overline{\bm{\omega}}_F]=[\overline{\bm{\Lambda}_{\Phi}}]$, and $c=\tilde{c}_F$. Thus, the $G$-invariant flux form enforces both $\Gamma$ and $\Gamma_F$ to be the unique nonsymmorphic group in the corresponding arithmetic class.
	
\begin{figure}
		\centering
		\includegraphics[width=\columnwidth]{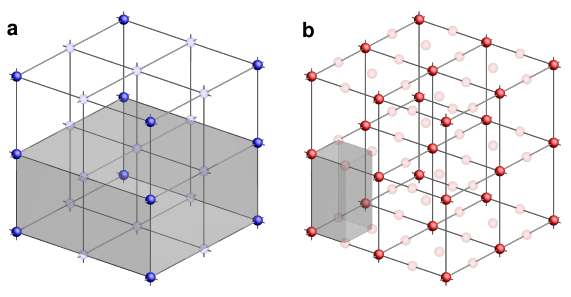}
		\caption{(a) The lattice $L$ and the central sublattice $Z$ (filled lattice points) for real-space $Pc$ with the given $\Phi$. (b) The dual lattices $L_F$ and $Z_F$ in momentum space. $L_F$ (filled lattice points) is a sublattice of $Z_F$. $Pc$ preserves $Z_F$. The shaded regions in (a) and (b) correspond to the unit cells of $Z$ and $Z_F$, respectively.}
		\label{fig-pc}
	\end{figure}

Before presenting the systematic formulation and the proof of Eq.~\eqref{eq:duality}, let us consider a simple example: $Pc$ acting on a monoclinic primitive lattice with $\pi$ flux through each plaquette in the $x$-$y$ plane. The flux form is
	\begin{equation}
		\Phi = \begin{bmatrix}
			0 & \frac{1}{2} & 0 \\
			-\frac{1}{2} & 0 & 0 \\
			0 & 0 & 0
		\end{bmatrix}.
	\end{equation}
	The point group is $D_1=\{E,M_z\}$ and $\bm{\omega}(M_z,M_z)=\bm{e}_x$. Accordingly, $Z$ is spanned by $2\bm{e}_x$, $2\bm{e}_y$, and $\bm{e}_z$, while $Z_F$ is spanned by $\bm{Q}_x=\bm{G}_x/2$, $\bm{Q}_y=\bm{G}_y/2$, and $\bm{Q}_z=\bm{G}_z$ (see Fig.~\ref{fig-pc}). Then, from Eq.~\eqref{eq:duality}, we obtain $\bm{\omega}_F(M_z,M_z)=\bm{G}_y/2=\bm{Q}_y$, which clearly corresponds to a momentum-space glide reflection that inverts $k_z$ and translates $\bm{k}$ by $\bm{Q}_y/2$. Thus, $\Gamma_F$ cannot be symmorphic. Here, $c=\tilde{c}=c_F=\tilde{c}_F=mP$, in which there are only two crystallographic groups, the symmorphic group $Pm$ and the nonsymmorphic group $Pc$. Therefore, $\Gamma_F$ is enforced to be $Pc$.
	
	\textit{The Brillouin zone} The momentum-space crystallographic group corresponds to the $G$ action on the Brillouin zone. We therefore begin by constructing the Brillouin zone in the presence of the flux form $\Phi$.
	
	The key observation is that, for rational $\Phi$, we can perform the congruence transformation
	\begin{equation}\label{eq:flux_form}
		\Omega^T \Phi \Omega=\frac{q}{p}\begin{bmatrix}
			0 & 1 & 0\\
			-1 & 0 & 0\\
			0 & 0 & 0
		\end{bmatrix},
	\end{equation}
	where $q$ and $p$ are coprime positive integers and $\Omega$ is an integer matrix~\cite{newman1972integral}. Accordingly, the center $Z$ of the projective algebra $\mathcal{L}$ of $L$ is a sublattice of $L$, namely a free Abelian group, with basis vectors 
	$\bm{a}_1=p\bm{e}_j \Omega_{j1}$, $\bm{a}_2=p\bm{e}_j \Omega_{j2}$, and $\bm{a}_3=\bm{e}_j \Omega_{j3}$. This establishes the previous claim that $L/Z\cong \mathbb{Z}_p^2$. Then, the quotient algebra $\mathcal{L}/Z$ has two generators $P$ and $Q$, satisfying 
	\begin{equation}\label{eq:Heisen-Main}
		PQ=e^{\frac{2\pi i q}{p}}QP,\quad P^p=Q^p=1.
	\end{equation}
	Thus, $\mathcal{L}/Z$ is exactly the finite Heisenberg algebra $\mathrm{Heis}_{q}(\mathbb{Z}_p\times\mathbb{Z}_p)$, which has a unique irreducible representation $U_{H}$~\cite{weyl1950theory,Moore2022GroupTheory}. Consequently, an irreducible representation of $\mathcal{L}$ takes the form
	\begin{equation}
		U_{\bm{k}}(\bm{t})=U_{H}(\bar{\bm{t}}) e^{2\pi i\bm{k}\cdot \bm{t}}.
	\end{equation}
	Restricted to the central sublattice $Z$, the representation reduces to
	\begin{equation}\label{eq:rep_z}
		U_{\bm{k}}(\bm{z})=e^{2\pi i\bm{k}\cdot \bm{z}}.
	\end{equation}
	
	The reciprocal lattice $Z_F$ of $Z$ is generated by the dual basis $\mathbf{Q}_i$ of $\bm{a}_j$, with $\mathbf{Q}_i\cdot \bm{a}_j=\delta_{ij}$. For any $\mathbf{K}\in Z_F$, we can show that $U_{\bm{k}+\bm{K}}(\bm{t})=e^{i\bm{K}\cdot\bm{t}}U_{\bm{k}}(\bm{t})=V_{\bm{K}}U_{\bm{k}}(\bm{t})V^\dagger_{\bm{K}}$ for some unitary matrix $V_{\bm{K}}$ (see the SM~\cite{SM}). Thus, the Brillouin zone should be the fundamental domain of $Z_F$ in momentum space rather than that of $L_F$, recalling that $L_F$ is a sublattice of $Z_F$, i.e., $T^3_F\cong \mathbb{R}^3/Z_F$.
	
	\textit{Fractional translations} We next consider the action of $G$ on $T_F^3$. In any representation $U$ with multiplier $\nu$, the conjugation of $\bm{t}$ by $\gamma$ is given by $^\gamma U(\bm{t})=[U(\gamma)]^\dagger U(\bm{t}) U(\gamma)=\vartheta(\gamma|\bm{t})U(g^{-1}\bm{t})$, with $\vartheta(\gamma|\bm{t})={\nu(\gamma^{-1},\bm{t})\nu(\gamma^{-1}\bm{t},\gamma)}/{\nu(\gamma^{-1},\gamma)}$. Accordingly, the action of $G$ on $U_{\bm{k}}$ should be
	\begin{equation}\label{eq:g-on-irrep}
		^g U_{\bm{k}}(\bm{t})=\eta(g|\bm{t})U_{\bm{k}}(g^{-1}\bm{t})
	\end{equation}
	with $\eta(g|\bm{t})=\vartheta((g,0)|\bm{t})$. 
	
	As shown in the SM~\cite{SM}, $\eta$ and $\sigma=\nu|_{L\times L}$ satisfy two consistency equations:
	\begin{equation}\label{eq:const_1-Main}
		\frac{\sigma(g^{-1}\bm{t}_2,g^{-1}\bm{t}_1)}{\sigma(\bm{t}_2,\bm{t}_1)} = \frac{\eta(g|\bm{t}_2+ \bm{t}_1)}{\eta(g|\bm{t}_2)\eta(g|\bm{t}_1)}, 
	\end{equation}
	and
	\begin{equation}\label{eq:const_2-Main}
		\frac{\eta(g_2g_1|\bm{t})}{\eta(g_2|\bm{t})\eta(g_1|g_2^{-1}\bm{t})} = \frac{\sigma(\bm{\omega}(g_2,g_1),\bm{t})}{\sigma(\bm{t},\bm{\omega}(g_2,g_1))}.
	\end{equation}
	We can always choose $\sigma$ such that $\sigma|_{Z\times Z}=1$ (see the SM~\cite{SM}). Then these two equations reduce to
	\begin{equation}
		\eta(g|\bm{z}_2 + \bm{z}_1)=\eta(g|\bm{z}_2)\eta(g|\bm{z}_1)
	\end{equation}
	and
	\begin{equation}
		\eta(g_2g_1|\bm{z})=\eta(g_2|\bm{z})\eta(g_1|g_2^{-1}\bm{z}).
	\end{equation}
	The general solution of the first equation is 
	\begin{equation}\label{eq:eta-z}
		\eta(g|\bm{z})=e^{2\pi i \bm{\kappa}_g\cdot\bm{z}},
	\end{equation}
	and the second is then equivalent to Eq.~\eqref{eq:rec_cocycle}.
	
	With $\eta(g|\bm{z})=e^{2\pi i \bm{\kappa}_g\cdot\bm{z}}$ and Eq.~\eqref{eq:rep_z}, restricting Eq.~\eqref{eq:g-on-irrep} to $Z$ yields $^g U_{\bm{k}}(\bm{z})=\eta(g|\bm{z})U_{\bm{k}}(g^{-1}\bm{z})=e^{2\pi i (g\bm{k}+\bm{\kappa}_g)\cdot \bm{z}}=U_{g\bm{k}+\bm{\kappa}_g}(\bm{z})$. Thus, $\bm{\kappa}_g$ is precisely the fractional reciprocal translation associated with $g$. 
	
	\textit{The bi-nonsymmorphicity relation} The left-hand side of Eq.~\eqref{eq:const_1-Main} can be regarded as a coboundary transformation between $^g \sigma(\bm{t}_2,\bm{t}_1)=\sigma(g^{-1}\bm{t}_2,g^{-1}\bm{t}_1)$ and $\sigma(\bm{t}_2,\bm{t}_1)$, and therefore $^g \sigma$ and $\sigma$ correspond to the same $\mathcal{W}$ or $\Phi$. In other words, the flux form, or equivalently the cohomology class of $\sigma$, is invariant under the $G$ action. 
	
	
	In general, however, there need not exist a $G$-invariant $\sigma$ for a $G$-invariant $\mathcal{W}$, just as a connection configuration representing a $G$-invariant gauge-flux configuration is not, in general, itself $G$-invariant. Consequently, we need to analyze the variation of $^g\sigma/\sigma$. Since $^g\sigma$ and $\sigma$ correspond to the same $\mathcal{W}$, their ratio must be a symmetric form, i.e.,
	\begin{equation}\label{eq:def_S}
		{\sigma(g^{-1}\bm{t}_2,g^{-1}\bm{t}_1)}/{\sigma(\bm{t}_2,\bm{t}_1)}= e^{2\pi i t_2^T S(g) t_1 }
	\end{equation}
	with $S(g)$ a symmetric matrix.
	
	It is straightforward to verify that
	$\Delta(g|\bm{t})=e^{\pi i t^T S(g) t }$ is a solution for $\eta$ to Eq.~\eqref{eq:const_1-Main}. Then, since $\eta/\Delta$ is linear on $L$, the general solution for $\eta$ to Eq.~\eqref{eq:const_1-Main} can be written as
	\begin{equation}\label{eq:eta}
		\eta(g|\bm{t})=e^{\pi i t^T S(g) t } e^{2\pi i \bm{q}(g)\cdot \bm{t}}.
	\end{equation}
	The fact that $\eta(g|\bm{z})=e^{2\pi i \bm{\kappa}_g\cdot \bm{z}}$ implies
	\begin{equation}
		\tfrac{1}{2}\bm{D}{\tilde{S}}(g)+\bm{q}(g)\equiv \bm{\kappa}_{g} \mod Z_F.
	\end{equation}
	Here, $\bm{D}{\tilde{S}}(g)=\sum_{\alpha} [\tilde{S}(g)]_{\alpha\alpha} \bm{Q}_\alpha$, where $\tilde{S}(g)=V^TS(g)V$ with $\bm{a}_\alpha=\bm{e}_iV_{i\alpha}$. This follows essentially from the condition $\sigma|_{Z\times Z}=1$; see the SM~\cite{SM} for details.
	
	Applying the coboundary operator to both sides gives
	\begin{equation}\label{eq:q-kappa}
		\noindent \!\!\!\tfrac{1}{2}\delta \bm{D}\!{\tilde{S}}(g_2,\! g_1\!)\!+\!\delta\bm{q}(g_2,\!g_1\!)\equiv  \bm{\omega}_F(g_2,\!g_1\!)\, \! \!\!\! \mod B^2(\!G,\!Z_F\!).
	\end{equation}
	The coboundary operators and the group cohomologies used in this work are introduced systematically in the SM~\cite{SM}.
	
	Substituting Eq.~\eqref{eq:eta} into Eq.~\eqref{eq:const_2-Main} yields
	\begin{equation}\label{eq:reduce-2nd}
		e^{\pi i t^T \delta S(g_2,g_1) t }e^{2\pi i \delta\bm{q}(g_2,g_1)\cdot \bm{t}} =e^{2\pi i \bm{t}\cdot \bm{\Phi}\cdot \bm{\omega}(g_2,g_1) } .
	\end{equation}
	Here, $\delta S(g_2,g_1)=E_{g^{-1}_2}^T S(g_1) E_{g^{-1}_2}-S(g_2g_1)+S(g_2)$, with $E_g$ the integer matrix representation of $g$ under the basis $\bm{e}_i$, and $\delta \bm{q}(g_2,g_1)=g_2\bm{q}(g_1)-\bm{q}(g_2g_1)+\bm{q}(g_2)$. Using Eq.~\eqref{eq:def_S}, one can show that $\delta S(g_2,g_1)$ is an integer matrix. Since it is also symmetric, we have $e^{\pi i t^T \delta S(g_2,g_1) t }=e^{\pi i \sum_i t_i [\delta S(g_2,g_1)]_{ii} }$. Thus, we can rewrite Eq.~\eqref{eq:reduce-2nd} as
	\begin{equation}
		e^{\pi i \bm{D} \delta S(g_2,g_1) \cdot\bm{t} }e^{2\pi i \delta\bm{q}(g_2,g_1)\cdot \bm{t}} =e^{2\pi i \bm{t}\cdot \bm{\Phi}\cdot \bm{\omega}(g_2,g_1) } ,
	\end{equation}
	by introducing $\bm{D} \delta S(g_2,g_1)=\sum_{i} [\tilde{S}(g)]_{ii} \bm{G}_i$. In light of Eq.~\eqref{eq:q-kappa}, we can introduce the flux twist,
	\begin{equation}
		\bm{\Lambda}_{\Phi}(g_2,g_1)=\tfrac{1}{2}[\delta\bm{D}{\tilde{S}}(g_2,g_1)-\bm{D}{\delta S}(g_2,g_1)],
	\end{equation}
	and derive the bi-nonsymmorphicity relation \eqref{eq:duality}. The reasons why $\overline{\bm{\Lambda}_{\Phi}}$ and $\overline{\bm{\Phi}\cdot \bm{\omega}}$ are $Z_F/L_F$-valued cocycles can be found in the SM~\cite{SM}.
	
	We exhaust all arithmetic classes $c$ and all $G$-invariant flux forms $\Phi$ for each $c$, and find that almost all cases correspond to a trivial flux twist $\overline{\bm{\Lambda}_{\Phi}}$, except for the three arithmetic classes $mm2F$, $mmmF$, and $m\bar{3}F$. The full classification is given in the SM~\cite{SM}.
	
	\textit{Trivial flux twist} For the 70 arithmetic classes with trivial flux twist, the bi-nonsymmorphicity relation simplifies to Eq.~\eqref{eq:s-duality}. In practice, this relation strongly constrains the momentum-space nonsymmorphicity in terms of the flux form and the real-space nonsymmorphicity, and in many cases even enforces momentum-space nonsymmorphicity, as already seen in the example above. 
	
\begin{figure}
		\centering
		\includegraphics[width=\columnwidth]{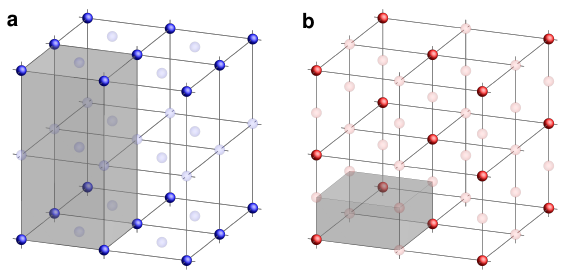}
		\caption{(a) The base-centered lattice $L$ and its central primitive sublattice $Z$ (filled points) for $Ccc2$ under the given $\Phi$. (b) The dual lattices: base-centered $L_F$ (filled points) and primitive $Z_F$. $Pba2$ preserves $Z_F$. The shaded regions indicate the unit cells of $Z$ and $Z_F$, respectively.}
		\label{fig-Ccc2}
	\end{figure}

	We now present another, richer example ($Ccc2\rightarrow Pba2$), considering $\Gamma=Ccc2$ with $G=C_{2v}$ in class $c=mm2C$ acting on an orthorhombic base-centered lattice $L$. As illustrated in Fig.~\ref{fig-Ccc2}(a), $L$ is generated by~\cite{bradley_book}
	\begin{equation}
		\bm{e}_1 = \tfrac{1}{2} (1,1,0),\, \bm{e}_2 = \tfrac{1}{2} (1,-1,0),\, \bm{e}_3 = (0,0,1),
	\end{equation}
	with dual reciprocal lattice $L_F$ generated by
	\begin{equation}
		\bm{G}_1 =  (1,1,0),\,\bm{G}_2 = (1,-1,0), \,\bm{G}_3 = (0,0,1).
	\end{equation}
	$G$ acts on $L_F$ according to the same arithmetic class, i.e., $c=mm2C$ is self-dual with $c=c_F=mm2C$.
	
	Let us consider the $G$-invariant flux form,
	\begin{equation}
		\Phi = \frac{1}{2}\begin{bmatrix}
			0 & 0 & 1 \\
			0 & 0 & 1 \\
			-1 & -1 & 0
		\end{bmatrix},
	\end{equation}
	which may be realized by inserting $\pi$ fluxes through plaquettes of the lattice. The corresponding central sublattice $Z$ is spanned by the basis
	$\bm{a}_1 = \bm{e}_1 + \bm{e}_2$, $ \bm{a}_2 = \bm{e}_1 - \bm{e}_2$, and $\bm{a}_3 = 2\bm{e}_3$, i.e.,
	\begin{align}
		\bm{a}_1 = (1,0,0),\  \bm{a}_2 = (0,1,0),\  \bm{a}_3 =(0,0,2),
	\end{align}
	with dual basis for $Z_F$ given by (Fig.~\ref{fig-Ccc2}(b))
	\begin{align}
		\bm{Q}_1=  (1,0,0), \, \bm{Q}_2=  (0,1,0),\, \bm{Q}_3= \tfrac{1}{2}(0,0,1) .
	\end{align}
	$Z$ and $Z_F$ belong to the arithmetic class $\tilde{c}=\tilde{c}_F=mm2P$ on orthorhombic primitive lattices, different from $c=c_F=mm2C$ on orthorhombic base-centered lattices.
	
	For $Ccc2$, $\bm{\omega}$ is encoded in $\bm{\tau}_{M_x} = \bm{\tau}_{M_y} =  \bm{e}_z/2$, corresponding to
	\begin{equation}
		\bm{\omega}(M_x,M_x) = \bm{\omega}(M_y,M_y) = \bm{e}_z .
	\end{equation}
	Since $\bm{\Phi}\cdot \bm{e}_z=\tfrac{1}{2}(\bm{G}_1+\bm{G}_2)=\bm{Q}_1$, Eq.~\eqref{eq:s-duality} yields
	\begin{equation}
		\bm{\omega}_F(M_x,M_x) = \bm{\omega}_F(M_y,M_y) = \bm{Q}_1 \mod {L_F}.
	\end{equation}
	We then traverse all crystallographic groups in the arithmetic class $\tilde{c}_F=mm2P$, namely Nos.~25--34, and find that only $Pba2$ satisfies the duality relation, with $\bm{\kappa}_{M_x} = \bm{Q}_2/2$ and $\bm{\kappa}_{M_y} = \bm{Q}_1/2$, or equivalently $\bm{\omega}_F(M_x,M_x)=\bm{Q}_2$ and $\bm{\omega}_F(M_y,M_y)=\bm{Q}_1$. Note that $\bm{Q}_1+\bm{Q}_2=\bm{G}_1$, and therefore $\bm{Q}_1=\bm{Q}_2 \mod L_F$. Thus, $\Gamma_F$ is enforced to be $Pba2$ by the flux form $\Phi$.
	
	In addition, an example illustrating how the bi-nonsymmorphicity relation constrains 
	$G$-invariant flux configurations for certain nonsymmorphic real-space groups can be found in the SM~\cite{SM}.
	\begin{figure}
		\centering
		\includegraphics[width=\columnwidth]{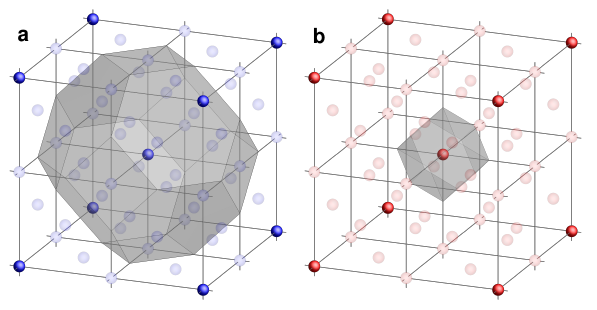}
		\caption{(a) The face-centered lattice $L$ and its body-centered central sublattice $Z$ (filled points) for $Fdd2$ under the given $\Phi$. (b) The dual lattices: body-centered $L_F$ (filled points) and face-centered $Z_F$. $Fdd2$ preserves $Z_F$. The shaded regions indicate the unit cells of $Z$ and $Z_F$, respectively.}
		\label{fig-fdd2}
	\end{figure}
	
	\textit{Nontrivial flux twist} The exceptional arithmetic classes with nontrivial flux twists are $mm2F$ on a face-centered orthorhombic lattice with point group $C_{2v}$, $mmmF$ on a face-centered orthorhombic lattice with point group $D_{2h}$, and $m\bar{3} F$ on a face-centered cubic lattice with point group $T_h$. The underlying reason is essentially the same in all cases and can be explained completely in the most elementary case, $mm2F$, since the other two can be regarded as including extra symmetries without additional complexity.
	
	We therefore focus on $mm2F$ on the face-centered orthorhombic lattice $L$ generated by~\cite{bradley_book}
	\begin{equation}
		\bm{e}_1 = \tfrac{1}{2}(0,1,1),\, \bm{e}_2 = \tfrac{1}{2}(1,0,1),\, \bm{e}_3 = \tfrac{1}{2}(1,1,0),
	\end{equation}
	with dual basis
	\begin{align}
		\bm{G}_1 = (-1,1,1),\,\bm{G}_2 = (1,-1,1),\, \bm{G}_3 = (1,1,-1),
	\end{align}
	generating $L_F$, where the $C_{2v}$ action on $L_F$ corresponds to $c_F = mm2I$.
	
	The $C_{2v}$ invariance requires $\Phi$ to take values in $\mathbb{Z}/4$, represented by $\{0, 1/4, 1/2, -1/4 \}$. It is sufficient to consider only the flux form
	\begin{equation}
		\Phi = \frac{1}{4}\begin{bmatrix}
			0 & 1 & -1 \\
			-1 & 0 & 1 \\
			1 & -1 & 0
		\end{bmatrix},
	\end{equation}
	since the only flux forms with nontrivial twists are $\Phi$ and $-\Phi$. The flux form $\Phi$ can be realized by $\pm \pi/2$ fluxes per plaquette on the lattice.
	Accordingly, the central sublattice $Z$ is spanned by the basis
	$\bm{a}_i = -(\bm{e}_1+\bm{e}_2+\bm{e}_3)+4\bm{e}_i$, i.e.,
	\begin{equation}
		\bm{a}_1 = (-1,1,1),\, \bm{a}_2 = (1,-1,1),\, \bm{a}_3 = (1,1,-1).
	\end{equation}
	The dual basis for $Z_F$ is then given by (Fig.~\ref{fig-fdd2}(b))
	\begin{align}
		\bm{Q}_1 = \tfrac{1}{2}(0,1,1),\, \bm{Q}_2 = \tfrac{1}{2}(1,0,1),\,\bm{Q}_3 = \tfrac{1}{2}(1,1,0).
	\end{align}
	Remarkably, with this flux form, the real-space and momentum-space arithmetic classes are interchanged. We now have $\tilde{c}=c_F=mm2I$ and $\tilde{c}_F=c=mm2F$. It is then straightforward to derive the flux twist $\bm{\Lambda}_{\Phi}$.
	
	The core part of the argument is to compute $H^{2,\tilde{c}_F}(G,Z_F/L_F)\cong \mathbb{Z}_2^3$, noting that $Z_F/L_F\cong L/Z\cong \mathbb{Z}_4^2$, and to present a complete set of cohomology invariants that fully determines whether a cocycle is nontrivial; these details are given in the SM~\cite{SM}. We then show that $[\overline{\bm{\Lambda}_{\Phi}}]$ is nontrivial.
	
	For $c=\tilde{c}_F=mm2F$, there are only two crystallographic groups: the symmorphic group $Fmm2$ and the nonsymmorphic group $Fdd2$. Using the cohomology invariants, it is straightforward to show that $[\overline{\bm{\Lambda}_{\Phi}}]$ differs from $[\overline{\bm{\omega}}_F]$ for $\Gamma_F=Fdd2$. Thus, in order to satisfy the bi-nonsymmorphicity relation \eqref{eq:duality}, the real-space crystallographic group $\Gamma$ must be the nonsymmorphic group $Fdd2$, and indeed it is. Therefore, the $C_{2v}$-invariant flux form enforces $\Gamma=\Gamma_F=Fdd2$. For the other two arithmetic classes, $mmmF$ and $m\bar{3}F$, the same $G$-invariant flux form similarly enforces $\Gamma=\Gamma_F=Fddd$ and $\Gamma=\Gamma_F=Fd\bar{3}$, respectively.
	
	
	
	
	\textit{Summary and Discussion} In summary, we have developed a theory of momentum-space crystallographic groups emerging from projective representations of real-space crystallographic groups in the presence of gauge flux. The theory features a flux-mediated bi-nonsymmorphicity relation, which efficiently determines the momentum-space nonsymmorphic group from the real-space nonsymmorphic group and the symmetric gauge flux, and therefore serves as a guiding principle for designing artificial crystals engineered to realize nonsymmorphic symmetry in momentum space.
	
	
	Once \(\Gamma_F\) is identified, all irreducible projective representations of \(\Gamma\) can be constructed as induced representations from the little cogroup \(G_{\boldsymbol{k}}\) for each \(\boldsymbol{k}\) in \(T_F^d\), by generalizing the conventional \(k\)-star theory---where a \(k\)-star is the $G$-orbit of \(\boldsymbol{k}\) in $T_F^d$---to the setting of projective representations~\cite{mackey1989unitary}.

	\bibliographystyle{apsrev4-2}
	\bibliography{Reciprocal_Ref}

\appendix
\clearpage
\newpage

\onecolumngrid
\begin{center}
	{\large \bfseries Supplementary Materials for ``Flux-Mediated Correspondence Between Real- and Momentum-Space Nonsymmorphicity''}
\end{center}
\vspace{2ex} 

\begingroup
\let\oldcontentsline\contentsline
\renewcommand\contentsline[4]{%
	\ifx\@currenvir\@currenvir 
	\else
	\fi
	\def\maybeReference{#2}%
	\def\ReferenceString{Reference}%
	\ifx\maybeReference\ReferenceString
	\else
	\oldcontentsline{#1}{#2}{#3}{#4}%
	\fi
}

\setcounter{secnumdepth}{2}
\tableofcontents

\setcounter{section}{0}
\renewcommand{\thesection}{S\Roman{section}}

\setcounter{equation}{0}
\renewcommand{\theequation}{S\arabic{equation}}

\setcounter{figure}{0}
\renewcommand{\thefigure}{S\arabic{figure}}

\setcounter{table}{0}
\renewcommand{\thetable}{S\arabic{table}}

\setcounter{page}{1}

\section{The canonical decomposition of multipliers}
Let us start with introducing the presentation of crystallographic groups. Each crystallographic group $\Gamma$ belongs to an arithmetic class $c$, which specifies how the point group $G$ of $\Gamma$ acts on the translation subgroup $L$. Here, $L$ is also referred to as a lattice. In more detail, the arithematic class is represented by two ingredients: 1) A set of primitive lattice vectors $\{\bm{e}_i\}$ that spans the lattice $L$, and 2) The point group $G$ is a finite subgroup of $\mathrm{O}(3)$. Thus, an element $g$ of $G$ acts naturally on a lattice vector $\bm{t}$ of $L$ as $g\bm{t}$. A crystallographic group $\Gamma$ is specified in the arithmetic class $c$ by a fractional lattice translation $\bm{\tau}_g$ associated to each point group element $g$. The fractional translations satisfy the relation:
\begin{equation}
	\bm{\omega}(g_2,g_1)=\delta \bm{\tau}(g_2,g_1)=g_2\bm{\tau}_{g_1}-\bm{\tau}_{g_2g_1}+\bm{\tau}_{g_2} \in L.
\end{equation}
Here, $\bm{\omega}$ as a function from $G\times G$ to $L$ is referred to as the 2-cocycle of the space group $\Gamma$. Accordingly, $\Gamma$ can be represented as the twisted semi-direct product, namely $\Gamma=L\rtimes_{(c,\bm{\omega})} G$, where each group element is represented as $\gamma=(\bm{t},g)$ with $\bm{t}\in L$ and $g \in G$ and the multiplication is given by
\begin{equation}
	(\bm{t}_2, g_2)(\bm{t}_1, g_1)=(\bm{t}_2+g_2\bm{t}_1+\bm{\omega}(g_2, g_1), g_2g_1).
\end{equation}

We can define  a $G$-action on the space of all $\nu$-representations of $L$. For a representation $U$, $\gamma$ transforms $U$ to $^\gamma U$ given by
\begin{equation}
	^\gamma U(\bm{t}):=\vartheta(\gamma|\bm{t})U(\gamma^{-1}\bm{t}\gamma)=\vartheta(\gamma|\bm{t})U(g^{-1}\bm{t})
\end{equation}
with
\begin{equation}\label{eq:theta}
	\vartheta(\gamma|\bm{t})=\frac{\nu(\gamma^{-1},\bm{t})\nu(\gamma^{-1}\bm{t},\gamma)}{\nu(\gamma^{-1},\gamma)}.
\end{equation}
where we have used the fact that $\gamma^{-1}(\bm{t},e)\gamma=(g^{-1}\bm{t}, e)$ with $e$ the identity of $G$. Here,  $\vartheta$ satisfies the two identities,
\begin{equation}\label{eq:multi-theta}
	\vartheta(\gamma|\bm{t}_1)\vartheta(\gamma|\bm{t}_2)=\frac{\sigma(\bm{t}_1,\bm{t}_2)}{\sigma(g^{-1}\bm{t}_1,g^{-1}\bm{t}_2)}\vartheta(\gamma|\bm{t}_1+\bm{t}_2)
\end{equation}
and
\begin{equation}\label{eq:equi-theta}
	\vartheta(\gamma_2|\bm{t})\vartheta(\gamma_1|g_2^{-1}\bm{t})=\vartheta(\gamma_2\gamma_1|\bm{t}),
\end{equation}
corresponding to the multiplications of the two arguments, respectively. Here, $\sigma$ is just the restriction of $\nu$ on the subgroup $L$, namely $\sigma=\nu|_{L\times L}$. From the first, one can see that $^\gamma U$ is indeed a $\sigma$-representation of $L$, and from the second it is clear that the above defined $\Gamma$-action is a left group action.

To focus on the point group $G$, let us introduce
\begin{equation}
	\eta(g|\bm{t}):=\vartheta((0,g)|\bm{t})
\end{equation}
by restricting the first argument of $\vartheta$ on the embedding of the point group $G$, and
\begin{equation}
	\alpha(g_2,g_1):=\nu((0,g_2), (0,g_1))
\end{equation}
by restricting both arguments of $\nu$ on the embedding of $G$ in $\Gamma$. Accordingly, we have the following two significant identities,
\begin{equation}\label{eq:const_1}
	\frac{\sigma(g^{-1}\bm{t}_1,g^{-1}\bm{t}_2)}{\sigma(\bm{t}_1,\bm{t}_2)} = \frac{\eta(g|\bm{t}_1 + \bm{t}_2)}{\eta(g|\bm{t}_1)\eta(g|\bm{t}_2)}, \\ 
\end{equation}
and
\begin{equation}\label{eq:const_2}
	\frac{\eta(g_2g_1|\bm{t})}{\eta(g_2|\bm{t})\eta(g_1|g_2^{-1}\bm{t})} = \frac{\sigma(\bm{\omega}(g_2,g_1),\bm{t})}{\sigma(\bm{t},\bm{\omega}(g_2,g_1))}.
\end{equation}
The first is clear from Eq.~\eqref{eq:multi-theta}. The second can be derived from Eq.~\eqref{eq:equi-theta} immediately with this identity,
\begin{equation}
	\vartheta((\bm{t}',g)|\bm{t})=\frac{\sigma(\bm{t},\bm{t}')}{\sigma(\bm{t}',\bm{t})}\eta(g|\bm{t}).
\end{equation}
The identity is derived as follows. First, plugging $(\bm{t},e)(0,g)=(\bm{t},g)$ into Eq.~\eqref{eq:equi-theta}, we obtain
\begin{equation}
	\vartheta((\bm{t}',g)|\bm{t})=\vartheta((\bm{t}',e)|\bm{t})\eta(g|\bm{t}).
\end{equation}
Then, from the definition of $\vartheta$, Eq.~\eqref{eq:theta}, it is clear that $ \vartheta((\bm{t}',e)|\bm{t})={\sigma(\bm{t},\bm{t}')}/{\sigma(\bm{t}',\bm{t})}$.

In fact, any multiplier can be transformed to the canonical form
\begin{equation}\label{eq:decomposition}
	\nu(\gamma_2,\gamma_1) = \sigma(\bm{t}_2, g_2 \bm{t}_1) \sigma(\bm{t}_2 + g_2\bm{t}_1, \bm{\omega}(g_2, g_1))\eta^{-1}(g_2|g_2 \bm{t}_1)\alpha(g_2, g_1). 
\end{equation}
In addition to \eqref{eq:const_1} and \eqref{eq:const_2}, the third consistency equation,
\begin{equation}
	\frac{\alpha(g_3,g_2)\alpha(g_3g_2,g_1)}{\alpha(g_3,g_2g_1)\alpha(g_2,g_1)}  =  \eta^{-1}(g_3,g_3 \bm{\omega}(g_2,g_1)) \frac{\sigma(g_3 \bm{\omega}(g_2,g_1),\bm{\omega}(g_3,g_2 g_1))}{\sigma(\bm{\omega}(g_3,g_2),\bm{\omega}(g_3g_2,g_1))}, \label{decomposed-cocycle-3}
\end{equation}
should be satisfied so that \eqref{eq:decomposition} is an eligible multiplier.

\section{Rationality of multipliers}
Let the calligraphic $\mathcal{L}$ denote the algebra of translation operators $U_{\bm{t}}$ and the italic $L$ be the Abelian group formed by translations $\bm{t}$.

The commutation relation of two translation operators $U_{\t_1}$ and $U_{\t_2}$ is given by
\begin{equation}
	U(\t_1)U(\t_2)=\mathcal{W}(\t_1,\t_2)U(\t_2)U(\t_1),
\end{equation}
where
\begin{equation}
	\mathcal{W}(\t_1,\t_2)=\frac{\sigma(\t_1,\t_2)}{\sigma(\t_2,\t_1)}.
\end{equation}
with
\begin{equation}
	\mathcal{W}(\t_1,\t_2)=\frac{1}{\mathcal{W}(\t_2,\t_1)}.
\end{equation}
From the cocycle equation of $\sigma$, one can show $\mathcal{W}$ is bi-linear for both arguments,
\begin{equation}
	\mathcal{W}(\t_1+\t_2,\t_3)=\mathcal{W}(\t_1,\t_3)\mathcal{W}(\t_2,\t_3),\quad  \mathcal{W}(\t_1,\t_2+\t_3)=\mathcal{W}(\t_1,\t_2)\mathcal{W}(\t_1,\t_3).
\end{equation}

Then, we introduce the flux form $\bm{\Phi}$ through
\begin{equation}
	\mathcal{W}(\t_1,\t_2)=e^{2\pi i \bm{\Phi}(\t_1,\t_2)}
\end{equation}
Clearly, $\bm{\Phi}$ is skew-symmetric and bi-linear modulo $\mathbb{Z}$, i.e., for any $\t_1,\t_2,\t_3 \in \L$,
\begin{equation}
	\bm{\Phi}(\t_1,\t_2)=-\bm{\Phi}(\t_2,\t_1) \mod \mathbb{Z}
\end{equation}
and
\begin{equation}
	\bm{\Phi}(\t_1+\t_2,\t_3)=\bm{\Phi}(\t_1,\t_3)+\bm{\Phi}(\t_2,\t_3) \mod \mathbb{Z},\quad \bm{\Phi}(\t_1,\t_2+\t_3)=\bm{\Phi}(\t_1,\t_2)+\bm{\Phi}(\t_1,\t_3) \mod \mathbb{Z}.
\end{equation}

Choosing a basis $\{\bm{e}_1,\bm{e}_2,\bm{e}_3\}$ of the lattice $L$, we can represent the flux form $\bm{\Phi}$ as the matrix $\Phi$ with
\begin{equation}
	\Phi_{ij}=\bm{\Phi}(\bm{e}_i,\bm{e}_j),
\end{equation}
The matrix $\Phi$ is referred to as the flux matrix and can be chosen in the skew-symmetric form,
\begin{equation}
	\Phi=\begin{bmatrix}
		0 & \phi^3 & -\phi^2\\
		-\phi^3 & 0 & \phi^1\\
		\phi^2 & -\phi^1 & 0
	\end{bmatrix},
\end{equation}
where $\phi^i $ may be interpreted as the flux through the area spanned by $\bm{e}_j$ and $\bm{e}_k$ if $\epsilon_{ijk}=1$.

We assume that all $\phi^i$ are rational numbers. If a multiplier of a space group corresponds to rational fluxes $\phi^i$, projective representations with this multiplier are called rational projective representations.

\section{The central sublattice}
Let us consider the center $\mathcal{Z}$ of the algebra $\mathcal{L}$ of translation operators, and let $Z\subset L$ be the corresponding underlying subgroup of $L$.  That is, each element $U_{\bm{z}}$ with $\bm{z}\in Z$ commutes with all translation operators $U_{\t}$ for all $\bm{t}\in L$. In terms of the flux form, $\bm{z}\in Z$ if and only if $e^{2\pi i \bm{\Phi}(\t,\bm{z})}=1$ for all $\bm{t}\in L$. Specifically,
\begin{equation}
	\Phi z=0\mod \mathbb{Z} \quad \mathrm{or} \quad \epsilon_{ijk}\phi^j z^k=0\mod \mathbb{Z}.
\end{equation}
Here, $z$ denotes the integral column vector representing $\bm{z}$ in the basis $\{\bm{e}_i\}$. 

It is clear from the linearity of the equation that $Z$ is a three-dimensional lattice, and it is referred to as the central sublattice. To explicitly characterize $Z$, we proceed to exhibit a set of primitive translations $\bm{a}_i$ with $i=1,2,3$ for the center sublattice $Z$.

We can assume that $0< \phi_i <1$ and express each $\phi^i$ as a ratio of two positive integers $m_i$ and $n_i$,
\begin{equation}
	\phi_i =\frac{m_i}{n_i}.
\end{equation}
Here, we require that $m_i$ and $n_i$ are coprime, i.e., $(m_i, n_i)=1$. Let $p$ be the least common multiple of $n_1$, $n_2$ and $n_3$ and $q$ be the greatest common divisor of $pm_i/n_i$, i.e., $q=\mathrm{gcd}(pm_1/n_1,pm_2/n_2,pm_3/n_3)$. Then, the flux matrix $\Phi$ can be written as
\begin{equation}
	\Phi=\frac{q}{p}\begin{bmatrix}
		0 & \xi^3 & -\xi^2\\
		-\xi^3 & 0 & \xi^1\\
		\xi^2 & -\xi^1 & 0
	\end{bmatrix}.
\end{equation}
Here, $\xi^i$ are three coprime integers, with $\mathrm{gcd}(\xi^1,\xi^2,\xi^3)=1$, and $q$ and $p$ are coprime with $(q,p)=1$.
Let us introduce $\Xi=p\Phi/q$, namely
\begin{equation}
	\Xi=\begin{bmatrix}
		0 & \xi^3 & -\xi^2\\
		-\xi^3 & 0 & \xi^1\\
		\xi^2 & -\xi^1 & 0
	\end{bmatrix},
\end{equation}
and call $\Xi$ integer-normalized flux matrix. Accordingly, $\bm{z}\in Z$ if and only if
\begin{equation}
	\Xi z=0 \mod p.
\end{equation}
To solve the linear congruence equations, since $\Xi$ is skew symmetric, we can perform the congruence transformation,
\begin{equation}
	\Omega^T\Xi\Omega=\begin{bmatrix}
		0 & 1 & 0\\
		-1 & 0 & 0\\
		0 & 0 & 0
	\end{bmatrix}
\end{equation}
with $\Omega\in \mathrm{GL}(3,\mathbb{Z})$.
Here, the entry $1$ of the matrix on the right-hand side originates from $\mathrm{gcd}(\xi^1,\xi^2,\xi^3)=1$.  Accordingly, the flux matrix can be written in the canonical form,
\begin{equation}
	\Omega^T \Phi \Omega=\frac{q}{p}\begin{bmatrix}
		0 & 1 & 0\\
		-1 & 0 & 0\\
		0 & 0 & 0
	\end{bmatrix}.
\end{equation}
Then, it is obvious that the primitive translations $\bm{a}_i=a_i^{~j}\bm{e}_j$ of the central sublattice $Z$ can be chosen as
\begin{equation}
	a_1=\Omega\begin{bmatrix}
		p \\
		0\\
		0
	\end{bmatrix},\quad a_2=\Omega\begin{bmatrix}
		0 \\
		p \\
		0
	\end{bmatrix},\quad a_3=\Omega\begin{bmatrix}
		0 \\
		0\\
		1
	\end{bmatrix}.
\end{equation}
If one or two of $\phi$'s are zero, the flux matrix can also be written in the canonical form with the basis $\bm{a}_i$ for $Z$ defined accordingly.

\section{Construction of translation multipliers trivial on the central sublattice}
Our next task is to construct a multiplier $\sigma$ for the flux form $\bm{\Phi}$ with $\sigma|_{Z\times Z}=1$. Let us assume the form of the multiplier,
\begin{equation}
	\sigma(\t_1,\t_2)=e^{2\pi i \bm{A}(\t_1,\t_2)}=e^{2\pi i\,  t^T_1 At_2}.
\end{equation}
Here, $\bm{A}$ is a bilinear form, referred to as the connection form, and the connection matrix $A$ is given by $A_{ij}=\bm{A}(\bm{e}_i,\bm{e}_j)$. Then, the flux matrix is related to the connection matrix by
\begin{equation}
	\Phi=A-A^T \mod \mathbb{Z}.
\end{equation}
A possible choice of $A$ is $A=\Phi^+$. Here, for any matrix $M$, $M^+$ denotes the upper-right triangular matrix of $M$. It is obvious that $\Phi=\Phi^+-(\Phi^+)^T$ as $\Phi$ is skew symmetric. However, such an $A$ is not trivial on the normal sublattice.

Let us introduce the matrix $V=(a_1,a_2,a_3)$ with $a_i$ being the column vector of $\bm{a}_i$, namely
\begin{equation}
	V=\Omega\begin{bmatrix}
		p & 0 & 0\\
		0 & p & 0\\
		0 & 0 & 1
	\end{bmatrix}.
\end{equation}
Then,  $\sigma$ is trivial on the normal sublattice if and only if
\begin{equation}
	V^TAV\in M_3(\mathbb{Z}).
\end{equation}
Clearly, $A=\Phi^+$ does not satisfy the requirement. To construct a connection with $V^TAV\in M_3(\mathbb{Z})$, it is significant to observe that
\begin{equation}
	V^T\Phi V\in M_3(\mathbb{Z}),
\end{equation}
which is equivalent to the definition of the central sublattice. This motivates us to propose
\begin{equation}
	A=(V^{-1})^T(V^T\Phi V)^+ V^{-1}.
\end{equation}
Noticing that $V^T\Phi V=(V^T\Phi V)^+-[(V^T\Phi V)^+]^T$,
we can immediately verify that $A$ is indeed an eligible connection matrix with $\Phi=A-A^T$.

\section{The finite Heisenberg algebra from a projective representation of translations}
From the expression of the exhibited base vectors $\bm{a}_i$ of the central sublattice $Z$, we can see that if we transform the primitive translations $\bm{e}_i$ of the lattice $L$ by $\Omega$ to obtain a new basis of $L$, 
\begin{equation}
	(\bm{e}'_1,\bm{e}'_2,\bm{e}'_3)=(\bm{e}_1,\bm{e}_2,\bm{e}_3)\Omega,
\end{equation}
the primitive translations of the normal sublattice can be expressed as $\bm{a}_1=p\bm{e}'_1$, $\bm{a}_2=p\bm{e}'_2$ and $\bm{a}_3=\bm{e}'_3$. Since $\Omega$ is an element of $GL(3,\mathbb{Z})$, $\{\bm{e}'_i\}$ is also a set of primitive lattice vectors for $L$ and may be referred to as proper primitive lattice vectors under the flux matrix $\Phi$. Under this basis, the connection form and the flux form are, respectively, represented by the matrices,
\begin{equation}
	A'=\Omega^T A \Omega=\begin{bmatrix}
		0 & q/p & 0\\
		0 & 0 & 0\\
		0 & 0 & 0
	\end{bmatrix},\quad \Phi'=\Omega^T \Phi \Omega=\begin{bmatrix}
		0 & q/p & 0\\
		-q/p & 0 & 0\\
		0 & 0 & 0
	\end{bmatrix}.
\end{equation}

Consequently, the quotient  of the lattice by factoring out the central sublattice is isomorphic to the finite Heisenberg algebra $\mathrm{Heis}_{q}(\mathbb{Z}_p\times\mathbb{Z}_p)$ at level $q$,
\begin{equation}
	\mathcal{L}/\mathcal{Z}\cong \mathrm{Heis}_{q}(\mathbb{Z}_p\times\mathbb{Z}_p),
\end{equation}
which is elucidated in the following.

Each $\bm{t}\in L$ can be uniquely decomposed as
\begin{equation}
	\bm{t}=\bm{z}+\bar{\bm{t}}
\end{equation}
with $\bm{z}\in Z$. Here,
\begin{equation}
	\bar{\bm{t}}=\tilde{\bar{t}}^{i}\bm{e}'_i
\end{equation}
with $\tilde{\bar{t}}^1,\tilde{\bar{t}}^2\in\{0,1,2,\cdots,p-1\}$ and $\tilde{\bar{t}}^3=0$. Thus, $(\tilde{\bar{t}}^1,\tilde{\bar{t}}^2)$ or $\bar{\bm{t}}$ labels an element of $L/Z$, and it is natural to introduce
\begin{equation}
	\bar{U}_{\bar{\bm{t}}}=U_{\bar{\bm{t}}}.
\end{equation}
where $\bar{U}_{\bar{\bm{t}}}$ represents an element of $\mathcal{L}/\mathcal{Z}$. Accordingly, we can introduce the generators,
\begin{equation}
	P=\bar{U}_{\bar{\bm{e}}'_1},\quad Q=\bar{U}_{\bar{\bm{e}}'_2}.
\end{equation}

We observe the following multiplication relations,
\begin{equation}
	U_{\bm{e}'_1}U_{\bm{e}'_2}=e^{\frac{2\pi i q}{p}}U_{\bm{e}'_1+\bm{e}'_2}, \quad U_{\bm{e}'_2}U_{\bm{e}'_1}=U_{\bm{e}'_1+\bm{e}'_2},
\end{equation}
and
\begin{equation}
	U_{\bm{e}'_i}U_{\bm{e}'_i}=U_{2\bm{e}'_i}
\end{equation}
with $i=1,2$.

Thus, the algebraic relations of the generators of $\mathcal{L}/\mathcal{Z}$ are given by
\begin{equation}\label{eq:Heisen}
	PQ=e^{\frac{2\pi i q}{p}}QP,\quad P^p=Q^p=1,
\end{equation}
Note that $\overline{p\bm{e}'_i}=0$ and therefore $P^q=Q^p=1$. These algebraic relations exactly correspond to the definition of the finite Heisenberg algebra $\mathrm{Heis}_{q}(\mathbb{Z}_p\times\mathbb{Z}_p)$ at level $q$. That is, $\mathrm{Heis}_{q}(\mathbb{Z}_p\times\mathbb{Z}_p)$ is generated by $P$ and $Q$ satisfying the above algebraic relations.

The finite Heisenberg algebra has the remarkable property that it has a unique  irreducible unitary representation, which is $p$ dimensional if $(p,q)=1$. For a concrete matrix representation, we can choose $P$ as the unit cyclic permutation of $p$ elements, and $Q$ a diagonal matrix with the $A$th diagonal entry being $e^{\frac{2\pi i q A}{p}}$, i.e.,
\begin{equation}
	P_{AB}=\delta_{\overline{A+1},\overline{B}},\quad Q_{AB}=e^{\frac{2\pi i q A}{p}}\delta_{AB},
\end{equation}
where $A,B\in \{1,2,\cdots, p-1\}$ and $\overline{A}$ is the corresponding element of $A$ in $\mathbb{Z}_p$. Accordingly, we specify the Heisenberg algebra as
\begin{equation}
	U_H(\bar{\bm{t}})=Q^{\tilde{\bar{t}}^2}P^{\tilde{\bar{t}}^1}=Q^{\tilde{t}^2}P^{\tilde{t}^1}
\end{equation}
for all $\bar{\bm{t}}\in L/Z$. Since the Heisenberg algebra has a unique irreducible representation, any matrix representation is a multiple of this matrix representation up to a unitary transformation.

\section{The Brillouin zone as the space of irreducible representations of lattice translations}
In the ordinary representation theory of space groups, the Brillouin zone is defined as the collection of all irreducible representations of the translation group $L$. This is still true for projective representation of the space group. With the multiplier $\sigma$, all irreducible representations of $\mathcal{L}$ have a one-to-one correspondence with all irreducible representations of the center $\mathcal{Z}$. Since $\sigma|_{Z\times Z}=1$, the Brillouin zone is just the connection of all irreducible representations of $Z$ as an Abelian group.

Let us recall that  $Z$ is spanned by the primitive lattice vectors of $Z$,
\begin{equation}
	\bm{a}_1=p\tilde{\bm{e}}_1,\quad  \bm{a}_2=p\tilde{\bm{e}}_2,\quad  \bm{a}_3=\tilde{\bm{e}}_3,
\end{equation}
i.e.,
\begin{equation}
	Z=\{ \bm{z}=\sum_{i=1}^3 n^\alpha\bm{a}_\alpha,~ n^\alpha\in \mathbb{Z}\}.
\end{equation}
Accordingly, the reciprocal lattice $Z_F$ of $Z$ is spanned by the primitive reciprocal lattice vectors
\begin{equation}
	\bm{Q}^\alpha=\frac{\epsilon^{\alpha\beta\gamma}\bm{a}_\beta\times\bm{a}_\gamma}{\bm{a}_1\cdot (\bm{a}_2\times\bm{a}_3)},
\end{equation}
with the property
\begin{equation}
	\bm{Q}^\alpha\cdot \bm{a}_\beta=\delta^\alpha_\beta.
\end{equation}
Explicitly, $Z_F$ can be presented as
\begin{equation}
	Z_F=\{ \bm{K}=\sum_{i=1}^3K_\alpha\bm{Q}^\alpha,~ K_\alpha\in \mathbb{Z}\}.
\end{equation}
The Brillouin zone is just the fundamental domain under the reciprocal lattice translations $Z_F$, namely
\begin{equation}
	\widehat{Z}=\mathbb{R}^3/Z_F\approx T^3_F,
\end{equation}
where it is indicated that $\widehat{Z}$ is  topologically the 3-torus $T^3_F$.

For each wave vector $\bm{k}$, the corresponding irreducible $\sigma$-representation $U_{\bm{k}}(\bm{t})$ of $\mathcal{L}$ is given by 
\begin{equation}
	U_{\bm{k}}(\bm{t})=e^{2\pi i \bm{k}\cdot\bm{t}} U_H(\bar{\bm{t}}).
\end{equation}
For any $\bm{K}\in Z_F$, let us show that $U_{\bm{k}+\bm{K}}$ is equivalent to $U_{\bm{k}}$ by a unitary transformation. First, the translation by the reciprocal lattice vector $\bm{K}$ leads to an additional phase, i.e.,
\begin{equation}
	U_{\bm{k}+\bm{K}}(\bm{t})=e^{2\pi i\bm{K}\cdot \bm{t}}U_{\bm{k}}(\bm{t}).
\end{equation}
It is straightforward to see the phase can be expressed as
\begin{equation}
	e^{2\pi i\bm{K}\cdot \bm{t}}=e^{\frac{2\pi i}{p}(K_1 \tilde{t}^1+K_2\tilde{t}^2)}.
\end{equation}
Then, one can verify that the phase can be induced from the unitary transformation $V_{\bm{K}} U_{\bm{k}}(\bm{t})V^\dagger_{\bm{K}}$ with
\begin{equation}
	V_{\bm{K}}=Q^{sK_1}P^{s K_2 }
\end{equation}
by using the algebraic relations of the finite Heisenberg algebra in Eq.~\eqref{eq:Heisen}. 
Recall that $(p,q)=1$ if and only there exist integers $s$ and $r$ so that  
\begin{equation}
	rp+sq=1,
\end{equation}
and $s$ can be chosen as any integer satisfying this equation.

The equivalence between $U_{\bm{k}+\bm{K}}$ and $U_{\bm{k}}$ is consistent with our claim that $\widehat{Z}=\mathbb{R}^3/Z_F$ is the space of all irreducible $\sigma$-representations of  the translation group $L$. The fact that these representations are indeed all irreducible representations can be justified by Mackey's representation theory. 

\section{The $G$ action  on the Brillouin zone }
The Brillouin zone is solely determined by the central sublattice $Z$, as seen from the following. The irreducible representation $
U_{\bm{k}}(\bm{t})=e^{2\pi i \bm{k}\cdot\bm{t}} U_H(\bar{\bm{t}})$ is reduced to 
\begin{equation}
	\rho_{\bm{k}}(\bm{z})=U_{\bm{k}}(\bm{z})=e^{2\pi i \bm{k}\cdot \bm{z}}
\end{equation}
on the central sublattice because $U_{H}(\bm{z})=1$ for all $\bm{z}\in Z$. Then, all $\bm{k}$ in the Brillouin zone have a one-to-one correspondence with all the irreducible representations of the central lattice $Z$ as obviously $\rho_{\bm{k}}=\rho_{\bm{k}+\bm{K}}$ for all $\bm{K}\in Z_F$ from $\bm{K}\cdot \bm{z}\in \mathbb{Z}$.

Furthermore, it is significant to notice that the action of the point group $G$ on the Brillouin zone only depends on its action on the central sublattice $Z$. This relies on the fact that $Z$ is a normal subgroup of  the space group $\Gamma$, i.e., the central sublattice is invariant under all $g\in G$. This is a consequence of the invariance of the flux form $\bm{\Phi}$ under the $G$ action. In more detail, for any $\bm{z}\in Z$, $g\in G$ and $\bm{t}\in L$, the invariance implies
\begin{equation}
	e^{2\pi i \bm{\Phi}(g\bm{z},\bm{t})}=e^{2\pi i \bm{\Phi}(\bm{z},g^{-1}\bm{t})}=1,
\end{equation}
from which we see $g\bm{z}\in Z$ for all $\bm{z}\in Z$.

Then, the $G$-action on the Brillouin zone can be derived from the $G$-action on the irreducible representations $\rho_{\bm{k}}$, which is given by
\begin{equation}\label{eq:G-rho}
	^g \rho_{\bm{k}}(\bm{z})=\eta(g|\bm{z})\rho_{\bm{k}}(g^{-1}\bm{z})
\end{equation}
as can be inferred from the definition of $\eta$. Thus, the key step is to derive a more concrete form of $\eta(g|\bm{z})$ for $g\in G$ and $\bm{z}\in Z$. The first two consistency equations are significantly simplified on the central lattice to be 
\begin{eqnarray}
	&& \eta(g|\bm{z}_1 + \bm{z}_2) = \eta(g|\bm{z}_1)\eta(g|\bm{z}_2), \\ \label{eq:eta_1}
	&& \eta(g_2 g_1|\bm{z})=\eta(g_2|\bm{z})\eta(g_1|g_2^{-1}\bm{z}) ,
\end{eqnarray}
as we can choose $A$ with $\sigma|_{Z\times Z}=1$. The first equation just states that the phase $\eta(g|\bm{z})$ is multiplicatively linear in the second argument, and therefore $\eta(g|\bm{z})$ takes the general form,
\begin{equation}
	\eta(g|\bm{z})=e^{2\pi i \bm{\kappa}_g\cdot \bm{z}},
\end{equation}
where $\bm{\kappa}_g$ is a vector depending on $g$. Substituting the general form into the second equation, we obtain
\begin{equation}
	\exp {\left[2\pi i ( g_2\bm{\kappa}_{g_1}- \bm{\kappa}_{g_2g_1}+\bm{\kappa}_{g_2})\cdot \bm{z}\right]}=1
\end{equation}
for all $\bm{z}\in Z$. Thus, the exponent should be valued in the reciprocal lattice $Z_F$ and therefore can be defined as the $2$-cocycle $\bm{\omega}_F(g_2,g_1)$ valued in $Z_F$, namely
\begin{equation}
	\bm{\omega}_F(g_2,g_1)=g_2\bm{\kappa}_{g_1}- \bm{\kappa}_{g_2g_1}+\bm{\kappa}_{g_2} \in Z_F.
\end{equation}
Plugging $\eta(g|\bm{z})=e^{2\pi i \bm{\kappa}_g\cdot \bm{z}}$ into Eq.~\eqref{eq:G-rho}, we find that $g$ acts on the momentum space as
\begin{equation}
	g:~\bm{k}\mapsto g\bm{k}+\bm{\kappa}_{g}.
\end{equation}
We observe that $\bm{\kappa}_{g}$ can be interpreted as the momentum-space fractional translation on the reciprocal lattice $Z_F$. 


\section{$G$ representations and group cohomologies}
In this section, we explicitly formulate the matrix representations of the point group under various bases for real space and momentum space and also for tensor spaces in this work. For each representation, we explicitly describe the corresponding coboundary operators for the group cohomology with various natural coefficients.

In real space, the basis $\bm{a}_\alpha$ for $Z$ and the basis $\bm{e}_i$ are related by
\begin{equation}
	\bm{a}_\alpha=\bm{e}_i V_{i\alpha}.
\end{equation}
In momentum space, the dual bases $\bm{Q}_\alpha$ and $\bm{G}_i$ are related by
\begin{equation}
	\bm{Q}_\alpha= \bm{G}_{i}[V^{-1}]^T_{i\alpha}.
\end{equation}

We denote the representation of $G$ on the lattice $L$ by $E$ and on the central sublattice $Z$ by $\tilde{E}$, i.e.,
\begin{equation}
	g\bm{e}_i= \bm{e}_j [E(g)]_{ji},\quad g\bm{a}_\alpha= \bm{a}_\beta [\tilde{E}_g]_{\beta\alpha},
\end{equation}
for all $g\in G$.
They are related by
\begin{equation}
	\tilde{E}(g)=V^{-1}E_g V.
\end{equation}
Then, the representations on the dual bases $L_F$ and $Z_F$ are, respectively, given by
\begin{equation}
	g\bm{G}_i= \bm{G}_j [E^T_{g^{-1}}]_{ji},\quad g\bm{Q}_\alpha= \bm{Q}_\beta [\tilde{E}^T_{g^{-1}}]_{\beta\alpha}.
\end{equation}

Let $\bm{f}^{(n)}$ be a vector-valued function of $G^n$. Then, the coboundary operator $\delta$ is defined as
\begin{equation}
	\begin{split}
		\delta\bm{f}^{(n)}(g_{n+1},g_n,g_{n-1},\cdots, g_1)= & g_{n+1}\bm{f}^{(n)}(g_n,g_{n-1},\cdots, g_1)+(-1)^{n}\bm{f}^{(n)}(g_{n+1},g_{n},\cdots, g_2)\\ &+\sum_{i=0}^{n-1}(-1)^{i+1}\bm{f}^{(n)}(g_{n+1},\cdots, g_{n+1-i}g_{n-i},\cdots, g_1).
	\end{split}
\end{equation}
For instance, the low-order examples are listed below.
\begin{equation}
	\delta \bm{f}^{(0)} (g)=g\bm{f}^{(0)}-\bm{f}^{(0)}.
\end{equation}
\begin{equation}
	\delta \bm{f}^{(1)} (g_2,g_1)=g_2\bm{f}^{(1)}(g_1)-\bm{f}^{(1)}(g_2g_1)+\bm{f}^{(1)}(g_2).
\end{equation}
\begin{equation}
	\delta \bm{f}^{(2)} (g_3,g_2,g_1)=g_3\bm{f}^{(2)}(g_2,g_1)-\bm{f}^{(2)}(g_3g_2,g_1)+\bm{f}^{(2)}(g_3,g_2g_1)-\bm{f}^{(2)}(g_3,g_2).
\end{equation}
If we represent a vector $\bm{f}$ by a column vector $f$ in the basis $\bm{G}_i$, then the coboundary operator is given by
\begin{equation}
	\begin{split}
		\delta{f}^{(n)}(g_{n+1},g_n,g_{n-1},\cdots, g_1)= & E_{g^{-1}_{n+1}}^T{f}^{(n)}(g_n,g_{n-1},\cdots, g_1)+(-1)^{n}{f}^{(n)}(g_{n+1},g_{n},\cdots, g_2)\\ &+\sum_{i=0}^{n-1}(-1)^{i+1}{f}^{(n)}(g_{n+1},\cdots, g_{n+1-i}g_{n-i},\cdots, g_1).
	\end{split}
\end{equation}
If we represent $\bm{f}$ by a column vector $\tilde{f}$ in the basis $\bm{Q}_\alpha$, then
\begin{equation}
	\begin{split}
		\delta{\tilde{f}}^{(n)}(g_{n+1},g_n,g_{n-1},\cdots, g_1)= & \tilde{E}_{g^{-1}_{n+1}}^T{\tilde{f}}^{(n)}(g_n,g_{n-1},\cdots, g_1)+(-1)^{n}{\tilde{f}}^{(n)}(g_{n+1},g_{n},\cdots, g_2)\\ &+\sum_{i=0}^{n-1}(-1)^{i+1}{\tilde{f}}^{(n)}(g_{n+1},\cdots, g_{n+1-i}g_{n-i},\cdots, g_1).
	\end{split}
\end{equation}

Then, we can define cohomology groups by the coboundary operators. If we restrict the vector on the reciprocal lattice $L_F$, since $L_F$ is closed under the group action, we can define $H^n(G,L_F)$ as
\begin{equation}
	H^n(G,L_F)=\frac{Z^n(G,L_F)}{B^n(G,L_F)}.
\end{equation}
with $Z^n(G,L_F)$ consisting of all $n$-cocycles, namely all $L_F$-valued $n$-variable functions that vanish under $\delta$, and $B^n(G,L_F)$ consists of all coboundaries of $L_F$-valued $(n-1)$-variable functions. Since $\delta^2=0$ by construction, $B^n(G,L_F)$ is a subgroup of $Z^n(G,L_F)$ and therefore we can define  $H^n(G,L_F)=\frac{Z^n(G,L_F)}{B^n(G,L_F)}$ as the quotient group. 

Similarly, we can define
\begin{equation}
	H^n(G,Z_F)=\frac{Z^n(G,Z_F)}{B^n(G,Z_F)},\quad H^n(G,Z_F/L_F)=\frac{Z^n(G,Z_F/L_F)}{B^n(G,Z_F/L_F)}.
\end{equation} 
Note that the $G$ action preserves both $Z_F$ and $L_F$, and therefore we can mod the sublattice $L_F$ of $Z_F$, i.e.,
\begin{equation}
	0\rightarrow L_F\rightarrow Z_F \rightarrow Z_F/L_F \rightarrow 0,
\end{equation}
to define $Z_F/L_F$-valued functions.  They satisfy the long exact sequence,
\begin{multline}
	0\rightarrow H^0(G,L_F)\rightarrow H^0(G,Z_F) \rightarrow H^0(G,Z_F/L_F)\rightarrow H^1(G,L_F)\rightarrow H^1(G,Z_F) \rightarrow H^1(G,Z_F/L_F)\rightarrow \\ H^2(G,L_F)\rightarrow H^2(G,Z_F) \rightarrow H^2(G,Z_F/L_F)\rightarrow H^3(G,L_F)\rightarrow \cdots .
\end{multline}
It is clear that $\bm{\omega}_F\in Z^2(G,Z_F)$ and the corresponding cohomology class is $[\bm{\omega}_F]\in H^2(G,Z_F)$. Then, its image modulo $L_F$ is denoted as $[\overline{\omega}_F]$ in $H^2(G,Z_F/L_F)$.

In real space, we have
\begin{equation}
	0\rightarrow Z \rightarrow L \rightarrow L/Z \rightarrow 0,
\end{equation}
with $L/Z\cong Z_F/L_F\cong \mathbb{Z}_p\times \mathbb{Z}_p$. The long exact sequence is given by
\begin{multline}
	0\rightarrow H^0(G,Z)\rightarrow H^0(G,L) \rightarrow H^0(G,L/Z)\rightarrow H^1(G, Z)\rightarrow H^1(G, L) \rightarrow H^1(G,L/Z)\rightarrow \\ H^2(G,Z)\rightarrow H^2(G,L) \rightarrow H^2(G,L/Z)\rightarrow H^3(G,Z)\rightarrow \cdots .
\end{multline}
Note that if we represent a vector $\bm{t}$ as the column vector $t$ in the basis $\bm{e}_i$, the coboundary operator is defined as
\begin{equation}
	\begin{split}
		\delta{t}^{(n)}(g_{n+1},g_n,g_{n-1},\cdots, g_1)= & E_{g_{n+1}}{t}^{(n)}(g_n,g_{n-1},\cdots, g_1)+(-1)^{n}{t}^{(n)}(g_{n+1},g_{n},\cdots, g_2)\\ &+\sum_{i=0}^{n-1}(-1)^{i+1}{t}^{(n)}(g_{n+1},\cdots, g_{n+1-i}g_{n-i},\cdots, g_1).
	\end{split}
\end{equation}
for an associated column-vector valued function ${t}^{(n)}$.

In this work, we also use tensor-valued functions. For instance, a bi-linear form $\bm{F}(\bm{t}_2,\bm{t}_1)$ over $L$ can be expressed as $\bm{F}=F_{i j} \bm{G}_i \otimes \bm{G}_j$. Accordingly, the point group action is given by
\begin{equation}
	g\bm{F}=F_{i j} g\bm{G}_i \otimes g\bm{G}_j=F_{i j} g\bm{G}_{i'} [E^T_{g^{-1}}]_{i'i}\otimes \bm{G}_{j'} [E^T_{g^{-1}}]_{j'j}=[E^T_{g^{-1}}FE_{g^{-1}}]_{ij}\bm{G}_i \otimes \bm{G}_j
\end{equation}
Therefore, the point group action on the matrix representation of the bilinear form is given by
\begin{equation}
	gF=E^T_{g^{-1}}FE_{g^{-1}}
\end{equation}

Then, the corresponding coboundary operator for the associated matrix-valued functions is given by
\begin{equation}
	\begin{split}
		\delta{F}^{(n)}(g_{n+1},g_n,g_{n-1},\cdots, g_1)= & E_{g^{-1}_{n+1}}^T{F}^{(n)}(g_n,g_{n-1},\cdots, g_1)E_{g^{-1}_{n+1}}+(-1)^{n}{F}^{(n)}(g_{n+1},g_{n},\cdots, g_2)\\ &+\sum_{i=0}^{n-1}(-1)^{i+1}{F}^{(n)}(g_{n+1},\cdots, g_{n+1-i}g_{n-i},\cdots, g_1).
	\end{split}
\end{equation}
Particularly, we can restrict to symmetric (skew-symmetric) forms or symmetric (skew symmetric) matrices, since the subspace of matrices is preserved by the group action. Then, we can formulate the cohomology groups
\begin{equation}
	H^{n,c_F}(G, M^{(s)}_d(\mathbb{\mathbb{Z}})),\quad H^{n,c_F}(G, M^{(s)}_d(\mathbb{\mathbb{R}})),\quad H^{n,c_F}(G, M^{(s)}_d(\mathbb{\mathbb{R}/\mathbb{Z}}))
\end{equation}
which form a long exact sequence for 
\begin{equation}
	0\rightarrow M^{(s)}_d(\mathbb{\mathbb{Z}})\rightarrow M^{(s)}_d(\mathbb{\mathbb{R}})\rightarrow M^{(s)}_d(\mathbb{\mathbb{R}/\mathbb{Z}})\rightarrow 0.
\end{equation}
Here, the group action is indicated by the superscript $c_F$. $M^{(s)}_d(R)$ denote symmetric and anti-symmetric matrices for $s=0$ and $s=1$, respectively, with each entry in the Abelian group $R$.

If we express the bilinear form as $\tilde{F}_{\alpha \beta} \bm{Q}_\alpha \otimes \bm{Q}_\beta$. Then, the point group action is given by
\begin{equation}
	g \tilde{F}=\tilde{E}_{g^{-1}}^T{\tilde{F}}\tilde{E}_{g^{-1}}.
\end{equation}
The corresponding coboundary operator is given by
\begin{equation}
	\begin{split}
		\delta{\tilde{F}}^{(n)}(g_{n+1},g_n,g_{n-1},\cdots, g_1)= & \tilde{E}_{g^{-1}_{n+1}}^T{\tilde{F}}^{(n)}(g_n,g_{n-1},\cdots, g_1)\tilde{E}_{g^{-1}_{n+1}}+(-1)^{n}{\tilde{F}}^{(n)}(g_{n+1},g_{n},\cdots, g_2)\\ &+\sum_{i=0}^{n-1}(-1)^{i+1}{\tilde{F}}^{(n)}(g_{n+1},\cdots, g_{n+1-i}g_{n-i},\cdots, g_1).
	\end{split}
\end{equation}
To distinguish the group action in the definition of the cohomology groups, we may denote the cohomology groups as
\begin{equation}
	H^{n,\tilde{c}_F}(G, {M}^{(s)}_d(\mathbb{\mathbb{Z}})),\quad H^{n,\tilde{c}_F}(G, {M}^{(s)}_d(\mathbb{\mathbb{R}})),\quad H^{n,\tilde{c}_F}(G, {M}^{(s)}_d(\mathbb{\mathbb{R}/\mathbb{Z}})),
\end{equation}
which also satisfy the corresponding long exact sequence.

\section{$G$-invariant flux form and the cocycle $\overline{\Phi \omega}$ }

Accordingly, the point group $G$ acts on a multiplier $\sigma(\bm{t}_2,\bm{t}_1)$ as $\sigma_g(\bm{t}_2,\bm{t}_1)=\sigma(g^{-1}\bm{t}_2,g^{-1}\bm{t}_1)$. Then, Eq.~\eqref{eq:const_1}, which is copied  below for readers' convenience,

\begin{equation}
	\frac{\sigma(g^{-1}\bm{t}_2,g^{-1}\bm{t}_1)}{\sigma(\bm{t}_2,\bm{t}_1)} = \frac{\eta(g|\bm{t}_2 + \bm{t}_1)}{\eta(g|\bm{t}_2)\eta(g|\bm{t}_1)}, \label{decomposed-cocycle-1}
\end{equation}
can be interpreted as that the transformed multiplier $\sigma_R$ is related to the original multiplier $\sigma$ by a coboundary of $\eta(g|\bm{t})$. Consequently, the flux form $\bm{\Phi}$ is invariant under the action of the point group $G$. Thus,
\begin{equation}
	e^{2\pi i \bm{\Phi}(\bm{t}_2,\bm{t}_1)}=e^{2\pi i \bm{\Phi}(g\bm{t}_2, g\bm{t}_1)}
\end{equation}
for all $g\in G$. Alternatively, for the flux matrix $\Phi$, 
\begin{equation}
	E_g^T\Phi E_g-\Phi\in M_3(\mathbb{Z}),
\end{equation}
for all $g\in G$. 

We now verify that $\overline{\bm{\Phi}\bm{\omega}}$ is indeed a cocycle.
\begin{equation}
	\begin{split}
		\delta \Phi \omega(g_3,g_2,g_1) &=E_{g_3^{-1}}^T\Phi \omega(g_2,g_1)-\Phi \omega(g_3g_2,g_1)+\Phi\omega(g_3,g_2g_1)-\Phi\omega(g_3,g_2)\\
		&=E_{g_3^{-1}}^T\Phi \omega(g_2,g_1)-\Phi E_{g_3}\omega(g_2,g_1)+\Phi\delta^2\omega(g_3,g_2,g_1).
	\end{split}
\end{equation}

\begin{equation}
	\begin{split}
		\delta \Phi \omega(g_3,g_2,g_1) &=E_{g_3^{-1}}^T\Phi E_{g_3^{-1}} E_{g_3} \omega(g_2,g_1)-\Phi E_{g_3}\omega(g_2,g_1)\\
		&=(E_{g_3^{-1}}^T\Phi E_{g_3^{-1}}-\Phi) E_{g_3} \omega(g_2,g_1)
	\end{split}
\end{equation}

\section{The cocycle $\overline{\Lambda_{\Phi}}$ }

As having justified above, we can choose $\sigma$ to be trivial on the central sublattice $Z$, i.e., $\sigma|_{Z\times Z}=1$. Then, from the definition of symmetric $S$,
\begin{equation}\label{eq:S_def}
	\frac{\sigma(g^{-1}\bm{t}_2,g^{-1}\bm{t}_1)}{\sigma(\bm{t}_2,\bm{t}_1)}= e^{2\pi i t_2^T S(g) t_1 },
\end{equation}
we have
\begin{equation}
	e^{2\pi i z_2^T S(g) z_1 }=e^{2\pi i \tilde{z}_2^T \tilde{S}(g) \tilde{z}_1 }=1
\end{equation}
for any $\bm{z}_{1,2}$ in $Z$. Here, we introduced
\begin{equation}
	\tilde{S}(g)=V^TS(g)V,
\end{equation}
which is an integral symmetric matrix for each $g$ in $G$, since $\tilde{z}_{1,2}$ are arbitrary integral column vectors.

In the main text, we show the general solution to the first consistency equation \eqref{eq:const_1} is
\begin{equation}
	\eta(g|\bm{t})=e^{\pi i t^T S(g) t } e^{2\pi i \bm{q}(g)\cdot \bm{t}}.
\end{equation}
Then, the restriction of the general solution on the central sublattice $Z$ can be expressed as
\begin{equation}
	\eta(g|\bm{z})=e^{\pi i \tilde{z}^T \tilde{S}(g) \tilde{z} } e^{2\pi i \bm{q}(g)\cdot \bm{z}}=e^{\pi i \tilde{z}^TD\tilde{S}} e^{2\pi i \bm{q}(g)\cdot \bm{z}}.
\end{equation}
Here, $D\tilde{S}$ denote the integral column vector formed by diagonal entries of the integral symmetric matrix $\tilde{S}$, i.e., $[D\tilde{S}]_\alpha=\tilde{S}_{\alpha\alpha}$. Since $\tilde{S}$ is symmetric and integral, the parity of
\begin{equation}
	\sum_{\alpha\beta} \tilde{S}_{\alpha\beta}\tilde{z}_\alpha \tilde{z}_\beta =\tilde{S}_{\alpha\alpha} \tilde{z}_\alpha^2+2\sum_{\alpha <\beta} \tilde{S}_{\alpha\beta} \tilde{z}_\alpha \tilde{z}_\beta.
\end{equation}
is determined solely by the diagonal entries. Moreover, $\tilde{z}_\alpha^2=\tilde{z}_\alpha\mod 2$ leads to the second equality. To compare with $\eta(g|\bm{z})=e^{2\pi i \bm{\kappa}_g\cdot \bm{z}}$, let us introduce the vector-valued function,
\begin{equation}
	\bm{D}{\tilde{S}}(g)=[D\tilde{S}]_\alpha \bm{Q}_\alpha =\sum_{\alpha} [\tilde{S}(g)]_{\alpha\alpha} \bm{Q}_\alpha,
\end{equation}
and cast the above expression as
\begin{equation}
	\eta(g|\bm{z})=\exp 2\pi i \left[ \frac{1}{2}\bm{D}{\tilde{S}}(g)+\bm{q}(g) \right]\cdot \bm{z}.
\end{equation}
Thus, we have
\begin{equation}
	\frac{1}{2}\bm{D}{\tilde{S}}(g)+\bm{q}(g)=\bm{\kappa}_{g} \mod Z_F.
\end{equation}
Performing the coboundary operation on both sides leads to
\begin{equation} \label{eq:delta_q}
	\frac{1}{2}\delta \bm{D}{\tilde{S}}(g_2,g_1)+\delta\bm{q}(g_2,g_1)=\bm{\omega}_F(g_2,g_1) \mod B^2(G,Z_F),
\end{equation}
where
\begin{equation}
	\delta \bm{D}{\tilde{S}}(g_2,g_1)=g_2\bm{D}{\tilde{S}}(g_1)-\bm{D}{\tilde{S}}(g_2g_1)+\bm{D}{\tilde{S}}(g_2).
\end{equation}

Substituting the general solution $\eta(g|\bm{t})$ to the second consistency equation \eqref{eq:const_2}, we obtain
\begin{equation} \label{eq: concrete_eq2}
	e^{\pi i t^T \delta S(g_2,g_1) t} e^{2\pi i \delta \bm{q}(g_2,g_1)\cdot \bm{t}}= e^{2\pi i t^T \Phi \omega(g_2,g_1)},
\end{equation}
with
\begin{equation}
	\delta S(g_2,g_1)=E_{g^{-1}_2}^T S(g_1) E_{g_2^{-1}}-S(g_2g_1)+S(g_2)
\end{equation}
and 
\begin{equation}
	\delta \bm{q}(g_2,g_1)=g\bm{q}(g_1)-\bm{q}(g_2g_1)+\bm{q}(g_2).
\end{equation}
Then, it is significant to notice that $\delta S(g_2,g_1)$ is an integral symmetric matrix even though $S(g)$ is not integral. To see this, we recognize the left-hand side of \eqref{eq:S_def} can be interpreted as $\delta \sigma (g)$ regarding $\sigma$ as a function on $L\times L$. Then, applying the coboundary operator to both sides leads to
\begin{equation}
	\delta S(g_2,g_1)=0 \mod M_d(\mathbb{Z}),
\end{equation}
since $\delta^2=0$. Similar to the case of $\tilde{S}$, $e^{\pi i t^T \delta S(g_2,g_1) t} $ depends only on the diagonal entries of $\delta S(g_2,g_1)$, i.e.,
\begin{equation}
	e^{\pi i t^T \delta S(g_2,g_1) t}= e^{\pi i t^T D\delta S(g_2,g_1)}
\end{equation}
with $D\delta S(g_2,g_1)$ the column vector of diagonal entries of $\delta S(g_2,g_1)$. For later use, we also introduce the vector version,
\begin{equation}
	\bm{D}{\delta S}(g_2,g_1)=[D\delta S(g_2,g_1)]_i \bm{G}_i=\sum_i[\delta S(g_2,g_1)]_{ii}\bm{G}_i.
\end{equation}
Then, in light of \eqref{eq:delta_q}, we can substitute $\delta\bm{q}$ in \eqref{eq: concrete_eq2} by $\bm{\omega}_F-\tfrac{1}{2}\delta\bm{D}\tilde{S}$ up to a coboundary in $B^2(G, Z_F)$ and derive the bi-nonsymmorphicity relation in the main text. The $\bm{\Lambda}_{\Phi}$ defined by
\begin{equation}\label{eq:def_Lambda}
	\bm{\Lambda}_{\Phi}(g_2,g_1)=\frac{1}{2}[\delta\bm{D}{\tilde{S}}(g_2,g_1)-\bm{D}{\delta S}(g_2,g_1)].
\end{equation}

We now show that from the definition $\bm{\Lambda}_{\Phi}$ is automatically valued in $Z_F$ for $G$-invariant $\Phi$, i.e., this is not a requirement by the second consistency equation. Let us look into the identity,
\begin{equation}
	e^{\pi i \tilde{z}^T\delta D \tilde{S}(g_2,g_1)}=\exp \pi i \tilde{z}^T \left[\tilde{E}^T_{g_2^{-1}}D\tilde{S}(g_1)-D(\tilde{E}^T_{g_2^{-1}}\tilde{S}(g_1)\tilde{E}_{g_2^{-1}})\right] e^{\pi i \tilde{z}^T D \delta\tilde{S}(g_2,g_1)}
\end{equation}
which can be straightforwardly verified. We evoke the theorem that for any symmetric integral matrix $S$, $D(A^TS A)\equiv A^TDS \mod 2$ for any integral matrix $A$, which shall be proved below. Then,
\begin{equation}
	\tilde{E}^T_{g_2^{-1}}D\tilde{S}(g_1)\equiv D(\tilde{E}^T_{g_2^{-1}}\tilde{S}(g_1)\tilde{E}_{g_2^{-1}}). \mod 2
\end{equation}
Thus, the first factor on the right-hand side equals $1$. But, in the second, $\tilde{z}^T D \delta\tilde{S}(g_2,g_1)$ is just the representation of $\bm{D} \delta{S}(g_2,g_1)\cdot \bm{z}$ in the basis $\bm{Q}_{\alpha}$, i.e., $\tilde{z}^T D \delta\tilde{S}(g_2,g_1)=\bm{D} \delta{S}(g_2,g_1)\cdot \bm{z}$. Thus, $\bm{z}\cdot \overline{\bm{\Lambda}_{\Phi}}$ is integral for all $\bm{z}$ in $Z$, or equivalently $\overline{\bm{\Lambda}_{\Phi}}$ is valued in $Z_F$.

We proceed to show that $\overline{\bm{\Lambda}_{\Phi}}$, namely $\overline{\bm{\Lambda}_{\Phi}} \mod L_F$, is a cocycle, and therefore 
\begin{equation}
	[\overline{\bm{\Lambda}_{\Phi}}]\in H^2(G,Z_F/L_F).
\end{equation}
The first term in the definition \eqref{eq:def_Lambda} of $\overline{\bm{\Lambda}_{\Phi}}$ is already a coboundary that vanishes under the coboundary operator $\delta$. Therefore, we only need to analyze the second. It is straightforward to derive that
\begin{equation}
	\begin{split}
		\delta D\delta S(g_3,g_2,g_1)&=E_{g_3^{-1}}^T D\delta S(g_2,g_1)-D E_{g_3^{-1}}^T \delta S(g_2,g_1) E_{g_3^{-1}}+D\delta^2S(g_3,g_2,g_1)\\
		&=E_{g_3^{-1}}^T D\delta S(g_2,g_1)-D E_{g_3^{-1}}^T \delta S(g_2,g_1) E_{g_3^{-1}}.
	\end{split}
\end{equation}
Then, evoking the theorem $D(A^TS A)\equiv A^TDS \mod 2$ again, we see $\delta D\delta S(g_3,g_2,g_1)\equiv 0\mod 2$, i.e., $e^{2\pi i \delta \bm{\Lambda}_{\Phi}(g_3,g_2,g_1)\cdot \bm{t}}=1$ for all $\bm{t}\in L$. Thus, $\overline{\bm{\Lambda}_{\Phi}}$ is a cocyle in $Z^2(G,Z_F/L_F)$.

In conclusion, we have justified that the bi-nonsymmorphicity relation,
\begin{equation}
	[\overline{\bm{\omega}_F}]=[\overline{\bm{\Phi}\bm{\omega}}]+[\overline{\bm{\Lambda}_{\Phi}}],
\end{equation}
is well defined in $H^2(G,Z_F/L_F)$.

For completeness, we prove the following theorem that has been used twice above.
Let $S \in M_n(\mathbb{Z})$ be symmetric, and let $A \in M_n(\mathbb{Z})$. Denote by $D(S)$ the column vector of diagonal entries of $S$. Then
\[
D(A^TSA)= A^T D(S) \mod 2.
\]

\begin{proof}
	Write
	\[
	S=(s_{pq})_{1\le p,q\le n}, \qquad A=(a_{pi})_{1\le p,i\le n}.
	\]
	Let $S' = A^TSA$.
	We compute the $i$-th diagonal entry of $S'$. By matrix multiplication,
	\[
	[S']_{ii}
	\equiv \sum_{p,q=1}^n a_{pi}s_{pq}a_{qi}.
	\]
	Since $S$ is symmetric, we have $s_{pq}=s_{qp}$, so we may separate the diagonal and off-diagonal terms:
	\[
	[S']_{ii}
	= \sum_{p=1}^n s_{pp}a_{pi}^2
	+ \sum_{\substack{p,q=1, p\neq q}}^n s_{pq}a_{pi}a_{qi}.
	\]
	Grouping the off-diagonal terms into pairs $(p,q)$ and $(q,p)$ with $p<q$, we obtain
	\[
	[S']_{ii}
	= \sum_{p=1}^n s_{pp}a_{pi}^2
	+ 2\sum_{1\le p<q\le n} s_{pq}a_{pi}a_{qi}.
	\]
	Reducing modulo $2$, the second sum vanishes, since it is multiplied by $2$. Thus
	\[
	[S']_{ii}\equiv \sum_{p=1}^n s_{pp}a_{pi}^2 \mod 2.
	\]
	Now for every integer $m$ one has $m^2 \equiv m \mod 2 $,
	so
	\[
	[S']_{ii}\equiv \sum_{p=1}^n s_{pp}a_{pi} \mod 2.
	\]
	But this is exactly the $i$-th entry of the vector $A^T D(S)$, since
	\[
	[A^T D(S)]_i = \sum_{p=1}^n a_{pi}s_{pp}.
	\]
	Therefore, for each $i=1,\dots,n$,
	\[
	[S']_{ii}\equiv [A^T D(S)]_i \mod 2.
	\]
	Hence, $D(A^TSA)\equiv A^T D(S)\mod 2$ as claimed.
\end{proof}

\section{Restriction of flux by the bi-nonsymmorphicity relation}

The bi-nonsymmorphicity relation can also restrict the allowed flux in addition to the constraints from the first cocycle equation
\begin{equation*}
	\frac{\sigma(g^{-1}\bm{t}_1,g^{-1}\bm{t}_2)}{\sigma(\bm{t}_1,\bm{t}_2)} = \frac{\eta(g|\bm{t}_1+\bm{t}_2)}{\eta(g|\bm{t}_1)\eta(g|\bm{t}_2)}.
\end{equation*}
A quick example is provided by the two‑dimensional lattice of arithmetic class $mP$. The point group is $D_{1} = \langle M_x \mid M_x^2 = 1 \rangle = \left\{E,M_x\right\}$. The basis of the lattice is
\begin{equation}
	\begin{split}
		&\bm{e}_1 = (1,0),\quad \bm{e}_2 = (0,1),\\
		&\bm{G}_1 = (1,0),\quad \bm{G}_2 = (0,1).
	\end{split}
\end{equation}
The first cocycle equation restricts the flux to be either $0$ or $\pi$. When a $\pi$ flux per plaquette is present, the flux form is
\begin{equation}
	\Phi = \begin{bmatrix}
		0 & \frac{1}{2} \\
		-\frac{1}{2} & 0
	\end{bmatrix}.
\end{equation}
A basis of the center $Z$ is then
\[
\bm{a}_1 = 2\bm{e}_1,\quad \bm{a}_2 = 2\bm{e}_2,
\]
and the bases of $Z$ and $Z_F$ are
\begin{equation}
	\begin{split}
		&\bm{a}_1 = (2,0),\quad \bm{a}_2 = (0,2),\\
		&\bm{Q}_1 = \tfrac{1}{2}(1,0),\quad \bm{Q}_2 = \tfrac{1}{2}(0,1).
	\end{split}
\end{equation}
The quotient $Z_F/L_F$ is isomorphic to $\mathbb{Z}_2 \times \mathbb{Z}_2$, with a convenient basis given by the cosets of $\bm{Q}_1$ and $\bm{Q}_2$. The group element $M_x$ acts trivially on $\mathbb{Z}_2 \times \mathbb{Z}_2$. A direct computation shows that $\bm{\Lambda}_{\Phi} = \bm{0}$.

Consider the nonsymmorphic group $Pg$ in this arithmetic class. Taking the fractional translation $\bm{\tau}_{M_x} = \frac{1}{2}(0,1)$, the cocycle $\bm{\omega}$ (listed in order $E, M_x$) is
\begin{equation}
	\bm{\omega}(g_2,g_1) = \begin{bmatrix}
		\bm{0} & \bm{0} \\
		\bm{0} & \bm{e}_2
	\end{bmatrix}.
\end{equation}
Consequently,
\begin{equation}
	\bm{\Phi}\cdot\bm{\omega}(g_2,g_1) = \begin{bmatrix}
		\bm{0} & \bm{0} \\
		\bm{0} & \bm{Q}_1
	\end{bmatrix}.
\end{equation}
Thus $\overline{\bm{\Phi}\cdot\bm{\omega}}$ is a nontrivial cocycle on $\mathbb{Z}_2 \times \mathbb{Z}_2$.

However, a suitable $\bm{\omega}_F$ satisfying the bi-nonsymmorphicity relation $[\overline{\bm{\omega}_F}] = [\overline{\bm{\Phi}\bm{\omega}}] + [\overline{\bm{\Lambda}_{\Phi}}]$ does **not** exist, because the momentum‑space nonsymmorphic cocycle would have to be of the form
\begin{equation}
	\bm{\omega}_F(g_2,g_1) = \begin{bmatrix}
		\bm{0} & \bm{0} \\
		\bm{0} & \bm{Q}_2
	\end{bmatrix},
\end{equation}
which only has a nontrivial component along $\bm{Q}_2$ on $\bm{\omega}_F(M_x,M_x)$. This means that the nonsymmorphic group $Pg$ cannot accommodate a $\pi$ flux. This result is consistent with $H^2(Pg,U(1))=1$.

\section{Computation of $H^2(C_{2v},\mathbb{Z}_4 \times \mathbb{Z}_4)$}\label{section:Z4_cocycle}

Consider the point group $C_{2v} = \langle M_x, M_y \mid M_x^2 = M_y^2 = (M_x M_y)^2 = 1 \rangle$ acting on a lattice $Z_F$ spanned by the vectors
\begin{align}
	\bm{Q}_1 = \tfrac{1}{2}(0,1,1),\, \bm{Q}_2 = \tfrac{1}{2}(1,0,1),\,\bm{Q}_3 = \tfrac{1}{2}(1,1,0),
\end{align} 
via the matrices
\begin{equation}
	\begin{split}
		M_x = \begin{bmatrix} 1 &1 & 1 \\ 0 & 0 & -1  \\ 0 & -1 &0\end{bmatrix}, \qquad
		M_y = \begin{bmatrix} 0 &0 & -1 \\ 1 & 1 & 1  \\ -1 & 0 &1\end{bmatrix}.
	\end{split}
\end{equation}
Define a sublattice $L_F \subset Z_F$ generated by $\bm{G}_i = 4\bm{Q}_i - (\bm{Q}_1 + \bm{Q}_2 + \bm{Q}_3)$ for $i=1,2,3$.  One checks that $\bm{Q}_1 + \bm{Q}_2 + \bm{Q}_3,4 \bm{Q}_1,4 \bm{Q}_2 \in L_F $. Consequently the quotient $A = Z_F/L_F $ is isomorphic to $ \mathbb{Z}_4 \times \mathbb{Z}_4 $.  A convenient basis for $A$ is given by the cosets of $\bm{Q}_1$ and $\bm{Q}_2$.
The action of $C_{2v}$ on the $\mathbf{Q}_i$ is given by
\begin{align}
	M_x \bm{Q}_1 = \bm{Q}_1,\quad M_x \bm{Q}_2 = \bm{Q}_1 - \bm{Q}_3 ,\quad M_y \bm{Q}_1 = \bm{Q}_2 - \bm{Q}_3,\quad M_y \bm{Q}_2 = \bm{Q}_2
\end{align}
Passing to the quotient $A$ and using the relations $\mathbf{Q}_3 \equiv -\mathbf{Q}_1-\mathbf{Q}_2 \pmod{L_F}$ and $4\mathbf{Q}_i \equiv 0$, we obtain the induced action as matrices (mod~4):

\begin{equation}
	\begin{split}
		M_x = \begin{bmatrix} 1 & 2 \\ 0 & 1 \end{bmatrix}, \qquad
		M_y = \begin{bmatrix} 1 & 0 \\ 2 & 1 \end{bmatrix},
	\end{split}
\end{equation}
acting by left multiplication on column vectors of $A = \mathbb{Z}_4 \times \mathbb{Z}_4$.

A 2-cocycle $\bm{\mu}: C_{2v} \times C_{2v} \to A$ classifies extensions. Using the group presentation, we can encode the cocycle by the vectors
\[
\mathbf{l}_x = \bm{\mu}(M_x, M_x),\qquad 
\mathbf{l}_y = \bm{\mu}(M_y, M_y),\qquad 
\mathbf{l}_{xy} = \bm{\mu}(M_y, M_x) - \bm{\mu}(M_x, M_y).
\]
In the corresponding central extension, the lifts $\mathbf{M}_x, \mathbf{M}_y$ of the generators satisfy
\begin{align}
	\mathbf{M}_x^2 &= \mathbf{l}_x, \label{eq:rel1}\\
	\mathbf{M}_y^2 &= \mathbf{l}_y, \label{eq:rel2}\\
	\mathbf{M}_y \mathbf{M}_x &= \mathbf{l}_{xy} \,\mathbf{M}_x \mathbf{M}_y. \label{eq:rel3}
\end{align}

The consistency of the extension requires that the relations are preserved under conjugation by $\mathbf{M}_x$ and $\mathbf{M}_y$. Using the action of $C_{2v}$ on $A$ (which coincides with conjugation in the extension), we obtain the following constraints:
\begin{align}
	M_x \mathbf{l}_x &= \mathbf{l}_x, \label{eq:cond1}\\
	M_y \mathbf{l}_y &= \mathbf{l}_y, \label{eq:cond2}\\
	M_y \mathbf{l}_x &= \mathbf{l}_{xy} + M_x \mathbf{l}_{xy} + \mathbf{l}_x, \label{eq:cond3}\\
	M_x \mathbf{l}_y &= -\mathbf{l}_{xy} - M_y \mathbf{l}_{xy} + \mathbf{l}_y. \label{eq:cond4}
\end{align}
Solving these equations yields the conditions
\begin{equation}
	l_x^2 \in \{0,2\},\quad l_y^1 \in \{0,2\},\quad (l_x^1 - l_y^2) \in \{0,2\}, \quad (l_{xy}^1 - l_y^2) \in \{0,2\}, \quad (l_{xy}^2 - l_x^1) \in \{0,2\}, \label{eq:constraints}
\end{equation}
where we write $\mathbf{l}_x = (l_x^1, l_x^2)^T$, etc.

Two cocycles that differ by a coboundary give equivalent extensions. A coboundary is determined by a 1-cochain $\chi: C_{2v} \to A$ with $\chi(1)=0$; set $\boldsymbol{\chi}_x = \chi(M_x) = (\chi_x^1, \chi_x^2)^T$, $\boldsymbol{\chi}_y = \chi(M_y) = (\chi_y^1, \chi_y^2)^T$. The induced transformations on the parameters are
\begin{align}
	\mathbf{l}_x &\to \mathbf{l}_x + 2\begin{pmatrix} \chi_x^1+\chi_x^2 \\ \chi_x^2 \end{pmatrix}, \label{eq:cob1}\\
	\mathbf{l}_y &\to \mathbf{l}_y + 2\begin{pmatrix} \chi_y^1 \\ \chi_y^1+\chi_y^2 \end{pmatrix}, \label{eq:cob2}\\
	\mathbf{l}_{xy} &\to \mathbf{l}_{xy} + 2\begin{pmatrix} -\chi_y^2 \\ \chi_x^1 \end{pmatrix}. \label{eq:cob3}
\end{align}
By choosing
\[
\chi_x^2 = \tfrac12 l_x^2,\qquad \chi_y^1 = \tfrac12 l_y^1,
\]
we can set $l_x^2 = 0$ and $l_y^1 = 0$. The remaining parameters $l_x^1, l_y^2, l_{xy}^1, l_{xy}^2$ are then all of the same parity; their parity defines a $\mathbb{Z}_2$ invariant $p \in \mathbb{Z}_2$. Two additional $\mathbb{Z}_2$ invariants are $l_x^1 + l_x^2 + l_{xy}^2$ and $l_y^1 + l_y^2 + l_{xy}^1$ (taken modulo $2$). Therefore we have three invariants:
\begin{align}
	&l_x^1 \equiv l_y^2 \equiv l_{xy}^1 \equiv l_{xy}^2 \pmod{2}, \\
	&l_x^1 + l_x^2 + l_{xy}^2 \in \{0,2\} \pmod{4},\\
	& l_y^1 + l_y^2 + l_{xy}^1 \in \{0,2\} \pmod{4}. 
\end{align}
In terms of the cocycle $\bm{\mu}$ these become
\begin{align}
	&\mu^1(M_x,M_x) \equiv \mu^2(M_y,M_y) \equiv \bigl(\mu^1(M_x,M_y)-\mu^1(M_y,M_x)\bigr) \equiv \bigl(\mu^2(M_y,M_x)-\mu^2(M_x,M_y)\bigr) \pmod{2}, \\
	&\mu^1(M_x,M_x) + \mu^2(M_x,M_x) + \mu^2(M_y,M_x)-\mu^2(M_x,M_y) \in \{0,2\} \pmod{4},\\
	& \mu^1(M_y,M_y) + \mu^2(M_y,M_y) + \mu^1(M_y,M_x)-\mu^1(M_x,M_y) \in \{0,2\} \pmod{4}. 
\end{align}
Hence the second cohomology group is
\[
H^2(C_{2v}, \mathbb{Z}_4 \times \mathbb{Z}_4) \cong \mathbb{Z}_2^3.
\]

With these three invariants, we can obtain three generators of $H^2(C_{2v}, \mathbb{Z}_4 \times \mathbb{Z}_4)$ by taking
\begin{align}
	\mathbf{l}_x = \begin{pmatrix} 1 \\ 0 \end{pmatrix}, \quad \mathbf{l}_y = \begin{pmatrix} 0 \\ 1 \end{pmatrix},\quad \mathbf{l}_{xy} = \begin{pmatrix} 1 \\ 1 \end{pmatrix}, \\
	\mathbf{l}_x = \begin{pmatrix} 2\\ 0 \end{pmatrix}, \quad \mathbf{l}_y = \begin{pmatrix} 0 \\ 0 \end{pmatrix},\quad \mathbf{l}_{xy} = \begin{pmatrix} 0 \\ 0 \end{pmatrix}, \\
	\mathbf{l}_x = \begin{pmatrix} 0\\ 0 \end{pmatrix}, \quad \mathbf{l}_y = \begin{pmatrix} 0 \\ 2 \end{pmatrix},\quad \mathbf{l}_{xy} = \begin{pmatrix} 0 \\ 0 \end{pmatrix}. 
\end{align}
We can also reconstruct the whole cocycle $\bm{\mu}$ by
\begin{equation}
	\bm{\mu} (M_x^{a_2} M_y^{b_2} ,M_x^{a_1} M_y^{b_1} )  = a_2 a_1 \mathbf{l}_x + b_2 a_1 M_x^{a_2} \mathbf{l}_{xy} + b_2 b_1 M_x^{a_2 + a_1} \mathbf{l}_y .
\end{equation}

\section{Further details on the examples in the main context}
\subsection{$mP$}
For the arithmetic class $mP$ in the main text, we choose the reflection operation as $M_z$, which differs from the conventional choice $M_y$ for monoclinic crystal systems in crystallography; however, they are essentially isomorphic.
We can choose the basis of $L$ and $L_F$ as
\begin{equation}
	\begin{split}
		&\bm{e}_1 = (1,0,0),\quad \bm{e}_2 = (0,1,0),\quad \bm{e}_3 = (0,0,1),\\
		&\bm{G}_1 = (1,0,0),\quad \bm{G}_2 = (0,1,0),\quad \bm{G}_3 = (0,0,1).
	\end{split}
\end{equation}
When the flux form is given by
\begin{equation}
	\Phi = \begin{bmatrix}
		0 & \frac{1}{2} & 0 \\
		-\frac{1}{2} & 0 & 0 \\
		0 & 0 & 0
	\end{bmatrix},
\end{equation}
a basis of $Z$ is
\[
\bm{a}_1 = 2\bm{e}_1,\quad \bm{a}_2 = 2\bm{e}_2,\quad \bm{a}_3 = \bm{e}_3.
\]
Thus the bases of $Z$ and $Z_F$ are
\begin{equation}
	\begin{split}
		&\bm{a}_1 = (2,0,0),\quad \bm{a}_2 = (0,2,0),\quad \bm{a}_3 = (0,0,1),\\
		&\bm{Q}_1 = \tfrac{1}{2}(1,0,0),\quad
		\bm{Q}_2 = \tfrac{1}{2}(0,1,0),\quad
		\bm{Q}_3 = (0,0,1).
	\end{split}
\end{equation}
The quotient $Z_F/L_F$ is isomorphic to $\mathbb{Z}_2 \times \mathbb{Z}_2$. A convenient basis for $\mathbb{Z}_2 \times \mathbb{Z}_2$ is given by the cosets of $\bm{Q}_1$ and $\bm{Q}_2$ (since $\bm{Q}_3 \equiv 0 \pmod{L_F}$). The group element $M_z$ acts trivially on $\mathbb{Z}_2 \times \mathbb{Z}_2$.

With this basis, we obtain the connection matrix
\begin{equation}
	A = \begin{bmatrix}
		0 & \frac{1}{2} & 0 \\
		0 & 0 & 0 \\
		0 & 0 & 0
	\end{bmatrix}.
\end{equation}
A direct computation gives $\delta A(M_z) = E_{M_z^{-1}}^T A E_{M_z^{-1}} - A = 0$, hence
\begin{equation}
	S(M_z) = 0.
\end{equation}
Therefore, the flux twist $\bm{\Lambda}_\Phi$ vanishes.

Consider the nonsymmorphic group $Pc$ in arithmetic class $mP$. Taking the fractional translation in real space as $\bm{\tau}_{M_z} = \frac{1}{2}(1,0,0)$, the cocycle $\bm{\omega}$ is( listed in order of $E, M_z$)
\begin{equation}
	\bm{\omega}(g_2, g_1) = \begin{bmatrix}
		\bm{0} & \bm{0} \\
		\bm{0} & \bm{e}_1
	\end{bmatrix}.
\end{equation}
Consequently,
\begin{equation}
	\begin{split}
		\bm{\Phi}\cdot \bm{\omega}(g_2, g_1) = \begin{bmatrix}
			\bm{0} & \bm{0} \\
			\bm{0} & \bm{Q}_2
		\end{bmatrix}.
	\end{split}
\end{equation}

To verify that the reciprocal space group is $Pc$, we take $\bm{\kappa}_{M_z} = \frac{1}{4}(0,1,0)$, yielding the cocycle $\bm{\omega}_F$:
\begin{equation}
	\begin{split}
		\bm{\omega}_F(g_2, g_1) = \begin{bmatrix}
			\bm{0} & \bm{0} \\
			\bm{0} & \bm{Q}_2
		\end{bmatrix}.
	\end{split}
\end{equation}
Thus the bi-nonsymmorphicity relation $[\overline{\bm{\omega}_F}] = [\overline{\bm{\Phi}\bm{\omega}}] + [\overline{\bm{\Lambda}_{\Phi}}]$ holds.

The bi-nonsymmorphicity relation  can also been  view from projective algebra of $Pc$:
\begin{equation}
	\begin{split}
		& \rho_{\bm{e}_i} \rho_{\bm{e}_j} = e^{i2\pi\Phi_{ij}} \rho_{\bm{e}_j}\rho_{\bm{e}_i},\\
		& \rho_{M_z} \rho_{\bm{e}_x} \rho_{M_z}^{-1}  =  \rho_{\bm{e}_x},\\
		& \rho_{M_z} \rho_{\bm{e}_y} \rho_{M_z}^{-1}  = \eta_{zy} \rho_{\bm{e}_y},\\
		& \rho_{M_z} \rho_{\bm{e}_z} \rho_{M_z}^{-1}  =   \rho_{\bm{e}_z}^{-1},\\
		&\rho_{M_z}^2 = \rho_{\bm{e}_x}. 
	\end{split}
\end{equation}

Here, self-consistency requires $e^{i\pi\Phi_{xy}} = \eta_{zy}, e^{i2\pi\Phi_{xz}} =1,e^{i2\pi\Phi_{yz}}  =\pm 1,e^{i2\pi\Phi_{xy}} \in U(1)$. The result is consistent with $H^2(Pc,U(1)) = \mathbb{Z}_2 \oplus U(1)$. From the projective algebra we obtain
\begin{equation}
	\begin{split}
		&\rho_{M_z} \rho_{\bm{e}_x}^{2} \rho_{M_z}^{-1}  =     \rho_{\bm{e}_x}^{2} \\
		&\rho_{M_z} \rho_{\bm{e}_y}^{2} \rho_{M_z}^{-1}  =  e^{i2\pi\Phi_{xy}}  \rho_{\bm{e}_y}^{2} \\
		&\rho_{M_z} \rho_{\bm{e}_z}  \rho_{M_z}^{-1}  =     \rho_{\bm{e}_z}^{-1} 
	\end{split}
\end{equation}

Since  $\bm{a}_1,\bm{a}_2,\bm{a}_3$ are generated by  $\rho_{\bm{e}_x}^2,\rho_{\bm{e}_y}^2,\rho_{\bm{e}_z}$, and $\bm{\kappa}_{M_z} =  \kappa_{M_z}^{x} \bm{Q}_x+ \kappa_{M_z}^{y} \bm{Q}_y + \kappa_{M_z}^{z} \bm{Q}_z = \frac{1}{2}(\kappa_{M_z}^{x},\kappa_{M_z}^{y},2\kappa_{M_z}^{3})$. The projective algebra indecates that $e^{i2\pi\kappa_{M_z}^{x}} = 1,e^{i2\pi\kappa_{M_z}^{y}} = e^{i2\pi\Phi_{xy}},e^{i2\pi\kappa_{M_x}^{z}} = 1$. Therefore the fractional translation satisfies  $\bm{\kappa}_{M_z}+ M_z \bm{\kappa}_{M_z} = \frac{1}{2}(0,1,0)$, which indicates that the reciprocal space group is  nonsymmorphic.

\subsection{$mm2F$}\label{section:mm2F}
For arithmetic class $mm2F$, we choose the basis of $L$ and $L_F$ as
\begin{equation}
	\begin{split}
		&\bm{e}_1 = \tfrac{1}{2}(0,1,1),\quad \bm{e}_2 = \tfrac{1}{2}(1,0,1),\quad \bm{e}_3 = \tfrac{1}{2}(1,1,0),\\
		&\bm{G}_1 = (-1,1,1),\quad \bm{G}_2 = (1,-1,1),\quad \bm{G}_3 = (1,1,-1).
	\end{split}
\end{equation}
When the flux form is given by
\begin{equation}
	\Phi = \begin{bmatrix}
		0 & \frac{1}{4} & -\frac{1}{4} \\
		-\frac{1}{4} & 0 & \frac{1}{4} \\
		\frac{1}{4} & -\frac{1}{4} & 0
	\end{bmatrix},
\end{equation}
a basis of $Z$ is
\[
\bm{a}_1 = 3\bm{e}_1 - \bm{e}_2 - \bm{e}_3,\quad
\bm{a}_2 = -\bm{e}_1 + 3\bm{e}_2 - \bm{e}_3,\quad
\bm{a}_3 = -\bm{e}_1 - \bm{e}_2 + 3\bm{e}_3.
\]
Thus the bases of $Z$ and $Z_F$ are
\begin{equation}
	\begin{split}
		&\bm{a}_1 = (-1,1,1),\quad \bm{a}_2 = (1,-1,1),\quad \bm{a}_3 = (1,1,-1),\\
		&\bm{Q}_1 = \tfrac{1}{2}(0,1,1),\quad \bm{Q}_2 = \tfrac{1}{2}(1,0,1),\quad \bm{Q}_3 = \tfrac{1}{2}(1,1,0).
	\end{split}
\end{equation}
The quotient $Z_F/L_F$ is isomorphic to $\mathbb{Z}_4 \times \mathbb{Z}_4$. A convenient basis for $\mathbb{Z}_4 \times \mathbb{Z}_4$ is given by the cosets of $\bm{Q}_1$ and $\bm{Q}_2$ (since $\mathbf{Q}_3 \equiv -\mathbf{Q}_1-\mathbf{Q}_2 \pmod{L_F}$).  
The action of $C_{2v}$ on the $\mathbf{Q}_i$ is given by
\begin{align}
	M_x \bm{Q}_1 &= \bm{Q}_1, \quad M_x \bm{Q}_2 = \bm{Q}_1-\bm{Q}_3,\\
	M_y \bm{Q}_1 &= \bm{Q}_2 - \bm{Q}_3, \quad M_y \bm{Q}_2 = \bm{Q}_2.
\end{align}
Passing to the quotient $\mathbb{Z}_4 \times \mathbb{Z}_4$ and using the relations $\mathbf{Q}_3 \equiv -\mathbf{Q}_1-\mathbf{Q}_2 \pmod{L_F}$ and $4\mathbf{Q}_i \equiv 0$, we obtain the induced action as matrices (mod~4):
\begin{equation}
	M_x = \begin{bmatrix} 1 & 2 \\ 0 & 1 \end{bmatrix}, \qquad
	M_y = \begin{bmatrix} 1 & 0 \\ 2 & 1 \end{bmatrix},
\end{equation}
acting by left multiplication on column vectors of $\mathbb{Z}_4 \times \mathbb{Z}_4$.

With this basis, we obtain the connection matrix
\begin{equation}
	A = \begin{bmatrix}
		\frac{1}{4} & \frac{3}{4} & 0 \\
		\frac{1}{2} & \frac{3}{4} & \frac{3}{4} \\
		\frac{1}{4} & \frac{1}{2} & \frac{1}{4}
	\end{bmatrix},
\end{equation}
and the matrices $S$ are given by
\begin{equation}
	S(M_x) = -\begin{bmatrix}
		0 & \frac{1}{2} & \frac{1}{2} \\
		\frac{1}{2} & \frac{1}{2} & 0 \\
		\frac{1}{2} & 0 & \frac{1}{2}
	\end{bmatrix},\quad
	S(M_y) = -\begin{bmatrix}
		\frac{1}{2} & -\frac{1}{2} & 1 \\
		\frac{1}{2} & 0 & \frac{1}{2} \\
		1 & \frac{1}{2} & \frac{1}{2}
	\end{bmatrix},\quad
	S(M_z) = -\begin{bmatrix}
		\frac{1}{2} & 0 & \frac{1}{2} \\
		0 & \frac{1}{2} & \frac{1}{2} \\
		\frac{1}{2} & \frac{1}{2} & 0
	\end{bmatrix}.
\end{equation}

Then we can obtain $\bm{\Lambda}_\Phi$ (listed in the order $E, M_x, M_y, R_z$):
\begin{equation}
	\bm{\Lambda}_\Phi = \begin{bmatrix}
		\bm{0} & \bm{0} & \bm{0} & \bm{0} \\
		\bm{0} & \bm{Q}_1 - \bm{Q}_2 - \bm{Q}_3 & -\bm{Q}_1 - \bm{Q}_2 + \bm{Q}_3 & -\bm{Q}_1 + \bm{Q}_2 - \bm{Q}_3 \\
		\bm{0} & -\bm{Q}_1 - \bm{Q}_2 + \bm{Q}_3 & -\bm{Q}_1 + \bm{Q}_2 - \bm{Q}_3 & \bm{Q}_1 - \bm{Q}_2 - \bm{Q}_3 \\
		\bm{0} & -\bm{Q}_1 + \bm{Q}_2 - \bm{Q}_3 & \bm{Q}_1 - \bm{Q}_2 - \bm{Q}_3 & -\bm{Q}_1 - \bm{Q}_2 + \bm{Q}_3
	\end{bmatrix}.
\end{equation}
Thus
\begin{equation}
	\overline{\bm{\Lambda}_\Phi} = \begin{bmatrix}
		\begin{pmatrix}0\\0\end{pmatrix} & \begin{pmatrix}0\\0\end{pmatrix} & \begin{pmatrix}0\\0\end{pmatrix} & \begin{pmatrix}0\\0\end{pmatrix} \\
		\begin{pmatrix}0\\0\end{pmatrix} & \begin{pmatrix}2\\0\end{pmatrix} & \begin{pmatrix}2\\2\end{pmatrix} & \begin{pmatrix}0\\2\end{pmatrix} \\
		\begin{pmatrix}0\\0\end{pmatrix} & \begin{pmatrix}2\\2\end{pmatrix} & \begin{pmatrix}0\\2\end{pmatrix} & \begin{pmatrix}2\\0\end{pmatrix} \\
		\begin{pmatrix}0\\0\end{pmatrix} & \begin{pmatrix}0\\2\end{pmatrix} & \begin{pmatrix}2\\0\end{pmatrix} & \begin{pmatrix}2\\2\end{pmatrix}
	\end{bmatrix},
\end{equation}
or, in terms of the presentation $g = M_x^a M_y^b$,
\begin{equation}
	\overline{\bm{\Lambda}_\Phi}(M_x^{a_2}M_y^{b_2}, M_x^{a_1}M_y^{b_1})
	= a_2 a_1 \begin{pmatrix}1\\0\end{pmatrix}
	+ b_2 b_1 \begin{pmatrix}0\\1\end{pmatrix}
	+ 2(a_2 b_1 + b_2 a_1) \begin{pmatrix}1\\1\end{pmatrix},
\end{equation}
where the matrix takes values in column vectors of $\mathbb{Z}_4 \times \mathbb{Z}_4$ (generated by $\bm{Q}_1,\bm{Q}_2$). The group elements $M_x$, $M_y$, and $R_z$ act on $\mathbb{Z}_4 \times \mathbb{Z}_4$ via the matrices
\begin{equation}
	M_x = \begin{bmatrix} 1 & 2 \\ 0 & 1 \end{bmatrix},\quad
	M_y = \begin{bmatrix} 1 & 0 \\ 2 & 1 \end{bmatrix},\quad
	R_z = \begin{bmatrix} 1 & 2 \\ 2 & 1 \end{bmatrix} \pmod{4}.
\end{equation}
As analyzed in Section~\ref{section:Z4_cocycle}, the invariants that characterize $\overline{\bm{\Lambda}_\Phi}$ are
\begin{align}
	&\overline{\Lambda_\Phi}^1(M_x,M_x) + \overline{\Lambda_\Phi}^2(M_x,M_x) \equiv 0 \pmod{2},\\
	&\overline{\Lambda_\Phi}^1(M_x,M_x) + \overline{\Lambda_\Phi}^2(M_x,M_x) + \overline{\Lambda_\Phi}^2(M_y,M_x) - \overline{\Lambda_\Phi}^2(M_x,M_y) \equiv 2 \pmod{4},\\
	&\overline{\Lambda_\Phi}^1(M_y,M_y) + \overline{\Lambda_\Phi}^2(M_y,M_y) + \overline{\Lambda_\Phi}^1(M_y,M_x) - \overline{\Lambda_\Phi}^1(M_x,M_y) \equiv 2 \pmod{4}.
\end{align}

Furthermore, for arithmetic class $mm2F$, there are only two crystallographic groups: the symmorphic $Fmm2$ and the nonsymmorphic $Fdd2$.
Taking the fractional translations in real space as $\bm{\tau}_{M_x} = \bm{\tau}_{M_y} = \frac{1}{4}(1,1,1)$ and $\bm{\tau}_{R_z} = \bm{0}$, the cocycle $\bm{\omega}$ is given by
\begin{equation}
	\bm{\omega}(g_2,g_1) = \begin{bmatrix}
		\bm{0} & \bm{0} & \bm{0} & \bm{0} \\
		\bm{0} & \bm{e}_1 & \bm{e}_1 & \bm{0} \\
		\bm{0} & \bm{e}_2 & \bm{e}_2 & \bm{0} \\
		\bm{0} & -\bm{e}_3 & -\bm{e}_3 & \bm{0}
	\end{bmatrix}.
\end{equation}
Therefore,
\begin{equation}
	\bm{\Phi}\cdot\bm{\omega}(g_2,g_1) = \begin{bmatrix}
		\bm{0} & \bm{0} & \bm{0} & \bm{0} \\
		\bm{0} & \bm{Q}_3 - \bm{Q}_2 & \bm{Q}_3 - \bm{Q}_2 & \bm{0} \\
		\bm{0} & \bm{Q}_1 - \bm{Q}_3 & \bm{Q}_1 - \bm{Q}_3 & \bm{0} \\
		\bm{0} & \bm{Q}_1 - \bm{Q}_2 & \bm{Q}_1 - \bm{Q}_2 & \bm{0}
	\end{bmatrix}.
\end{equation}
Meanwhile, in momentum space, taking $\bm{\kappa}_{M_x} = \bm{\kappa}_{M_y} = \frac{1}{4}(1,1,1)$ and $\bm{\kappa}_{R_z} = \bm{0}$, the cocycle $\bm{\omega}_F$ is
\begin{equation}
	\bm{\omega}_F(g_2,g_1) = \begin{bmatrix}
		\bm{0} & \bm{0} & \bm{0} & \bm{0} \\
		\bm{0} & \bm{Q}_1 & \bm{Q}_1 & \bm{0} \\
		\bm{0} & \bm{Q}_2 & \bm{Q}_2 & \bm{0} \\
		\bm{0} & -\bm{Q}_3 & -\bm{Q}_3 & \bm{0}
	\end{bmatrix}.
\end{equation}
Thus,
\begin{align}
	\overline{\bm{\Phi}\cdot\bm{\omega}} = \begin{bmatrix}
		\begin{pmatrix}0\\0\end{pmatrix} & \begin{pmatrix}0\\0\end{pmatrix} & \begin{pmatrix}0\\0\end{pmatrix} & \begin{pmatrix}0\\0\end{pmatrix} \\
		\begin{pmatrix}0\\0\end{pmatrix} & \begin{pmatrix}3\\2\end{pmatrix} & \begin{pmatrix}3\\2\end{pmatrix} & \begin{pmatrix}0\\0\end{pmatrix} \\
		\begin{pmatrix}0\\0\end{pmatrix} & \begin{pmatrix}2\\1\end{pmatrix} & \begin{pmatrix}2\\1\end{pmatrix} & \begin{pmatrix}0\\0\end{pmatrix} \\
		\begin{pmatrix}0\\0\end{pmatrix} & \begin{pmatrix}1\\3\end{pmatrix} & \begin{pmatrix}1\\3\end{pmatrix} & \begin{pmatrix}0\\0\end{pmatrix}
	\end{bmatrix},\quad 
	\overline{\bm{\omega}}_F = \begin{bmatrix}
		\begin{pmatrix}0\\0\end{pmatrix} & \begin{pmatrix}0\\0\end{pmatrix} & \begin{pmatrix}0\\0\end{pmatrix} & \begin{pmatrix}0\\0\end{pmatrix} \\
		\begin{pmatrix}0\\0\end{pmatrix} & \begin{pmatrix}1\\0\end{pmatrix} & \begin{pmatrix}1\\0\end{pmatrix} & \begin{pmatrix}0\\0\end{pmatrix} \\
		\begin{pmatrix}0\\0\end{pmatrix} & \begin{pmatrix}0\\1\end{pmatrix} & \begin{pmatrix}0\\1\end{pmatrix} & \begin{pmatrix}0\\0\end{pmatrix} \\
		\begin{pmatrix}0\\0\end{pmatrix} & \begin{pmatrix}1\\1\end{pmatrix} & \begin{pmatrix}1\\1\end{pmatrix} & \begin{pmatrix}0\\0\end{pmatrix}
	\end{bmatrix}.
\end{align}
The invariants characterizing $\overline{\bm{\Phi}\cdot\bm{\omega}}$ are
\begin{align}
	&\overline{\bm{\Phi}\cdot\bm{\omega}}^1(M_x,M_x) + \overline{\bm{\Phi}\cdot\bm{\omega}}^2(M_x,M_x) \equiv 1 \pmod{2},\\
	&\overline{\bm{\Phi}\cdot\bm{\omega}}^1(M_x,M_x) + \overline{\bm{\Phi}\cdot\bm{\omega}}^2(M_x,M_x) + \overline{\bm{\Phi}\cdot\bm{\omega}}^2(M_y,M_x) - \overline{\bm{\Phi}\cdot\bm{\omega}}^2(M_x,M_y) \equiv 0 \pmod{4},\\
	&\overline{\bm{\Phi}\cdot\bm{\omega}}^1(M_y,M_y) + \overline{\bm{\Phi}\cdot\bm{\omega}}^2(M_y,M_y) + \overline{\bm{\Phi}\cdot\bm{\omega}}^1(M_y,M_x) - \overline{\bm{\Phi}\cdot\bm{\omega}}^1(M_x,M_y) \equiv 2 \pmod{4},
\end{align}
while those characterizing $\overline{\bm{\omega}_F}$ are
\begin{align}
	&\overline{\bm{\omega}_F}^1(M_x,M_x) + \overline{\bm{\omega}_F}^2(M_x,M_x) \equiv 1 \pmod{2},\\
	&\overline{\bm{\omega}_F}^1(M_x,M_x) + \overline{\bm{\omega}_F}^2(M_x,M_x) + \overline{\bm{\omega}_F}^2(M_y,M_x) - \overline{\bm{\omega}_F}^2(M_x,M_y) \equiv 2 \pmod{4},\\
	&\overline{\bm{\omega}_F}^1(M_y,M_y) + \overline{\bm{\omega}_F}^2(M_y,M_y) + \overline{\bm{\omega}_F}^1(M_y,M_x) - \overline{\bm{\omega}_F}^1(M_x,M_y) \equiv 0 \pmod{4}.
\end{align}
We observe that both $\bm{\omega}_F$ and $\bm{\Phi}\cdot\bm{\omega}$ have odd parity of the first invariant. Therefore, the class $[\overline{\bm{\Lambda}_\Phi}] \in H^2(G, Z_F/L_F)$ lies in neither
\[
\bigl\{ [\overline{\bm{\omega}_F}] \mid [\bm{\omega}_F] \in H^2(G,L_F) \bigr\}
\]
nor
\[
\bigl\{ [\overline{\bm{\Phi}\cdot\bm{\omega}}] \mid [\bm{\omega}] \in H^2(G,L) \bigr\}.
\]
From the invariants we can see that the bi-nonsymmorphicity relation $[\overline{\bm{\omega}_F}] = [\overline{\bm{\Phi}\bm{\omega}}] + [\overline{\bm{\Lambda}_{\Phi}}]$ holds.

The bi-nonsymmorphicity relation can also be verified directly by providing the coboundary:
\begin{equation}
	\begin{split}
		&( \overline{\bm{\omega}_{F}} - \overline{\bm{\Phi} \cdot \bm{\omega}} + \overline{\bm{\Lambda}_\Phi})(g_2, g_1) \\
		&= \begin{bmatrix}
			\begin{pmatrix}0\\0\end{pmatrix} & \begin{pmatrix}0\\0\end{pmatrix} & \begin{pmatrix}0\\0\end{pmatrix} & \begin{pmatrix}0\\0\end{pmatrix} \\
			\begin{pmatrix}0\\0\end{pmatrix} & \begin{pmatrix}0\\2\end{pmatrix} & \begin{pmatrix}0\\0\end{pmatrix} & \begin{pmatrix}0\\2\end{pmatrix} \\
			\begin{pmatrix}0\\0\end{pmatrix} & \begin{pmatrix}0\\2\end{pmatrix} & \begin{pmatrix}2\\2\end{pmatrix} & \begin{pmatrix}2\\0\end{pmatrix} \\
			\begin{pmatrix}0\\0\end{pmatrix} & \begin{pmatrix}0\\0\end{pmatrix} & \begin{pmatrix}2\\2\end{pmatrix} & \begin{pmatrix}2\\2\end{pmatrix}
		\end{bmatrix}
		= \delta \begin{bmatrix}
			\begin{pmatrix}0\\0\end{pmatrix} \\
			\begin{pmatrix}1\\1\end{pmatrix} \\
			\begin{pmatrix}1\\0\end{pmatrix} \\
			\begin{pmatrix}2\\1\end{pmatrix}
		\end{bmatrix}.
	\end{split}
\end{equation}

The bi-nonsymmorphicity relation can also be viewed from the projective algebra of $Fdd2$:
\begin{equation}
	\begin{split}
		& \rho_{\bm{e}_i} \rho_{\bm{e}_j} = e^{i2\pi\Phi_{ij}} \rho_{\bm{e}_j}\rho_{\bm{e}_i},\\
		& \rho_{M_x} \rho_{\bm{e}_1} \rho_{M_x}^{-1} = \rho_{\bm{e}_1}, \\
		& \rho_{M_x} \rho_{\bm{e}_2} \rho_{M_x}^{-1} = \eta_{x2} \rho_{\bm{e}_1} \rho_{\bm{e}_3}^{-1}, \\
		& \rho_{M_x} \rho_{\bm{e}_3} \rho_{M_x}^{-1} = \eta_{x3}\rho_{\bm{e}_1} \rho_{\bm{e}_2}^{-1}, \\
		& \rho_{M_y} \rho_{\bm{e}_1} \rho_{M_y}^{-1} = \eta_{y1} \rho_{\bm{e}_2} \rho_{\bm{e}_3}^{-1}, \\
		& \rho_{M_y} \rho_{\bm{e}_2} \rho_{M_y}^{-1} = \rho_{\bm{e}_2}, \\
		& \rho_{M_y} \rho_{\bm{e}_3} \rho_{M_y}^{-1} = \eta_{y3} \rho_{\bm{e}_2} \rho_{\bm{e}_1}^{-1}, \\
		& \rho_{M_y} \rho_{M_x} \rho_{M_y}^{-1} \rho_{M_x}^{-1} = \alpha \rho_{\bm{e}_2} \rho_{\bm{e}_1}^{-1}.
	\end{split}
\end{equation}
Here, self-consistency requires $\eta_{x2}=\eta_{x3}$, $\eta_{y1}=\eta_{y3}$, $\eta_{y1}=\alpha^2\eta_{x2}$, and $e^{i2\pi\Phi_{ij}}\in \mathbb{Z}_4$. The parameters $\eta_{x2}$ and $\alpha$ can be trivialized by a coboundary transformation. The result is consistent with $H^2(Fdd2,U(1)) = \mathbb{Z}_4$.

From the projective algebra we obtain
\begin{equation}
	\begin{split}
		& \rho_{M_x} \rho_{\bm{e}_1}^{3} \rho_{\bm{e}_2}^{-1} \rho_{\bm{e}_3}^{-1} \rho_{M_x}^{-1}
		= e^{i2\pi\Phi_{ij}}\eta_{x2}^{-2} \bigl(\rho_{\bm{e}_1}^{3} \rho_{\bm{e}_2}^{-1} \rho_{\bm{e}_3}^{-1}\bigr)
		\bigl(\rho_{\bm{e}_1}^{-1} \rho_{\bm{e}_2}^{3} \rho_{\bm{e}_3}^{-1}\bigr)
		\bigl(\rho_{\bm{e}_1}^{-1} \rho_{\bm{e}_2}^{-1} \rho_{\bm{e}_3}^{3}\bigr), \\
		& \rho_{M_x} \rho_{\bm{e}_1}^{-1} \rho_{\bm{e}_2}^{3} \rho_{\bm{e}_3}^{-1} \rho_{M_x}^{-1}
		= \eta_{x2}^2 e^{i2\pi\Phi_{ij}} \bigl(\rho_{\bm{e}_1}^{-1} \rho_{\bm{e}_2}^{-1} \rho_{\bm{e}_3}^{3}\bigr)^{-1}, \\
		& \rho_{M_x} \rho_{\bm{e}_1}^{-1} \rho_{\bm{e}_2}^{-1} \rho_{\bm{e}_3}^{3} \rho_{M_x}^{-1}
		= \eta_{x2}^2 e^{i2\pi\Phi_{ij}} \bigl(\rho_{\bm{e}_1}^{-1} \rho_{\bm{e}_2}^{3} \rho_{\bm{e}_3}^{-1}\bigr)^{-1}.
	\end{split}
\end{equation}

Since $\bm{a}_1,\bm{a}_2,\bm{a}_3$ are generated by
$\rho_{\bm{e}_1}^{3} \rho_{\bm{e}_2}^{-1} \rho_{\bm{e}_3}^{-1}$,
$\rho_{\bm{e}_1}^{-1} \rho_{\bm{e}_2}^{3} \rho_{\bm{e}_3}^{-1}$, and
$\rho_{\bm{e}_1}^{-1} \rho_{\bm{e}_2}^{-1} \rho_{\bm{e}_3}^{3}$, and
$\bm{\kappa}_{M_x} = \kappa_{M_x}^{1} \bm{Q}_1 + \kappa_{M_x}^{2} \bm{Q}_2 + \kappa_{M_x}^{3} \bm{Q}_3
= \frac{1}{2}\bigl(\kappa_{M_x}^{2}+\kappa_{M_x}^{3},\; \kappa_{M_x}^{1}+\kappa_{M_x}^{3},\; \kappa_{M_x}^{1}+\kappa_{M_x}^{2}\bigr)$,
the projective algebra indicates that
$e^{i2\pi\kappa_{M_x}^{1}} = e^{i2\pi\Phi_{ij}}\eta_{x2}^{-2}$,
$e^{i2\pi\kappa_{M_x}^{2}} = e^{i2\pi\Phi_{ij}}\eta_{x2}^{2}$,
$e^{i2\pi\kappa_{M_x}^{3}} = e^{i2\pi\Phi_{ij}}\eta_{x2}^{2}$.
Therefore the fractional translation satisfies
$\bm{\kappa}_{M_x} + M_x \bm{\kappa}_{M_x} = \frac{1}{2}(0,1,1)$,
which indicates that the reciprocal-space group is also nonsymmorphic.

\subsection{$mm2C$}
For arithmetic class $mm2C$, we choose the basis of $L$ and $L_F$ as
\begin{equation}
	\begin{split}
		&\bm{e}_1 = \tfrac{1}{2}(1,-1,0),\quad \bm{e}_2 = \tfrac{1}{2}(1,1,0),\quad \bm{e}_3 = (0,0,1),\\
		&\bm{G}_1 = (1,-1,0),\quad \bm{G}_2 = (1,1,0),\quad \bm{G}_3 = (0,0,1).
	\end{split}
\end{equation}
When the flux form is given by
\begin{equation}
	\Phi = \begin{bmatrix}
		0 & 0 & \frac{1}{2} \\
		0 & 0 & \frac{1}{2} \\
		-\frac{1}{2} & -\frac{1}{2} & 0
	\end{bmatrix},
\end{equation}
a basis of $Z$ is
\[
\bm{a}_1 = \bm{e}_1 + \bm{e}_2,\quad
\bm{a}_2 = \bm{e}_1 - \bm{e}_2,\quad
\bm{a}_3 = 2\bm{e}_3.
\]
Thus the bases of $Z$ and $Z_F$ are
\begin{equation}
	\begin{split}
		&\bm{a}_1 = (1,0,0),\quad \bm{a}_2 = (0,1,0),\quad \bm{a}_3 = (0,0,2),\\
		&\bm{Q}_1 = (1,0,0),\quad \bm{Q}_2 = (0,1,0),\quad \bm{Q}_3 = \tfrac{1}{2}(0,0,1).
	\end{split}
\end{equation}
The quotient $Z_F/L_F$ is isomorphic to $\mathbb{Z}_2 \times \mathbb{Z}_2$. A convenient basis for $\mathbb{Z}_2 \times \mathbb{Z}_2$ is given by the cosets of $\bm{Q}_1$ and $\bm{Q}_3$ (since $\mathbf{Q}_1 + \bm{Q}_2 \equiv 0 \pmod{L_F}$).  
The action of $C_{2v}$ is trivial on $\mathbb{Z}_2 \times \mathbb{Z}_2$.

With this basis, we obtain the connection matrix
\begin{equation}
	A = \begin{bmatrix}
		0 & 0 & -\frac{1}{2} \\
		0 & 0 & \frac{1}{2} \\
		0 & 0 & 0
	\end{bmatrix}.
\end{equation}
The matrix $S$ satisfies
\begin{equation}
	S(M_x) = S(M_y) = S(M_z) = 0.
\end{equation}
Therefore, the flux twist $\bm{\Lambda}_\Phi$ vanishes.

Consider the nonsymmorphic group $Ccc2$ in arithmetic class $mm2C$. Taking the fractional translations in real space as $\bm{\tau}_{M_x} = \bm{\tau}_{M_y} = \frac{1}{2}(0,0,1)$ and $\bm{\tau}_{R_z} = \bm{0}$, the cocycle $\bm{\omega}$ is given by
\begin{equation}
	\bm{\omega}(g_2, g_1) = \begin{bmatrix}
		\bm{0} & \bm{0} & \bm{0} & \bm{0} \\
		\bm{0} & \bm{e}_3 & \bm{e}_3 & \bm{0} \\
		\bm{0} & \bm{e}_3 & \bm{e}_3 & \bm{0} \\
		\bm{0} & \bm{0} & \bm{0} & \bm{0}
	\end{bmatrix}.
\end{equation}
Therefore,
\begin{equation}
	\bm{\Phi}\cdot\bm{\omega}(g_2, g_1) = \begin{bmatrix}
		\bm{0} & \bm{0} & \bm{0} & \bm{0} \\
		\bm{0} & \bm{Q}_1 & \bm{Q}_1 & \bm{0} \\
		\bm{0} & \bm{Q}_1 & \bm{Q}_1 & \bm{0} \\
		\bm{0} & \bm{0} & \bm{0} & \bm{0}
	\end{bmatrix}.
\end{equation}

To verify that the reciprocal space group is $Pba2$, we take $\bm{\kappa}_{M_x} = \bm{\kappa}_{M_y} = \frac{1}{2}(1,1,0)$ and $\bm{\kappa}_{R_z} = \bm{0}$. The cocycle $\bm{\omega}_F$ is then
\begin{equation}
	\bm{\omega}_F(g_2, g_1) = \begin{bmatrix}
		\bm{0} & \bm{0} & \bm{0} & \bm{0} \\
		\bm{0} & \bm{Q}_2 & \bm{Q}_2 & \bm{0} \\
		\bm{0} & \bm{Q}_1 & \bm{Q}_1 & \bm{0} \\
		\bm{0} & -\bm{Q}_1-\bm{Q}_2 & -\bm{Q}_1-\bm{Q}_2 & \bm{0}
	\end{bmatrix}.
\end{equation}

To verify the bi-nonsymmorphicity relation, we project $\bm{\omega}_{F} - \Phi \bm{\omega}$ onto $\mathbb{Z}_2 \times \mathbb{Z}_2$:
\begin{equation}
	\overline{\bm{\omega}_{F}} - \overline{\bm{\Phi} \cdot \bm{\omega}} = 
	\begin{bmatrix}
		\begin{pmatrix}0\\0\end{pmatrix} & \begin{pmatrix}0\\0\end{pmatrix} & \begin{pmatrix}0\\0\end{pmatrix} & \begin{pmatrix}0\\0\end{pmatrix} \\
		\begin{pmatrix}0\\0\end{pmatrix} & \begin{pmatrix}1\\0\end{pmatrix} & \begin{pmatrix}1\\0\end{pmatrix} & \begin{pmatrix}0\\0\end{pmatrix} \\
		\begin{pmatrix}0\\0\end{pmatrix} & \begin{pmatrix}1\\0\end{pmatrix} & \begin{pmatrix}1\\0\end{pmatrix} & \begin{pmatrix}0\\0\end{pmatrix} \\
		\begin{pmatrix}0\\0\end{pmatrix} & \begin{pmatrix}0\\0\end{pmatrix} & \begin{pmatrix}0\\0\end{pmatrix} & \begin{pmatrix}0\\0\end{pmatrix}
	\end{bmatrix}
	-
	\begin{bmatrix}
		\begin{pmatrix}0\\0\end{pmatrix} & \begin{pmatrix}0\\0\end{pmatrix} & \begin{pmatrix}0\\0\end{pmatrix} & \begin{pmatrix}0\\0\end{pmatrix} \\
		\begin{pmatrix}0\\0\end{pmatrix} & \begin{pmatrix}1\\0\end{pmatrix} & \begin{pmatrix}1\\0\end{pmatrix} & \begin{pmatrix}0\\0\end{pmatrix} \\
		\begin{pmatrix}0\\0\end{pmatrix} & \begin{pmatrix}1\\0\end{pmatrix} & \begin{pmatrix}1\\0\end{pmatrix} & \begin{pmatrix}0\\0\end{pmatrix} \\
		\begin{pmatrix}0\\0\end{pmatrix} & \begin{pmatrix}0\\0\end{pmatrix} & \begin{pmatrix}0\\0\end{pmatrix} & \begin{pmatrix}0\\0\end{pmatrix}
	\end{bmatrix}
	= 0.
\end{equation}
Thus the bi-nonsymmorphicity relation holds, consistent with $\bm{\Lambda}_\Phi = 0$.

The bi-nonsymmorphicity relation can also be viewed from the projective algebra of $Ccc2$:
\begin{equation}
	\begin{split}
		& \rho_{\bm{e}_i} \rho_{\bm{e}_j} = e^{i 2\pi \Phi_{ij}} \rho_{\bm{e}_j}\rho_{\bm{e}_i},\\
		& \rho_{M_x} \rho_{\bm{e}_1} \rho_{M_x}^{-1} = \eta_{x1} \rho_{\bm{e}_2}^{-1}, \\
		& \rho_{M_x} \rho_{\bm{e}_2} \rho_{M_x}^{-1} = \eta_{x2} \rho_{\bm{e}_1}^{-1}, \\
		& \rho_{M_x} \rho_{\bm{e}_3} \rho_{M_x}^{-1} = \rho_{\bm{e}_3}, \\
		& \rho_{M_y} \rho_{\bm{e}_1} \rho_{M_y}^{-1} = \eta_{y1} \rho_{\bm{e}_2}, \\
		& \rho_{M_y} \rho_{\bm{e}_2} \rho_{M_y}^{-1} = \eta_{y2} \rho_{\bm{e}_1}, \\
		& \rho_{M_y} \rho_{\bm{e}_3} \rho_{M_y}^{-1} = \rho_{\bm{e}_3}, \\
		& \rho_{M_y} \rho_{M_x} \rho_{M_y}^{-1} \rho_{M_x}^{-1} = \alpha .
	\end{split}
\end{equation}
Here, self-consistency requires $e^{i 2\pi \Phi_{12}} = \pm 1$, $e^{i 2\pi \Phi_{13}} = \pm 1$, $\alpha = \pm 1$, and $\eta_{x1}\eta_{x2}^{-1} = \eta_{y1}\eta_{y2} = e^{i 2\pi \Phi_{13}} = e^{i 2\pi \Phi_{23}}$. The parameters $\eta_{x1},\eta_{y1}$ can be trivialized by a coboundary transformation. The result is consistent with $H^2(Ccc2,U(1)) = \mathbb{Z}_2^3$.

From the projective algebra we obtain
\begin{equation}
	\begin{split}
		& \rho_{M_x} \rho_{\bm{e}_1} \rho_{\bm{e}_2}^{-1} \rho_{M_x}^{-1} = e^{i 2\pi \Phi_{13}} \rho_{\bm{e}_1} \rho_{\bm{e}_2}^{-1},\\
		& \rho_{M_y} \rho_{\bm{e}_1} \rho_{\bm{e}_2} \rho_{M_y}^{-1} = e^{i 2\pi \Phi_{13}} \rho_{\bm{e}_1} \rho_{\bm{e}_2}.
	\end{split}
\end{equation}

Since $\bm{a}_1,\bm{a}_2,\bm{a}_3$ are generated by $\rho_{\bm{e}_1} \rho_{\bm{e}_2}$, $\rho_{\bm{e}_1} \rho_{\bm{e}_2}^{-1}$, and $\rho_{\bm{e}_3}$, and $\bm{\kappa}_{M_x} = \kappa_{M_x}^{1} \bm{Q}_1 + \kappa_{M_x}^{2} \bm{Q}_2 + \kappa_{M_x}^{3} \bm{Q}_3 = (\kappa_{M_x}^{1},\kappa_{M_x}^{2},\frac{1}{2}\kappa_{M_x}^{3})$, the projective algebra indicates that $e^{i2\pi\kappa_{M_x}^{1}} = \eta_{x1}\eta_{x2}$, $e^{i2\pi\kappa_{M_x}^{2}} = e^{i2\pi\Phi_{13}}$, $e^{i2\pi\kappa_{M_x}^{3}} = 1$. Therefore the fractional translation satisfies $\bm{\kappa}_{M_x} + M_x \bm{\kappa}_{M_x} =  (0,1,0) = \bm{Q}_2$. Similarly we also have $\bm{\kappa}_{M_y} + M_y \bm{\kappa}_{M_y} =   (1,0,0) = \bm{Q}_1$, which indicates that the reciprocal-space group is $Pba2$.

\section{Exhaustive computation of $\overline{\Lambda_{\Phi}}$ for the 73 arithmetic classes}

The exhaustive computation of $\overline{\bm{\Lambda}_{\Phi}}$ for 73 arithmetic classes proceeds in the following steps.

\begin{enumerate}
	\item \textbf{Basis of \(L\).} We first list the basis of \(L\) for each arithmetic class. The basis depends only on the Bravais lattice type. The basis of \(L\) for each Bravais lattice is shown in Fig.~\ref{fig:braivs}.
	
	\item \textbf{Basis of \(Z\).} For each arithmetic class, we choose a basis of the center \(Z\) for every possible flux that can appear in that class. The bases of \(Z\) are listed in Table~\ref{table:center}.  
	
	For a given Bravais lattice type, we first enumerate the fluxes that are invariant under the lowest‑symmetry group \(G\); these fluxes belong to \((H^2(L,U(1)))^G\). For higher‑symmetry groups the allowed fluxes form a subset, so the corresponding bases of \(Z\) are also included.  
	
	Each flux type is labelled by the denominators \(n_{ij}\) of the flux form entries \(\Phi_{ij}\). The basis $\bm{a}_i$ of \(Z\) is expressed as a linear combination of the basis vectors \(\bm{e}_1,\bm{e}_2,\bm{e}_3\) of \(L\); we list the coefficients of this combination. Flux types that are equivalent under exchange of basis vectors are omitted. For example, in the orthorhombic primitive lattice (oP), the type with \(n_{12}=1,\; n_{23}=1,\; n_{31}=2\) is equivalent to \(n_{12}=2,\; n_{23}=1,\; n_{31}=1\).
	
	\item \textbf{Connection matrix and standard choice of \(S(g)\).} Using the basis $\bm{a}_i$ of \(Z\) we define the connection matrix  
	\[
	A = (V^{-1})^T (V^T \Phi V)^+ V^{-1},
	\]  
	where $V=(a_1,a_2,a_3)$ with $a_i$ being the column vector of $\bm{a}_i$ is the transformation matrix from  $L$ to $Z$.
	A standard choice for \(S(g)\) is then  
	\begin{equation}
		S(g) = \delta A(g) - \bigl(E_{g^{-1}}^T \Phi E_{g^{-1}} - \Phi\bigr)^+
		= E_{g^{-1}}^T A E_{g^{-1}} - A - \bigl(E_{g^{-1}}^T \Phi E_{g^{-1}} - \Phi\bigr)^+,
	\end{equation}
	which allows us to fix the coboundary contribution to \(\overline{\bm{\Lambda}_{\Phi}} \).
	
	\item \textbf{Computation results.} By explicit calculation we find that \(\overline{\bm{\Lambda}_{\Phi}} = 0  \) for most arithmetic classes; hence \(\overline{\bm{\Lambda}_{\Phi}}\) is automatically trivial in those cases.  
	
	For 29 arithmetic classes, \(\overline{\bm{\Lambda}_{\Phi}}\) is non‑zero for certain flux forms. These classes are:  
	\(2C\), \(mC\), \(2m/C\), \(222I\), \(222F\), \(mm2F\), \(mm2I\), \(mmmF\), \(mmmI\), \(4P\), \(\bar{4}P\), \(4/mP\), \(422P\), \(422I\), \(4mmP\), \(\bar{4}m2P\), \(\bar{4}2mP\), \(321P\), \(312P\), \(3m1P\), \(31mP\), \(\bar{3}1mP\), \(\bar{3}m1P\), \(622P\), \(\bar{6}m2P\), \(\bar{6}2mP\), \(6/mmmP\), \(23F\), \(m\bar{3}F\).  
	
	Among these, the only classes where \(\overline{\bm{\Lambda}_{\Phi}}\) is genuinely non‑trivial (i.e., not a coboundary) are \(mm2F\), \(mmmF\), and \(m\bar{3}F\) . For the remaining 26 classes, \(\overline{\bm{\Lambda}_{\Phi}}\) is non‑zero but equals a coboundary.
	
	\item \textbf{Coboundary representation.} We explicitly show that for the 26 non‑zero yet trivial cases, \( \overline{\bm{\Lambda}_{\Phi}} = \delta \bm{\chi}\) for some \(\bm{\chi}\).
\end{enumerate}

\begin{figure}[ht]
	\centering
	\includegraphics[width=18cm]{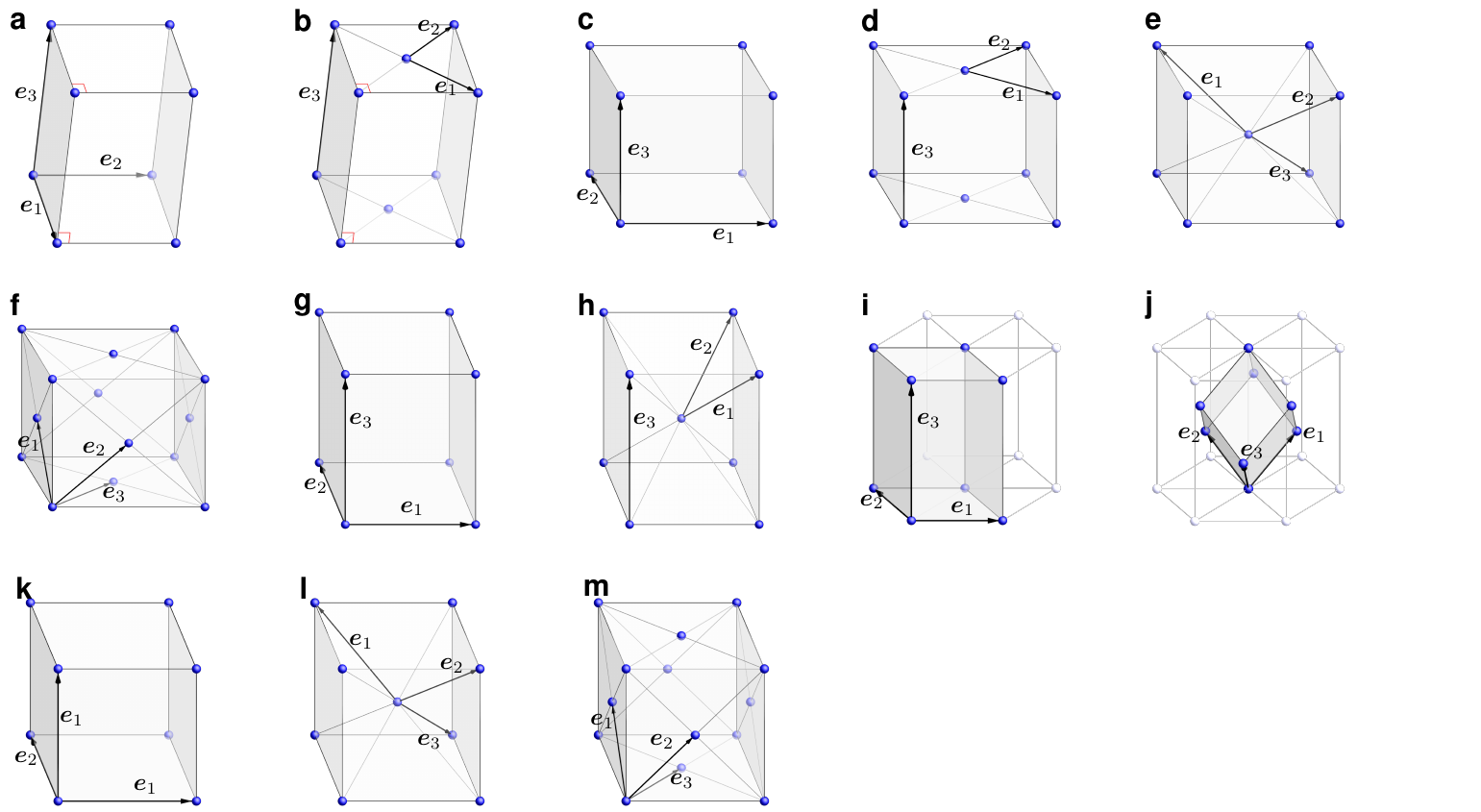}
	\caption{Basis choice of bravis lattice. We omit the triclinic lattice since it does not restrict the orientation of the basis vectors. (a)Monoclinc primitve. (b)Monoclinc base-centered. (c)Orthorhombic primitive. (d)Orthorhombic base-centered (e)Orthorhombic body-centered (f)Orthorhombic face-centered (g)Tetragonal primitive (h)Tetragonal body-centered (i)Hexagonal  primitive (j)Hexagonal  rhombohedral (k)Cubic primitive (l)Cubic body-centered (m)Cubic face-centered.}
	\label{fig:braivs}
\end{figure}

\begin{table}[H]
	\centering
	\includegraphics[width=18cm]{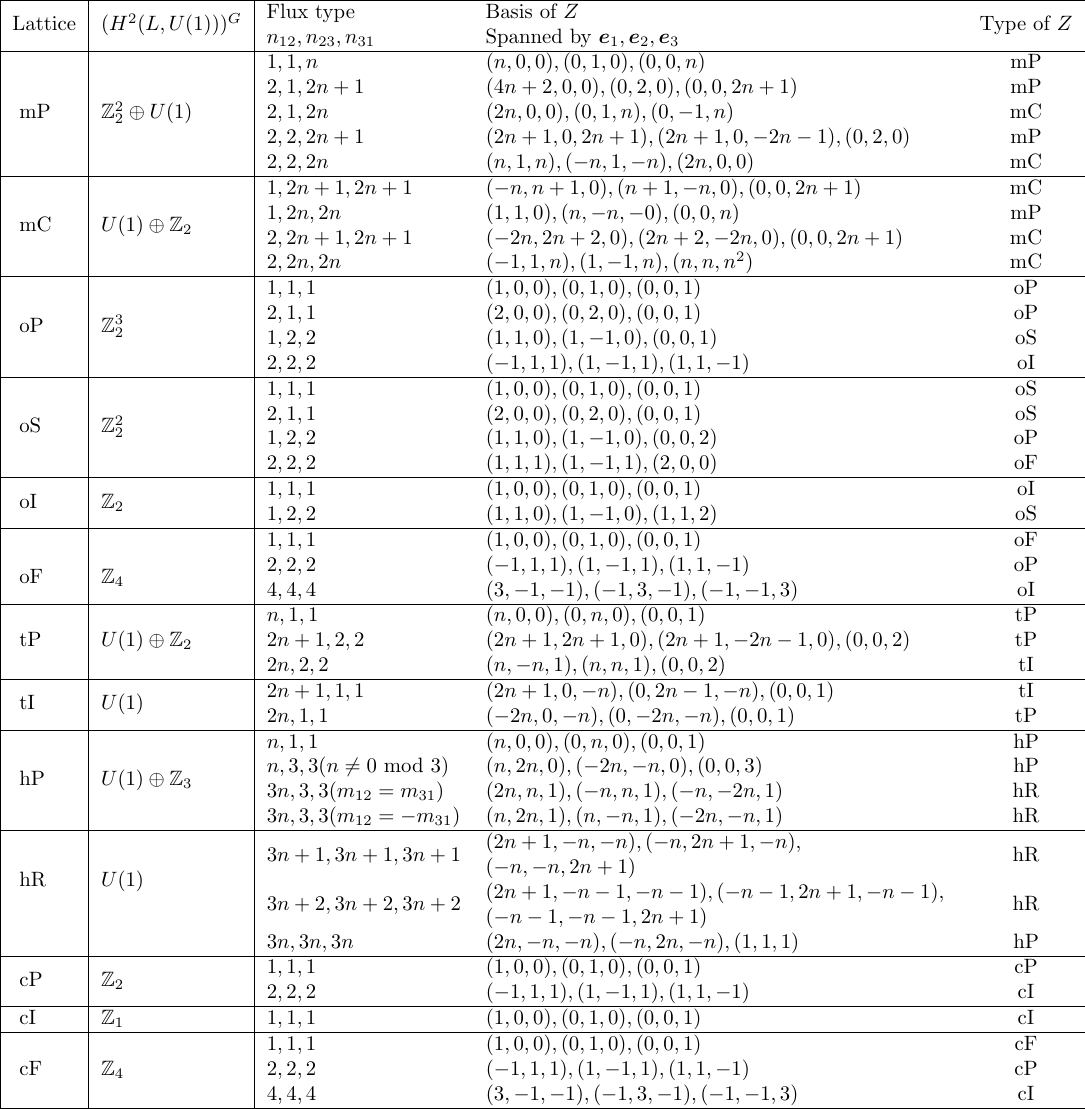}
	\caption{Basis choice of $Z$ for admissible fluxes in each type of Bravis lattice. The Bravais lattice types are abbreviated as follows: mP (monoclinic primitive), mC (monoclinic base-centered), oP (orthorhombic primitive), oS (orthorhombic base-centered), oI (orthorhombic body-centered), oF (orthorhombic face-centered), tP (tetragonal primitive), tI (tetragonal body-centered), hP (hexagonal primitive), hR (hexagonal rhombohedral), cP (cubic primitive), cI (cubic body-centered), cF (cubic face-centered). We omit the triclinic lattice since it does not restrict any flux.
	}
	\label{table:center}
\end{table}

\newpage
\section{Nonzero $\overline{\Lambda_\Phi}$ with trivial cohomology class}

In this section, we explicitly show that for the 26 arithmetic classes with \(\overline{\bm{\Lambda}_{\Phi}} \neq 0\) is a coboundary, i.e., \(\overline{\bm{\Lambda}_{\Phi}} = \delta \bm{\chi}\) for some \(\bm{\chi}\). For a given arithmetic class, we only consider the flux form \(\Phi\) when \(\overline{\bm{\Lambda}_{\Phi}} \neq 0\); cases where it vanishes are omitted.

\subsection{$2C$}\label{section:2C}
Consider arithmetic class $2C$. We choose the basis of $L$ and $L_F$ as
\begin{equation}
	\begin{split}
		&\bm{e}_1 =\frac{1}{2} (1,-1,0),\quad \bm{e}_2 = \frac{1}{2}(1,1,0),\quad \bm{e}_3 = (0,0,1),\\
		&\bm{G}_1 = (1,-1,0),\quad \bm{G}_2 = (1,1,0),\quad \bm{G}_3 = (0,0,1).
	\end{split}
\end{equation}
When the flux form takes the value
\begin{equation}
	\Phi = \begin{bmatrix}
		0 & \frac{1}{2} & \frac{1}{2n} \\
		-\frac{1}{2} & 0 & \frac{1}{2n} \\
		-\frac{1}{2n} & -\frac{1}{2n} & 0
	\end{bmatrix},
\end{equation}
a basis of $Z$ is given by
\[
\bm{a}_1 = -\bm{e}_1 + \bm{e}_2 + n\bm{e}_3,\quad
\bm{a}_2 = \bm{e}_1 - \bm{e}_2 + n\bm{e}_3,\quad
\bm{a}_3 = n\bm{e}_1 + n\bm{e}_2 + n^2\bm{e}_3.
\]
Thus the bases of $Z$ and $Z_F$ are
\begin{equation}
	\begin{split}
		&\bm{a}_1 = (0,1,n),\quad \bm{a}_2 = (0,-1,n),\quad \bm{a}_3 = (n,0,n^2),\\
		&\bm{Q}_1 = \frac{1}{2} (-1,1,\tfrac{1}{n}),\quad
		\bm{Q}_2 = \frac{1}{2} (-1,-1,\tfrac{1}{n}),\quad
		\bm{Q}_3 = \frac{1}{n}(1,0,0).
	\end{split}
\end{equation}
The quotient $Z_F/L_F$ is isomorphic to $\mathbb{Z}_{2n} \times \mathbb{Z}_{2n}$. A convenient basis for $\mathbb{Z}_{2n} \times \mathbb{Z}_{2n}$ is given by the cosets of $\bm{Q}_1$ and $\bm{Q}_3$ (since $\bm{Q}_2 \equiv \bm{Q}_1 + n\bm{Q}_3 \pmod{L_F}$).

The action of $C_{2}$ on the $\bm{Q}_i$ is given by
\begin{align}
	R_y \bm{Q}_1 = -\bm{Q}_2,\quad R_y \bm{Q}_3 = -\bm{Q}_3.
\end{align}
Passing to the quotient $Z_F/L_F$ and using the relations $\bm{Q}_2 \equiv \bm{Q}_1 + n\bm{Q}_3 \pmod{L_F}$ and $2n\bm{Q}_i \equiv 0$, we obtain the induced action as matrices (mod~$2n$):
\begin{equation}
	R_y = \begin{bmatrix}
		-1 & 0 \\
		n & -1
	\end{bmatrix}
\end{equation}
acting by left multiplication on column vectors of $\mathbb{Z}_{2n} \times \mathbb{Z}_{2n}$.

When $n$ is odd, $\overline{\bm{\Lambda}_\Phi} = 0$.
When $n$ is even, $\overline{\bm{\Lambda}_\Phi}$ (listed in the order $E$, $R_y$) is given by
\begin{equation}
	\overline{\bm{\Lambda}_\Phi} = \begin{bmatrix}
		\begin{pmatrix}0\\0\end{pmatrix} & \begin{pmatrix}0\\0\end{pmatrix} \\
		\begin{pmatrix}0\\0\end{pmatrix} & \begin{pmatrix}0\\n\end{pmatrix}
	\end{bmatrix} = \delta \begin{bmatrix}
		\begin{pmatrix}0\\0\end{pmatrix} \\
		\begin{pmatrix}1\\0\end{pmatrix}
	\end{bmatrix}.
\end{equation}

\subsection{$mC$}
For arithmetic class $mC$ with the flux form  $\Phi =\begin{bmatrix}
	0 & \frac{1}{2} & \frac{1}{2n}  \\
	-\frac{1}{2} & 0 & \frac{1}{2n} \\
	-\frac{1}{2n}  & -\frac{1}{2n}  &0
\end{bmatrix},$ 
the choice of basis is the same  as in Section.\ref{section:2C}, 

The quotient $Z_F/L_F$ is isomorphic to $\mathbb{Z}_{2n} \times \mathbb{Z}_{2n}$. A convenient basis for $\mathbb{Z}_{2n} \times \mathbb{Z}_{2n}$ is given by the cosets of $\bm{Q}_1$ and $\bm{Q}_3$(since $\bm{Q}_2 \equiv \bm{Q}_1 + n\bm{Q}_3 \pmod{L_F}$).

The action of $D_{1}$ on the $\bm{Q}_i$ is given by
\begin{align}
	M_y \bm{Q}_1 =  \bm{Q}_2,\quad M_y \bm{Q}_3 =  \bm{Q}_3.
\end{align}
Passing to the quotient $Z_F/L_F$ and using the relations $\bm{Q}_2 \equiv \bm{Q}_1 + n\bm{Q}_3 \pmod{L_F}$ and $2n\bm{Q}_i \equiv 0$, we obtain the induced action as matrices (mod~$2n$):
\begin{equation}
	M_y = \begin{bmatrix}
		1 & 0 \\
		n & 1
	\end{bmatrix}.
\end{equation}

The two‑cocycle  $\overline{\bm{\Lambda}_\Phi}$( listed in order of $E, M_y$) is given by
\begin{equation}
	\begin{split}
		&\overline{\bm{\Lambda}_\Phi}  = \begin{bmatrix}
			\begin{pmatrix}	0 \\ 0	\end{pmatrix} & \begin{pmatrix}	0 \\ 0	\end{pmatrix} \\
			\begin{pmatrix}	0 \\ 0	\end{pmatrix} & \begin{pmatrix}	2(n-1) \\ 0	\end{pmatrix}    
		\end{bmatrix}  =  \delta \begin{bmatrix}
			\begin{pmatrix}	0 \\ 0	\end{pmatrix} \\
			\begin{pmatrix}	n-1 \\ -\frac{n(n-1)}{2}	\end{pmatrix} 
		\end{bmatrix}.
	\end{split} 
\end{equation}

\subsection{$2/mC$}
Consider arithmetic class $2/mC$ with the flux form  $\Phi =\begin{bmatrix}
	0 & \frac{1}{2} & \frac{1}{2n}  \\
	-\frac{1}{2} & 0 & \frac{1}{2n} \\
	-\frac{1}{2n}  & -\frac{1}{2n}  &0
\end{bmatrix}.$
The choice of basis is the same  as in Section.\ref{section:2C}.
The quotient $Z_F/L_F$ is isomorphic to $\mathbb{Z}_{2n} \times \mathbb{Z}_{2n}$. A convenient basis for $\mathbb{Z}_{2n} \times \mathbb{Z}_{2n}$ is given by the cosets of $\bm{Q}_1$ and $\bm{Q}_3$.

The group elements  $R_y, M_y$ act  on $\mathbb{Z}_{2n} \times \mathbb{Z}_{2n}$ via  matrices (mod $2n$)
\begin{equation}
	R_y = \begin{bmatrix}
		-1 & 0 \\
		n & -1
	\end{bmatrix},	M_y = \begin{bmatrix}
		1 & 0 \\
		n & 1
	\end{bmatrix}  .
\end{equation}

a. When $n$ is even, the two‑cocycle  $\overline{\bm{\Lambda}_\Phi} $( listed in order of $E, R_y, M_y, M_y R_y   $) is given by
\begin{equation}
	\begin{split}
		&\overline{\bm{\Lambda}_\Phi}   = \begin{bmatrix}
			\begin{pmatrix}	0 \\ 0	\end{pmatrix} & \begin{pmatrix}	0 \\ 0	\end{pmatrix} & \begin{pmatrix}	0 \\ 0	\end{pmatrix}  & \begin{pmatrix}	0 \\ 0	\end{pmatrix}  \\
			\begin{pmatrix}	0 \\ 0	\end{pmatrix} & \begin{pmatrix}	0 \\ n	\end{pmatrix}   & \begin{pmatrix}	0 \\ n	\end{pmatrix} & \begin{pmatrix}	0 \\ 0	\end{pmatrix} \\ 
			\begin{pmatrix}	0 \\ 0	\end{pmatrix} & \begin{pmatrix}	2(n-1) \\ 0	\end{pmatrix}   & \begin{pmatrix}	2(n-1)  \\ 0	\end{pmatrix} & \begin{pmatrix}	0 \\ 0	\end{pmatrix} \\ 
			\begin{pmatrix}	0 \\ 0	\end{pmatrix} & \begin{pmatrix}	2 \\ n	\end{pmatrix}   & \begin{pmatrix}	2 \\ n 	\end{pmatrix} & \begin{pmatrix}	0 \\ 0	\end{pmatrix} \\ 
		\end{bmatrix}  =  \delta \begin{bmatrix}
			\begin{pmatrix}	0 \\ 0	\end{pmatrix} \\
			\begin{pmatrix}	1 \\ 0	\end{pmatrix} \\
			\begin{pmatrix}	n-1 \\ -\frac{n(n-1)}{2}	\end{pmatrix} \\
			\begin{pmatrix}	2-n \\ -\frac{n(n+1)}{2}	\end{pmatrix} \\
		\end{bmatrix}.
	\end{split} 
\end{equation}

b. When $n$ is odd, the two‑cocycle   $\bm{\Lambda}_\Phi $ ( listed in order of $E, R_y, M_y,  M_y R_y  $) is given by
\begin{equation}
	\begin{split}
		&\overline{\bm{\Lambda}_\Phi}  = \begin{bmatrix}
			\begin{pmatrix}	0 \\ 0	\end{pmatrix} & \begin{pmatrix}	0 \\ 0	\end{pmatrix} & \begin{pmatrix}	0 \\ 0	\end{pmatrix}  & \begin{pmatrix}	0 \\ 0	\end{pmatrix}  \\
			\begin{pmatrix}	0 \\ 0	\end{pmatrix} & \begin{pmatrix}	0 \\ 0	\end{pmatrix}   & \begin{pmatrix}	0 \\0	\end{pmatrix} & \begin{pmatrix}	0 \\ 0	\end{pmatrix} \\ 
			\begin{pmatrix}	0 \\ 0	\end{pmatrix} & \begin{pmatrix}	2(n-1) \\ 0	\end{pmatrix}   & \begin{pmatrix}	2(n-1)  \\ 0	\end{pmatrix} & \begin{pmatrix}	0 \\ 0	\end{pmatrix} \\ 
			\begin{pmatrix}	0 \\ 0	\end{pmatrix} & \begin{pmatrix}	2 \\ 0	\end{pmatrix}   & \begin{pmatrix}	2 \\ 0	\end{pmatrix} & \begin{pmatrix}	0 \\ 0	\end{pmatrix} \\ 
		\end{bmatrix}  =  \delta \begin{bmatrix}
			\begin{pmatrix}	0 \\ 0	\end{pmatrix} \\
			\begin{pmatrix}	0 \\ 0	\end{pmatrix} \\
			\begin{pmatrix}	n-1 \\ -\frac{n(n-1)}{2}	\end{pmatrix} \\
			\begin{pmatrix}	1-n \\ -\frac{n(n-1)}{2}	\end{pmatrix} \\
		\end{bmatrix}.
	\end{split} 
\end{equation}

\subsection{$222F$}\label{section:222F}

Consider arithmetic class $222F$. We choose the basis of $L$ and $L_F$ as
\begin{equation}
	\begin{split}
		&\bm{e}_1 =\frac{1}{2} (0,1,1),\quad \bm{e}_2 = \frac{1}{2}(1,0,1),\quad \bm{e}_3 = \frac{1}{2}(1,1,0),\\
		&\bm{G}_1 = (-1,1,1),\quad \bm{G}_2 = (1,-1,1),\quad \bm{G}_3 = (1,1,-1).
	\end{split}
\end{equation}
When the flux form takes the value
\begin{equation}
	\Phi = \begin{bmatrix}
		0 & \frac{1}{4} & -\frac{1}{4} \\
		-\frac{1}{4} & 0 & \frac{1}{4} \\
		\frac{1}{4} & -\frac{1}{4} & 0
	\end{bmatrix},
\end{equation}
a basis of $Z$ is given by
\[
\bm{a}_1 = 3\bm{e}_1 -\bm{e}_2 -\bm{e}_3,\quad
\bm{a}_2 = -\bm{e}_1+ 3\bm{e}_2 + -\bm{e}_3,\quad
\bm{a}_3 = -\bm{e}_1 - \bm{e}_2 + 3\bm{e}_3.
\]
Thus the bases of $Z$ and $Z_F$ are
\begin{equation}
	\begin{split}
		&\bm{a}_1 = (-1,1,1),\quad \bm{a}_2 = (1,-1,1),\quad \bm{a}_3 = (1,1,-1),\\
		&\bm{Q}_1 =\frac{1}{2} (0,1,1),\quad \bm{Q}_2 = \frac{1}{2}(1,0,1),\quad \bm{Q}_3 = \frac{1}{2}(1,1,0).
	\end{split}
\end{equation}
The quotient $Z_F/L_F$ is isomorphic to $\mathbb{Z}_{4} \times \mathbb{Z}_{4}$. A convenient basis for $\mathbb{Z}_{4} \times \mathbb{Z}_{4}$ is given by the cosets of $\bm{Q}_1$ and $\bm{Q}_2$(since $\mathbf{Q}_3 \equiv -\mathbf{Q}_1-\mathbf{Q}_2 \pmod{L_F}$).
The action of $D_{2}$ on the $\mathbf{Q}_i$ is given by
\begin{align}
	R_x \bm{Q}_1 =- \bm{Q}_1,\quad R_x \bm{Q}_2 = -\bm{Q}_1+ \bm{Q}_3 ,\quad R_y \bm{Q}_1 = -\bm{Q}_2 + \bm{Q}_3,\quad R_y \bm{Q}_2 = -\bm{Q}_2
\end{align}
Passing to the quotient $\mathbb{Z}_{4} \times \mathbb{Z}_{4}$ and using the relations $\mathbf{Q}_3 \equiv -\mathbf{Q}_1-\mathbf{Q}_2 \pmod{L_F}$ and $4\mathbf{Q}_i \equiv 0$, we obtain the induced action as matrices (mod~4):

\begin{equation}
	\begin{split}
		R_x = \begin{bmatrix} -1 & 2 \\ 0 & -1 \end{bmatrix}, \qquad
		R_y = \begin{bmatrix} -1 & 0 \\ 2 & -1 \end{bmatrix},
	\end{split}
\end{equation}
acting by left multiplication on column vectors of $  \mathbb{Z}_4 \times \mathbb{Z}_4$.

The two‑cocycle  $\overline{\bm{\Lambda}_\Phi}$( listed in order of $E, R_x, R_y, R_z$) is given by
\begin{equation}
	\begin{split}
		&\overline{\bm{\Lambda}_\Phi}  = \begin{bmatrix}
			\begin{pmatrix}	0 \\ 0	\end{pmatrix} & \begin{pmatrix}	0 \\ 0	\end{pmatrix}  & \begin{pmatrix}	0 \\ 0	\end{pmatrix}  & \begin{pmatrix}	0 \\ 0	\end{pmatrix} \\
			\begin{pmatrix}	0 \\ 0	\end{pmatrix} & \begin{pmatrix}	2 \\ 0	\end{pmatrix}  & \begin{pmatrix}	2 \\ 2	\end{pmatrix}  & \begin{pmatrix}	0 \\ 2	\end{pmatrix} \\
			\begin{pmatrix}	0 \\ 0	\end{pmatrix} & \begin{pmatrix}	2 \\ 2	\end{pmatrix}  & \begin{pmatrix}	0 \\ 2	\end{pmatrix}  & \begin{pmatrix}	2 \\ 0	\end{pmatrix} \\
			\begin{pmatrix}	0 \\ 0	\end{pmatrix} & \begin{pmatrix}	0 \\ 2	\end{pmatrix}  & \begin{pmatrix}	2 \\ 0	\end{pmatrix}  & \begin{pmatrix}	2 \\ 2	\end{pmatrix}  
		\end{bmatrix}  =  \delta \begin{bmatrix}
			\begin{pmatrix}	0 \\ 0	\end{pmatrix} \\
			\begin{pmatrix}	2 \\ 1	\end{pmatrix} \\
			\begin{pmatrix}	1 \\ 1	\end{pmatrix} \\
			\begin{pmatrix}	1 \\ 2	\end{pmatrix}   
		\end{bmatrix}.
	\end{split} 
\end{equation}

\subsection{$222I$}\label{section:222I}
For arithmetic  class $222I$, we choose the basis of $L$ and $L_F$ as
\begin{equation}
	\begin{split}
		&\bm{e}_1 =\frac{1}{2}(-1,1,1),\quad \bm{e}_2 = \frac{1}{2}(1,-1,1),\quad \bm{e}_3 = \frac{1}{2}(1,1,-1),\\
		&\bm{G}_1 =(0,1,1) ,\quad \bm{G}_2 = (1,0,1),\quad \bm{G}_3 = (1,1,0).
	\end{split}
\end{equation}
When the flux form takes the value
\begin{equation}
	\Phi = \begin{bmatrix}
		0 & 0 & \frac{1}{2}  \\
		0 & 0 & \frac{1}{2} \\
		-\frac{1}{2}  & -\frac{1}{2}  &0
	\end{bmatrix},
\end{equation}
a basis of $Z$ is given by
\[
\bm{a}_1 =   \bm{e}_1 +   \bm{e}_2,\quad \bm{a}_2 =   \bm{e}_1  -  \bm{e}_2,\quad \bm{a}_3 =   \bm{e}_1+ \bm{e}_2 +2 \bm{e}_3
\]
Thus the bases of $Z$ and $Z_F$ are
\begin{equation}
	\begin{split}
		&\bm{a}_1 = (0,0,1),\quad \bm{a}_2 = (-1,1,0),\quad \bm{a}_3 = (1,1,0),\\
		&\bm{Q}_1 =  (0,0,1),\quad \bm{Q}_2 = \frac{1}{2}(-1,1,0),\quad \bm{Q}_3 =\frac{1}{2} (1,1,0).
	\end{split}
\end{equation}
The quotient $Z_F/L_F$ is isomorphic to $\mathbb{Z}_{2} \times \mathbb{Z}_{2}$. A convenient basis for $\mathbb{Z}_{2} \times \mathbb{Z}_{2}$ is given by the cosets of $\bm{Q}_2$ and $\bm{Q}_3$(since $\bm{Q}_1 \equiv \bm{Q}_2 + \bm{Q}_3 \pmod{L_F}$).
The action of $D_{2}$ on the $\mathbf{Q}_i$ is given by
\begin{align}
	R_x \bm{Q}_2 =- \bm{Q}_3,\quad R_x \bm{Q}_3 = - \bm{Q}_2 ,\quad R_y \bm{Q}_2 =  \bm{Q}_3  ,\quad R_y \bm{Q}_3 =  \bm{Q}_2.
\end{align}
Passing to the quotient $\mathbb{Z}_{2} \times \mathbb{Z}_{2}$ and using the relations  $2\mathbf{Q}_i \equiv 0$, we obtain the induced action as matrices (mod~2):

\begin{equation}
	\begin{split}
		R_x = \begin{bmatrix} 0 & 1 \\ 1 & 0 \end{bmatrix}, \qquad
		R_y = \begin{bmatrix} 0 & 1 \\ 1& 0 \end{bmatrix},
	\end{split}
\end{equation}
acting by left multiplication on column vectors of $  \mathbb{Z}_2 \times \mathbb{Z}_2$.

The two‑cocycle   $\overline{\bm{\Lambda}_\Phi}$( listed in order of $E, R_x, R_y, R_z$) is given by
\begin{equation}
	\begin{split}
		&\overline{\bm{\Lambda}_\Phi}    = \begin{bmatrix}
			\begin{pmatrix}	0 \\ 0	\end{pmatrix} & \begin{pmatrix}	0 \\ 0	\end{pmatrix}  & \begin{pmatrix}	0 \\ 0	\end{pmatrix}  & \begin{pmatrix}	0 \\ 0	\end{pmatrix} \\
			\begin{pmatrix}	0 \\ 0	\end{pmatrix} & \begin{pmatrix}0 \\ 0	\end{pmatrix}  & \begin{pmatrix}	0 \\ 0	\end{pmatrix}  & \begin{pmatrix}	0 \\ 0	\end{pmatrix} \\
			\begin{pmatrix}	0 \\ 0	\end{pmatrix} & \begin{pmatrix}	1\\ 1	\end{pmatrix}  & \begin{pmatrix}	1 \\ 1	\end{pmatrix}  & \begin{pmatrix}	0 \\ 0	\end{pmatrix} \\
			\begin{pmatrix}	0 \\ 0	\end{pmatrix} & \begin{pmatrix}	1 \\ 1	\end{pmatrix}  & \begin{pmatrix}	1 \\ 1	\end{pmatrix}  & \begin{pmatrix}	0 \\ 0	\end{pmatrix}  
		\end{bmatrix}  =  \delta \begin{bmatrix}
			\begin{pmatrix}	0 \\ 0	\end{pmatrix} \\
			\begin{pmatrix}	0 \\ 0	\end{pmatrix} \\
			\begin{pmatrix}	0 \\ 1	\end{pmatrix} \\
			\begin{pmatrix}	1\\ 0	\end{pmatrix}   
		\end{bmatrix}.
	\end{split} 
\end{equation}

\subsection{$mm2I$}
For arithmetic class $mm2I$ with the flux form $\Phi = \begin{bmatrix}
	0 & 0 & \frac{1}{2}  \\
	0 & 0 & \frac{1}{2} \\
	-\frac{1}{2}  & -\frac{1}{2}  &0
\end{bmatrix}$, the choice of basis is the same  as in Section.\ref{section:222I}.

The quotient $Z_F/L_F$ is isomorphic to $\mathbb{Z}_{2} \times \mathbb{Z}_{2}$. A convenient basis for $\mathbb{Z}_{2} \times \mathbb{Z}_{2}$ is given by the cosets of $\bm{Q}_2$ and $\bm{Q}_3$(since $\bm{Q}_1 \equiv \bm{Q}_2 + \bm{Q}_3 \pmod{L_F}$).
The action of $C_{2v}$ on the $\mathbf{Q}_i$ is given by
\begin{align}
	M_x \bm{Q}_2 =  \bm{Q}_3,\quad M_x \bm{Q}_3 =   \bm{Q}_2 ,\quad M_y \bm{Q}_2 =  -\bm{Q}_3  ,\quad M_y \bm{Q}_3 =  -\bm{Q}_2.
\end{align}
Passing to the quotient $\mathbb{Z}_{2} \times \mathbb{Z}_{2}$ and using the relations  $2\mathbf{Q}_i \equiv 0$, we obtain the induced action as matrices (mod~2):

\begin{equation}
	\begin{split}
		M_x = \begin{bmatrix} 0 & 1 \\ 1 & 0 \end{bmatrix}, \qquad
		M_y = \begin{bmatrix} 0 & 1 \\ 1& 0 \end{bmatrix},
	\end{split}
\end{equation}
acting by left multiplication on column vectors of $  \mathbb{Z}_2 \times \mathbb{Z}_2$.

The two‑cocycle   $\overline{\bm{\Lambda}_\Phi}$( listed in order of $E, M_x, M_y, R_z$) is given by
\begin{equation}
	\begin{split}
		&\overline{\bm{\Lambda}_\Phi} =\begin{bmatrix}
			\begin{pmatrix}	0 \\ 0	\end{pmatrix} & \begin{pmatrix}	0 \\ 0	\end{pmatrix}  & \begin{pmatrix}	0 \\ 0	\end{pmatrix}  & \begin{pmatrix}	0 \\ 0	\end{pmatrix} \\
			\begin{pmatrix}	0 \\ 0	\end{pmatrix} & \begin{pmatrix}0 \\ 0	\end{pmatrix}  & \begin{pmatrix}	0 \\ 0	\end{pmatrix}  & \begin{pmatrix}	0 \\ 0	\end{pmatrix} \\
			\begin{pmatrix}	0 \\ 0	\end{pmatrix} & \begin{pmatrix}	1\\ 1	\end{pmatrix}  & \begin{pmatrix}	1 \\ 1	\end{pmatrix}  & \begin{pmatrix}	0 \\ 0	\end{pmatrix} \\
			\begin{pmatrix}	0 \\ 0	\end{pmatrix} & \begin{pmatrix}	1 \\ 1	\end{pmatrix}  & \begin{pmatrix}	1 \\ 1	\end{pmatrix}  & \begin{pmatrix}	0 \\ 0	\end{pmatrix}  
		\end{bmatrix}  =  \delta \begin{bmatrix}
			\begin{pmatrix}	0 \\ 0	\end{pmatrix} \\
			\begin{pmatrix}	0 \\ 0	\end{pmatrix} \\
			\begin{pmatrix}	0 \\ 1	\end{pmatrix} \\
			\begin{pmatrix}	1\\ 0	\end{pmatrix}   
		\end{bmatrix}.
	\end{split} 
\end{equation}

\subsection{$mmmI$}
For arithmetic  class $mmmI$ with flux form $\Phi = \begin{bmatrix}
	0 & 0 & \frac{1}{2}  \\
	0 & 0 & \frac{1}{2} \\
	-\frac{1}{2}  & -\frac{1}{2}  &0
\end{bmatrix}$ ,the choice of basis is the same  as in Section.\ref{section:222I}.
The quotient $Z_F/L_F$ is isomorphic to $\mathbb{Z}_{2} \times \mathbb{Z}_{2}$. A convenient basis for $\mathbb{Z}_{2} \times \mathbb{Z}_{2}$ is given by the cosets of $\bm{Q}_2$ and $\bm{Q}_3$.

The group elements $M_x$, $M_y$, and $M_z$ act on $\mathbb{Z}_2 \times \mathbb{Z}_2$ via  matrices (mod $2$)
\begin{equation}
	M_x = \begin{bmatrix}
		0 & 1 \\
		1 &0
	\end{bmatrix},M_y = \begin{bmatrix}
		0 & 1 \\
		1& 0
	\end{bmatrix},M_z = \begin{bmatrix}
		1 &0 \\
		0 & 1
	\end{bmatrix} .
\end{equation}

The two‑cocycle   $\overline{\bm{\Lambda}_\Phi}$( listed in order of $E, M_x, M_y, M_x M_y,M_z, M_x M_z, M_y M_z, M_x M_y M_z$) is given by
\begin{equation}
	\begin{split}
		&\overline{\bm{\Lambda}_\Phi}  = \begin{bmatrix}
			\begin{pmatrix}	0 \\ 0	\end{pmatrix} & \begin{pmatrix}	0 \\ 0	\end{pmatrix}  & \begin{pmatrix}	0 \\ 0	\end{pmatrix}  & \begin{pmatrix}	0 \\ 0	\end{pmatrix} & 	\begin{pmatrix}	0 \\ 0	\end{pmatrix} & \begin{pmatrix}	0 \\ 0	\end{pmatrix}  & \begin{pmatrix}	0 \\ 0	\end{pmatrix}  & \begin{pmatrix}	0 \\ 0	\end{pmatrix} \\
			\begin{pmatrix}	0 \\ 0	\end{pmatrix} & \begin{pmatrix}0 \\ 0	\end{pmatrix}  & \begin{pmatrix}	0 \\ 0	\end{pmatrix}  & \begin{pmatrix}	0 \\ 0	\end{pmatrix}  & 	\begin{pmatrix}	0 \\ 0	\end{pmatrix} & \begin{pmatrix}	0 \\ 0	\end{pmatrix}  & \begin{pmatrix}	0 \\ 0	\end{pmatrix}  & \begin{pmatrix}	0 \\ 0	\end{pmatrix}\\
			\begin{pmatrix}	0 \\ 0	\end{pmatrix} & \begin{pmatrix}	1\\ 1	\end{pmatrix}  & \begin{pmatrix}	1 \\ 1	\end{pmatrix}  & \begin{pmatrix}	0 \\ 0	\end{pmatrix} & \begin{pmatrix}	0 \\ 0	\end{pmatrix} & \begin{pmatrix}	1\\ 1	\end{pmatrix}  & \begin{pmatrix}	1 \\ 1	\end{pmatrix}  & \begin{pmatrix}	0 \\ 0	\end{pmatrix} \\
			\begin{pmatrix}	0 \\ 0	\end{pmatrix} & \begin{pmatrix}	1 \\ 1	\end{pmatrix}  & \begin{pmatrix}	1 \\ 1	\end{pmatrix}  & \begin{pmatrix}	0 \\ 0	\end{pmatrix}   & \begin{pmatrix}	0 \\ 0	\end{pmatrix} & \begin{pmatrix}	1\\ 1	\end{pmatrix}  & \begin{pmatrix}	1 \\ 1	\end{pmatrix}  & \begin{pmatrix}	0 \\ 0	\end{pmatrix} \\
			\begin{pmatrix}	0 \\ 0	\end{pmatrix} & \begin{pmatrix}	1\\ 1	\end{pmatrix}  & \begin{pmatrix}	1 \\ 1	\end{pmatrix}  & \begin{pmatrix}	0 \\ 0	\end{pmatrix} & \begin{pmatrix}	0 \\ 0	\end{pmatrix} & \begin{pmatrix}	1\\ 1	\end{pmatrix}  & \begin{pmatrix}	1 \\ 1	\end{pmatrix}  & \begin{pmatrix}	0 \\ 0	\end{pmatrix} \\
			\begin{pmatrix}	0 \\ 0	\end{pmatrix} & \begin{pmatrix}	1 \\ 1	\end{pmatrix}  & \begin{pmatrix}	1 \\ 1	\end{pmatrix}  & \begin{pmatrix}	0 \\ 0	\end{pmatrix}   & \begin{pmatrix}	0 \\ 0	\end{pmatrix} & \begin{pmatrix}	1\\ 1	\end{pmatrix}  & \begin{pmatrix}	1 \\ 1	\end{pmatrix}  & \begin{pmatrix}	0 \\ 0	\end{pmatrix} \\
			\begin{pmatrix}	0 \\ 0	\end{pmatrix} & \begin{pmatrix}	0 \\ 0	\end{pmatrix}  & \begin{pmatrix}	0 \\ 0	\end{pmatrix}  & \begin{pmatrix}	0 \\ 0	\end{pmatrix} & 	\begin{pmatrix}	0 \\ 0	\end{pmatrix} & \begin{pmatrix}	0 \\ 0	\end{pmatrix}  & \begin{pmatrix}	0 \\ 0	\end{pmatrix}  & \begin{pmatrix}	0 \\ 0	\end{pmatrix} \\
			\begin{pmatrix}	0 \\ 0	\end{pmatrix} & \begin{pmatrix}0 \\ 0	\end{pmatrix}  & \begin{pmatrix}	0 \\ 0	\end{pmatrix}  & \begin{pmatrix}	0 \\ 0	\end{pmatrix}  & 	\begin{pmatrix}	0 \\ 0	\end{pmatrix} & \begin{pmatrix}	0 \\ 0	\end{pmatrix}  & \begin{pmatrix}	0 \\ 0	\end{pmatrix}  & \begin{pmatrix}	0 \\ 0	\end{pmatrix}\\
		\end{bmatrix}  =  \delta \begin{bmatrix}
			\begin{pmatrix}	0 \\ 0	\end{pmatrix} \\
			\begin{pmatrix}	0 \\ 0	\end{pmatrix} \\
			\begin{pmatrix}	0 \\ 1	\end{pmatrix} \\
			\begin{pmatrix}	1\\ 0	\end{pmatrix}  \\ 
			\begin{pmatrix}	1\\ 0	\end{pmatrix}  \\ 
			\begin{pmatrix}	0 \\ 1	\end{pmatrix} \\
			\begin{pmatrix}	0 \\ 0	\end{pmatrix} \\
			\begin{pmatrix}	0 \\ 0	\end{pmatrix} 
		\end{bmatrix}.
	\end{split} 
\end{equation}

\subsection{$4P$, $\bar{4}P$}\label{section:4P}
For arithmetic  class $4P$, we choose the basis of $L$ and $L_F$ as
\begin{equation}
	\begin{split}
		&\bm{e}_1 =  (1,0,0),\quad \bm{e}_2 =  (0,1,0),\quad \bm{e}_3 =  (0,0,1),\\
		&\bm{G}_1 = (1,0,0),\quad \bm{G}_2 = (0,1,0),\quad \bm{G}_3 = (0,0,1).
	\end{split}
\end{equation}
When the flux form takes the value
\begin{equation}
	\Phi = \begin{bmatrix}
		0 & \frac{1}{2n}  & -\frac{1}{2}  \\
		-\frac{1}{2n}  & 0 & \frac{1}{2} \\
		\frac{1}{2}  & -\frac{1}{2}  &0
	\end{bmatrix},
\end{equation}
the basis of $Z$ is given by
\[
\bm{a}_1 = n\bm{e}_1- n\bm{e}_2 +\bm{e}_3,\quad
\bm{a}_2 = n\bm{e}_1+ n\bm{e}_2 + \bm{e}_3,\quad
\bm{a}_3 =  2\bm{e}_3.
\]
Thus the bases of $Z$ and $Z_F$ are
\begin{equation}
	\begin{split}
		&\bm{a}_1 = (n,-n,1),\quad \bm{a}_2 = (n,n,1),\quad \bm{a}_3 = (0,0,2),\\
		&\bm{Q}_1 =\frac{1}{2n} (1,-1,0),\quad \bm{Q}_2 = \frac{1}{2n}(1,1,0),\quad \bm{Q}_3 = \frac{1}{2n}(-1,0,n).
	\end{split}
\end{equation}
The quotient $Z_F/L_F$ is isomorphic to $\mathbb{Z}_{2n} \times \mathbb{Z}_{2n}$. Since $\mathbf{Q}_2 \equiv -\mathbf{Q}_1-2\mathbf{Q}_3 \pmod{L_F}$, a convenient basis for $\mathbb{Z}_{2n} \times \mathbb{Z}_{2n}$ is given by the cosets of $\bm{Q}_1$ and $\bm{Q}_3$.
The action of $C_{4}$ on the $\mathbf{Q}_i$ is given by
\begin{align}
	R_z \bm{Q}_1 =  \bm{Q}_2,\quad R_z \bm{Q}_3 = \bm{Q}_1+ \bm{Q}_3 
\end{align}
Passing to the quotient $\mathbb{Z}_{2n} \times \mathbb{Z}_{2n}$ and using the relations   $2n\mathbf{Q}_i \equiv 0$, we obtain the induced action as matrices (mod~$2n$):

\begin{equation}
	\begin{split}
		R_z = \begin{bmatrix} -1 & 1 \\-2 & 1 \end{bmatrix}
	\end{split}
\end{equation}
acting by left multiplication on column vectors of $  \mathbb{Z}_{2n} \times \mathbb{Z}_{2n}$.

The two‑cocycle  $\overline{\bm{\Lambda}_\Phi} $  is given by
\begin{equation}
	\begin{split}
		\overline{\bm{\Lambda}_\Phi}(R_z^{a_2}   ,R_z^{a_1}  )  = \begin{cases}
			\begin{pmatrix}	n \\ 0	\end{pmatrix} & a_2, a_1 \in \left\{1,3\right\}   \\
			\begin{pmatrix}	0 \\ 0	\end{pmatrix} & \text{Otherwise}
		\end{cases}
		= \delta \bm{\chi} (R_z^{a_2}   ,R_z^{a_1}  ).
	\end{split} 
\end{equation}
The coboundary  satisfying $\overline{\delta \bm{\chi}} = \bm{\Lambda}_\Phi$ is given by
\begin{equation}
	\bm{\chi}(R_z^a) = \begin{cases}
		\begin{pmatrix}	n \\ 0	\end{pmatrix} & a   \in \left\{2,3\right\}  \\
		\begin{pmatrix}	0 \\ 0	\end{pmatrix} & a   \in \left\{0,1\right\} 
	\end{cases}.
\end{equation}

When we replace the generator $R_z$ of $4P$ by the rotation inversion $\bar{R}_z$, we obtain $\bar{4}P$. The group element $\bar{R}_z$ acts on $\mathbb{Z}_{2n} \times \mathbb{Z}_{2n}$ via  matrix (mod $2n$)
\begin{equation}
	\bar{R}_z = \begin{bmatrix}   1 &- 1 \\
		2  &-1
	\end{bmatrix} .
\end{equation} 

The cocycle $\overline{\bm{\Lambda}^{\bar{4}P}_\Phi }$  is the same  as a function of $a_2,a_1$ ($a_2,a_1 = 0,1,2,3$)
\begin{equation}
	\overline{\bm{\Lambda}_\Phi^{4P}} (R_z^{a_2}   ,R_z^{a_1}  ) = \overline{\bm{\Lambda}_\Phi^{\bar{4}P}} (  \bar{R}_z^{a_2} ,  \bar{R}_z^{a_1} ) .
\end{equation}

\subsection{$4/mP$}
For arithmetic  class $4/mP$ with flux form $ 
\Phi = 	\begin{bmatrix}
	0 & \frac{1}{2n}  & -\frac{1}{2}  \\
	-\frac{1}{2n}  & 0 & \frac{1}{2} \\
	\frac{1}{2}  & -\frac{1}{2}  &0
\end{bmatrix}
$, the choice of basis is the same  as in Section.\ref{section:4P}.
The quotient $Z_F/L_F$ is isomorphic to $\mathbb{Z}_{2n} \times \mathbb{Z}_{2n}$. A convenient basis for $\mathbb{Z}_{2n} \times \mathbb{Z}_{2n}$ is given by the cosets of $\bm{Q}_1$ and $\bm{Q}_3$.

The group elements $R_z$ and $M_z$ act on $\mathbb{Z}_{2n} \times \mathbb{Z}_{2n}$ via  matrices (mod $2n$)

\begin{equation}
	R_z = \begin{bmatrix}   -1 & 1 \\
		-2  &1
	\end{bmatrix},M_z = \begin{bmatrix}   1 & 0 \\
		0  &1
	\end{bmatrix}  .
\end{equation}

The two‑cocycle   $\overline{\bm{\Lambda}_\Phi} $ is given by
\begin{equation}
	\begin{split}
		&\overline{\bm{\Lambda}_\Phi}  (M_z^{a_2} R_z^{b_2} ,M_z^{a_1} R_z^{b_1} )   =  \begin{cases}
			\begin{pmatrix}	n\\ 0	\end{pmatrix} & \text{ when }b_2,b_1 \in \left\{ 1,3\right\} \\
			\begin{pmatrix}	0\\ 0	\end{pmatrix} & \text{ Otherwise}  \\
		\end{cases} = \delta \bm{\chi}(g_2,g_1).
	\end{split} 
\end{equation}

The coboundary  satisfying $\overline{\delta \bm{\chi}} = \bm{\Lambda}_\Phi$ is given by
\begin{equation}
	\bm{\chi}(M_z^a R_z^b) = \begin{cases}
		\begin{pmatrix}	n \\ 0	\end{pmatrix} & b   \in \left\{2,3\right\}  \\
		\begin{pmatrix}	0 \\ 0	\end{pmatrix} & b   \in \left\{0,1\right\} 
	\end{cases}.
\end{equation}

\subsection{$422P$, $4mmP$, $\bar{4}2mP$, $\bar{4}m2P$}
For arithmetic  class $422P,4mmP,\bar{4}2mP,\bar{4}m2P$ with flux form $  
\Phi = 	\begin{bmatrix}
	0 & \frac{1}{2}  & -\frac{1}{2}  \\
	-\frac{1}{2}  & 0 & \frac{1}{2} \\
	\frac{1}{2}  & -\frac{1}{2}  &0
\end{bmatrix},
$ the choice of basis is the same  as in Section.\ref{section:4P}(set $n=1$).
The quotient $Z_F/L_F$ is isomorphic to $\mathbb{Z}_{2} \times \mathbb{Z}_{2}$. A convenient basis for $\mathbb{Z}_{2} \times \mathbb{Z}_{2}$ is given by the cosets of $\bm{Q}_1$ and $\bm{Q}_3$.

The group elements $R_z, \bar{R}_z,M_x,R_x $ act  on $\mathbb{Z}_{2} \times \mathbb{Z}_{2}$ via  matrices (mod $2 $)
\begin{equation}
	R_z =	\bar{R}_z = \begin{bmatrix}   1 & 1 \\
		0  &1
	\end{bmatrix},R_x=M_x = \begin{bmatrix}   1 & 0 \\
		0  &1
	\end{bmatrix}  .
\end{equation}

Their cocycle $\overline{\bm{\Lambda}_\Phi} $ have a similar structure:
\begin{equation}
	\begin{split}
		&\overline{\bm{\Lambda}_\Phi^{422P}}(R_x^{a_2} R_z^{b_2} ,R_x^{a_1} R_z^{b_1} ) = \overline{\bm{\Lambda}_\Phi^{4mmP}}(M_x^{a_2} R_z^{b_2} ,M_x^{a_1} R_z^{b_1} ) = \overline{\bm{\Lambda}_\Phi^{\bar{4}2mP}}(R_x^{a_2} \bar{R}_z^{b_2} ,R_x^{a_1} \bar{R}_z^{b_1} ) =\overline{\bm{\Lambda}_\Phi^{\bar{4}m2P}}(M_x^{a_2} \bar{R}_z^{b_2} ,M_x^{a_1} \bar{R}_z^{b_1} )\\
		= &\begin{cases}
			\begin{pmatrix}	n\\ 0	\end{pmatrix} & \text{ when }b_2,b_1 \in \left\{ 1,3\right\}\\
			\begin{pmatrix}	0\\ 0	\end{pmatrix} & \text{ Otherwise}  \\
		\end{cases} = \delta \bm{\chi}(g_2,g_1).
	\end{split}
\end{equation}
The coboundary  satisfying $\overline{\delta \bm{\chi}} = \bm{\Lambda}_\Phi$ is given by
\begin{equation}
	\bm{\chi}(b,a) = \begin{cases}
		\begin{pmatrix}	n\\ 0	\end{pmatrix} & \text{ when }b=2,3 \\
		\begin{pmatrix}	0\\ 0	\end{pmatrix} & \text{ Otherwise}  \\
	\end{cases}.
\end{equation}

\subsection{$321P$, $312P$, $31mP$, $3m1P$}\label{section:312P}
We first investigate arithmetic  class $321P,312P,3m1P,31mP$

we choose the basis of $L$ and $L_F$ as
\begin{equation}
	\begin{split}
		&\bm{e}_1 =  (1,0,0),\quad \bm{e}_2 =  \frac{1}{2} (-1,\sqrt{3},0),\quad \bm{e}_3 =  (0,0,1),\\
		&\bm{G}_1 = (1,\frac{ \sqrt{3}}{3},0),\quad \bm{G}_2 = (0,\frac{2 \sqrt{3}}{3},0),\quad \bm{G}_3 = (0,0,1).
	\end{split}
\end{equation}
When the flux form takes the value
\begin{equation}
	\Phi =  \begin{bmatrix}
		0 & \frac{1}{2} & 0  \\
		-\frac{1}{2} & 0 & 0 \\
		0 & 0  &0
	\end{bmatrix},
\end{equation}
the basis of $Z$ is given by
\[
\bm{a}_1 = 2\bm{e}_1 ,\quad
\bm{a}_2 =    2\bm{e}_2  ,\quad
\bm{a}_3 =   \bm{e}_3.
\]
Thus the bases of $Z$ and $Z_F$ are
\begin{equation}
	\begin{split}
		&\bm{a}_1 =  (2,0,0),\quad \bm{a}_2 =    (-1,\sqrt{3},0),\quad \bm{a}_3 =  (0,0,1),\\
		&\bm{Q}_1 = (\frac{1}{2},\frac{  \sqrt{3}}{6},0),\quad \bm{Q}_2 = (0,\frac{  \sqrt{3}}{3},0),\quad \bm{Q}_3 = (0,0,1).
	\end{split}
\end{equation}
The quotient $Z_F/L_F$ is isomorphic to $\mathbb{Z}_{2} \times \mathbb{Z}_{2}$. A convenient basis for $\mathbb{Z}_{2} \times \mathbb{Z}_{2}$ is given by the cosets of $\bm{Q}_1$ and $\bm{Q}_2$.
The action of $D_{3}$ on the $\mathbf{Q}_i$ is given by
\begin{align}
	R_z \bm{Q}_1 = -\bm{Q}_1+ \bm{Q}_2,\quad R_z \bm{Q}_2 = -\bm{Q}_1, \quad 	R_x \bm{Q}_1 =  \bm{Q}_1-\bm{Q}_2,\quad R_x \bm{Q}_2 = -\bm{Q}_2.
\end{align}
Passing to the quotient $\mathbb{Z}_{2} \times \mathbb{Z}_{2}$ and using the relations  $2\mathbf{Q}_i \equiv 0$, we obtain the induced action as matrices (mod~$2$):

\begin{equation}
	R_z = \begin{bmatrix}   1 & 1 \\
		1 &0
	\end{bmatrix},	R_x  = \begin{bmatrix}   1 & 0 \\
		1 &1
	\end{bmatrix}  ,
\end{equation}
acting by left multiplication on column vectors of $  \mathbb{Z}_{2} \times \mathbb{Z}_{2}$.
Similarly we can obtain
\begin{equation}
	R_x = R_y = M_x= M_y = \begin{bmatrix}   1 & 0 \\
		1 &1
	\end{bmatrix}  .
\end{equation}

Their  $\overline{\bm{\Lambda}_\Phi}$   have a similar structure:
\begin{equation}
	\begin{split}
		&	\overline{\bm{\Lambda}_\Phi^{321P}}(R_x^{a_2} R_z^{b_2} ,R_x^{a_1} R_z^{b_1} ) = \overline{\bm{\Lambda}_\Phi^{312P}}(R_y^{a_2} R_z^{b_2} ,R_y^{a_1} R_z^{b_1} ) = \overline{\bm{\Lambda}_\Phi^{3m1P}}(M_x^{a_2} R_z^{b_2} ,M_x^{a_1} R_z^{b_1} ) = \overline{\bm{\Lambda}_\Phi^{31mP}}(M_y^{a_2} R_z^{b_2} ,M_y^{a_1} R_z^{b_1} )\\
		=&\begin{cases}
			\begin{pmatrix}	1\\ 0	\end{pmatrix} & \text{ when }a_1 =1, b_2=1\\
			\begin{pmatrix}	0\\ 1	\end{pmatrix} & \text{ when }a_1 =1, a_2= 0, b_2=2 \text{ or } a_1 =1, a_2= 1, b_2=0 \\
			\begin{pmatrix}	0\\ 0	\end{pmatrix} & \text{ Otherwise}  \\
		\end{cases}  .
	\end{split} 
\end{equation}

The coboundary  satisfying $\overline{\delta \bm{\chi}} = \bm{\Lambda}_\Phi$ is given by
\begin{equation}
	\bm{\chi}(a,b) =a\begin{pmatrix}	1\\ 1	\end{pmatrix} .
\end{equation}

We proceed to investigate arithmetic  class  $31mP,312P$ with flux form $\Phi   
= \begin{bmatrix}
	0 & \frac{1}{2} & \frac{1}{3}  \\
	-\frac{1}{2} & 0 & \frac{1}{3} \\
	-\frac{1}{3}  & -\frac{1}{3}  &0
\end{bmatrix}
$ . The basis of $Z$ is given by
\[
\bm{a}_1 = 2\bm{e}_1 + 4\bm{e}_2 ,\quad
\bm{a}_2 =   -4 \bm{e}_1 -2\bm{e}_2  ,\quad
\bm{a}_3 =   3\bm{e}_3.
\]
Thus the bases of $Z$ and $Z_F$ are
\begin{equation}
	\begin{split}
		&\bm{a}_1 =  (0,2 \sqrt{3},0),\quad \bm{a}_2 =    (-3,-\sqrt{3},0),\quad \bm{a}_3 =  (0,0,3),\\
		&\bm{Q}_1 = (-\frac{1}{6},\frac{  \sqrt{3}}{6},0),\quad \bm{Q}_2 = ( -\frac{  \sqrt{3}}{3},0,0),\quad \bm{Q}_3 =  \frac{1}{3}(0,0,1).
	\end{split}
\end{equation}
The quotient $Z_F/L_F$ is isomorphic to $\mathbb{Z}_{2} \times \mathbb{Z}_{2}$. 
The action of  group elements on the $\mathbf{Q}_i$ is given by
\begin{align}
	&	R_z \bm{Q}_1 = -\bm{Q}_1+ \bm{Q}_2,\quad R_z \bm{Q}_2 = -\bm{Q}_1, \quad R_z \bm{Q}_3 =  \bm{Q}_3, \\
	&		R_y \bm{Q}_1 =  \bm{Q}_1-\bm{Q}_2,\quad R_y \bm{Q}_2 = -\bm{Q}_2,R_y \bm{Q}_3 = -\bm{Q}_3, \\
	& 	M_y \bm{Q}_1 =  -\bm{Q}_1+\bm{Q}_2 ,\quad M_z \bm{Q}_2 =  \bm{Q}_2,M_y \bm{Q}_3 = \bm{Q}_3 .
\end{align}
Notice that
\begin{equation}
	\begin{split}
		&	\bm{Q}_2 \equiv 2 \bm{Q}_1 +3 (\bm{Q}_2 + \bm{Q}_3) \pmod {L_F}, \\
		&	\bm{Q}_3 \equiv -2 \bm{Q}_1 - 2 (\bm{Q}_2 + \bm{Q}_3) \pmod {L_F},
	\end{split}
\end{equation}
a convenient basis for $\mathbb{Z}_{6} \times \mathbb{Z}_{6}$ is given by the cosets of $\bm{Q}_1$ and $\bm{Q}_2 + \bm{Q}_3$.
Passing to the quotient $\mathbb{Z}_{6} \times \mathbb{Z}_{6}$ and using the relations  $6\mathbf{Q}_i \equiv 0$, we obtain the induced action of group elements $R_z,R_y, M_y$  as matrices (mod $6 $)
\begin{equation}
	R_z = \begin{bmatrix}   1 & 3 \\
		3 &4
	\end{bmatrix},	  R_y  = \begin{bmatrix}   -1 & 0 \\
		3 &-1
	\end{bmatrix},M_y = \begin{bmatrix}   1 & 0 \\
		3 &1
	\end{bmatrix}  .
\end{equation}
acting by left multiplication on column vectors of $  \mathbb{Z}_{6} \times \mathbb{Z}_{6}$.

Their  $\overline{\bm{\Lambda}_\Phi}$   have a similar structures:
\begin{equation}
	\begin{split}
		&	 \overline{\bm{\Lambda}_\Phi^{312P}}(R_y^{a_2} R_z^{b_2} ,R_y^{a_1} R_z^{b_1} )    = \overline{\bm{\Lambda}_\Phi^{31mP}}(M_y^{a_2} R_z^{b_2} ,M_y^{a_1} R_z^{b_1} )\\
		=&\begin{cases}
			\begin{pmatrix}	3\\ 0	\end{pmatrix} & \text{ when }a_1 =1, b_2=1\\
			\begin{pmatrix}	0\\ 3	\end{pmatrix} & \text{ when }a_1 =1, a_2= 0, b_2=2 \text{ or } a_1 =1, a_2= 1, b_2=0 \\
			\begin{pmatrix}	0\\ 0	\end{pmatrix} & \text{ Otherwise}  \\
		\end{cases}  .
	\end{split} 
\end{equation}

The coboundary  satisfying $\delta \bm{\chi} = \overline{\bm{\Lambda}_\Phi}$ is given by
\begin{equation}
	\bm{\chi}(a,b) =a\begin{pmatrix}	3\\ 3	\end{pmatrix} .
\end{equation}

\subsection{ $\bar{3}1mP$, $\bar{3}m1P$}
For arithmetic  class $\bar{3}1mP,\bar{3}m1P$, the choice of basis is the same  as in Section.\ref{section:312P}.

When the forms takes $ 
\Phi = \begin{bmatrix}
	0 & \frac{1}{2} & 0  \\
	-\frac{1}{2} & 0 & 0 \\
	0 & 0  &0
\end{bmatrix}
$,  the quotient $Z_F/L_F$ is isomorphic to $\mathbb{Z}_{2} \times \mathbb{Z}_{2}$. A convenient basis for $\mathbb{Z}_{2} \times \mathbb{Z}_{2}$ is given by the cosets of $\bm{Q}_1$ and $\bm{Q}_2$.

The group elements $\bar{R}_z,M_y,M_x$ act on $\mathbb{Z}_2 \times \mathbb{Z}_2$   via  matrices (mod $2 $)
\begin{equation}
	\bar{R}_z = \begin{bmatrix}   1 & 1 \\
		1 &0
	\end{bmatrix},M_x=M_y = \begin{bmatrix}   1 & 0 \\
		1 &1
	\end{bmatrix}  .
\end{equation}

Their  $\overline{\bm{\Lambda}_\Phi}$ have a similar structures:

\begin{equation}
	\begin{split}
		&	\overline{\bm{\Lambda}_\Phi^{\bar{3}m1P}}(M_y^{a_2} \bar{R}_z^{b_2} ,M_y^{a_1} \bar{R}_z^{b_1} ) = \overline{\bm{\Lambda}_\Phi^{\bar{3}1mP}}(M_x^{a_2} \bar{R}_z^{b_2} ,M_x^{a_1} \bar{R}_z^{b_1} ) \\
		=&\begin{cases}
			\begin{pmatrix}	1\\ 0	\end{pmatrix} & \text{ when }a_1 =1, b_2 \in \left\{1,4\right\}\\
			\begin{pmatrix}	0\\ 1	\end{pmatrix} & \text{ when }a_1 =1, a_2= 0, b_2 \in \left\{2,5\right\} \text{ or } a_1 =1, a_2= 1, b_2 \in \left\{0,3\right\}  \\
			\begin{pmatrix}	0\\ 0	\end{pmatrix} & \text{ Otherwise}   
		\end{cases}  .
	\end{split} 
\end{equation}

The coboundary  satisfying $\delta \bm{\chi} = \overline{\bm{\Lambda}_\Phi}$ is given by
\begin{equation}
	\bm{\chi}(a,b) =a\begin{pmatrix}	1\\ 1	\end{pmatrix} .
\end{equation}

We proceed to investigate arithmetic  class $\bar{3}1mP$ with flux form $ 
\Phi = \begin{bmatrix}
	0 & \frac{1}{2} & \frac{1}{3}  \\
	-\frac{1}{2} & 0 & \frac{1}{3} \\
	-\frac{1}{3}  & -\frac{1}{3}  &0
\end{bmatrix}$.  The quotient $Z_F/L_F$ is isomorphic to $\mathbb{Z}_{6} \times \mathbb{Z}_{6}$. A convenient basis for $\mathbb{Z}_{6} \times \mathbb{Z}_{6}$ is given by the cosets of $\bm{Q}_1$ and $\bm{Q}_2 + \bm{Q}_3$.

The group elements $\bar{R}_z,M_y $ act on $\mathbb{Z}_6 \times \mathbb{Z}_6$   via  matrices (mod $6 $)
\begin{equation}
	\bar{R}_z = \begin{bmatrix}   1 & 3 \\
		3 &2
	\end{bmatrix},M_y = \begin{bmatrix}   1 & 0 \\
		3 &-1
	\end{bmatrix}  .
\end{equation}

The cocycle $\overline{\bm{\Lambda}_\Phi}$  is given by
\begin{equation}
	\begin{split}
		&	\overline{\bm{\Lambda}_\Phi^{\bar{3}m1P}}(M_y^{a_2} \bar{R}_z^{b_2} ,M_y^{a_1} \bar{R}_z^{b_1} ) =  \\
		=&\begin{cases}
			\begin{pmatrix}	3\\ 0	\end{pmatrix} & \text{ when }a_1 =1, b_2 \in \left\{1,4\right\}\\
			\begin{pmatrix}	0\\ 3	\end{pmatrix} & \text{ when }a_1 =1, a_2= 0, b_2 \in \left\{2,5\right\} \text{ or } a_1 =1, a_2= 1, b_2 \in \left\{0,3\right\}  \\
			\begin{pmatrix}	0\\ 0	\end{pmatrix} & \text{ Otherwise}   
		\end{cases}  .
	\end{split} 
\end{equation}

The coboundary  satisfying $\delta \bm{\chi} = \overline{\bm{\Lambda}_\Phi}$ is given by
\begin{equation}
	\bm{\chi}(a,b) =a\begin{pmatrix}	3\\ 3	\end{pmatrix} .
\end{equation}

\subsection{ $622P$, $6mmP$, $\bar{6}m2P$, $\bar{6}2mP$, $6/mmmP$} 

For arithmetic  class $6mmP,6mmP,\bar{6}m2P,\bar{6}2mP$ with flux form $ 
\Phi = \begin{bmatrix}
	0 & \frac{1}{2} & 0  \\
	-\frac{1}{2} & 0 & 0 \\
	0 & 0  &0
\end{bmatrix}
$, the choice of basis is the same  as in Section.\ref{section:312P}.
The quotient $Z_F/L_F$ is isomorphic to $\mathbb{Z}_{2} \times \mathbb{Z}_{2}$. A convenient basis for $\mathbb{Z}_{2} \times \mathbb{Z}_{2}$ is given by the cosets of $\bm{Q}_1$ and $\bm{Q}_2$.
The action of $D_{6}$ on the $\mathbf{Q}_i$ is given by
\begin{align}
	&	R_z \bm{Q}_1 =\bm{Q}_2,\quad R_z \bm{Q}_2 = -\bm{Q}_1 + \bm{Q}_2, \quad 	R_x \bm{Q}_1 =  \bm{Q}_1-\bm{Q}_2,\quad R_x \bm{Q}_2 = -\bm{Q}_2, \\
	&	\bar{R}_z \bm{Q}_1 =-\bm{Q}_2,\quad \bar{R}_z \bm{Q}_2 = \bm{Q}_1 - \bm{Q}_2, \quad 	M_x \bm{Q}_1 =  -\bm{Q}_1+\bm{Q}_2,\quad M_x \bm{Q}_2 = \bm{Q}_2 .
\end{align}
Passing to the quotient $\mathbb{Z}_{2} \times \mathbb{Z}_{2}$ and using the relations  $2\mathbf{Q}_i \equiv 0$, we obtain the induced action as matrices (mod~$2$):

\begin{equation}
	\bar{R}_z = R_z = \begin{bmatrix}   0 & 1 \\
		1 &1
	\end{bmatrix},	M_x = R_x  = \begin{bmatrix}   1 & 0 \\
		1 &1
	\end{bmatrix}  ,
\end{equation}
acting by left multiplication on column vectors of $  \mathbb{Z}_{2} \times \mathbb{Z}_{2}$.

Their   $\overline{\bm{\Lambda}_\Phi}$ have a similar structure:
\begin{equation}
	\begin{split}
		&\overline{\bm{\Lambda}_\Phi^{6mmP}}(M_x^{a_2} R_z^{b_2} ,M_x^{a_1} R_z^{b_1} ) = \overline{\bm{\Lambda}_\Phi^{622P}}(R_x^{a_2} R_z^{b_2} ,R_x^{a_1} R_z^{b_1} ) = \overline{\bm{\Lambda}_\Phi^{\bar{6}m2P}}(M_x^{a_2} \bar{R}_z^{b_2} ,M_x^{a_1} \bar{R}_z^{b_1} ) =\overline{\bm{\Lambda}_\Phi^{\bar{6}2mP}}(R_x^{a_2} \bar{R}_z^{b_2} ,R_x^{a_1} \bar{R}_z^{b_1} )\\
		=	&\begin{cases}
			\begin{pmatrix}	0\\ 1	\end{pmatrix} & \text{ when }a_1 =1, a_2=0, b_2 \in \left\{1,4\right\} \text{ or } a_1 =1, a_2=1, b_2 \in \left\{0,3\right\} \\
			\begin{pmatrix}	1\\ 0	\end{pmatrix} & \text{ when }a_1 =1, b_2 \in \left\{2,5\right\}  \\
			\begin{pmatrix}	0\\ 0	\end{pmatrix} & \text{ Otherwise}   
		\end{cases}  .
	\end{split} 
\end{equation}

The coboundary  satisfying $\delta \bm{\chi} = \overline{\bm{\Lambda}_\Phi}$ is given by
\begin{equation}
	\bm{\chi}(a,b) =a\begin{pmatrix}	1\\ 1	\end{pmatrix} .
\end{equation}

Furthermore, for arithmetic class $6/mmmP$ with the same flux form,  since the group element $M_z$ acts trivially on $\mathbb{Z}_2 \times \mathbb{Z}_2$, $\overline{\bm{\Lambda}^{6/mmmP}_\Phi}$ can be obtained from $\overline{\bm{\Lambda}_\Phi^{6mmP}}$ by
\begin{equation}
	\overline{\bm{\Lambda}_\Phi^{6/mmmP}}(M_z^{a_2} M_x^{b_2} R_z^{c_2}  ,M_z^{a_1} M_x^{b_1} R_z^{c_1} ) = \overline{\bm{\Lambda}_\Phi^{6mmP}}(M_y^{b_2} R_z^{c_2},M_y^{b_1} R_z^{c_1} )
\end{equation}
and  the coboundary  satisfying $\delta \bm{\chi} = \overline{\bm{\Lambda}_\Phi}$ is given by $\bm\chi(M_z^{a} M_x^{b} R_z^{c}  ) = b \begin{pmatrix}	1\\ 1	\end{pmatrix}   $.

\subsection{$23F$}

For arithmetic  class $23F$ with the flux form  $\Phi = \begin{bmatrix}
	0 & \frac{1}{4} & -\frac{1}{4} \\
	-\frac{1}{4} & 0 & \frac{1}{4} \\
	\frac{1}{4} & -\frac{1}{4} & 0
\end{bmatrix},$  the choice of basis is the same  as in Section.\ref{section:222F}.
The quotient $Z_F/L_F$ is isomorphic to $\mathbb{Z}_{4} \times \mathbb{Z}_{4}$. A convenient basis for $\mathbb{Z}_{4} \times \mathbb{Z}_{4}$ is given by the cosets of $\bm{Q}_1$ and $\bm{Q}_2$. The element of point group of $23F$ can be written as $R_{xyz}^{a} R_x^b R_y^c$, where $R_{xyz}$ is the generator of $C_3$, the action of  $R_{xzy}$ on the $\mathbf{Q}_i$ is given by 
\begin{align}
	&	R_{xzy} \bm{Q}_1 =\bm{Q}_2,\quad R_{xzy} \bm{Q}_2 =  \bm{Q}_3, \quad R_{xzy} \bm{Q}_3 =  \bm{Q}_1. 
\end{align}
Using the relations $\mathbf{Q}_3 \equiv -\mathbf{Q}_1-\mathbf{Q}_2 \pmod{L_F}$ and $4\mathbf{Q}_i \equiv 0$, we obtain the induced action as matrices (mod~4):

\begin{equation}
	R_{xyz} = \begin{bmatrix}
		0 & -1 \\
		1 & -1
	\end{bmatrix} ,
\end{equation}	 
acting by left multiplication on column vectors of $  \mathbb{Z}_{4} \times \mathbb{Z}_{4}$.

The two cocycle $\overline{\bm{\Lambda}^{23F}_\Phi}$ can be obtained from $\overline{\bm{\Lambda}_\Phi^{222F}}$  by
\begin{equation}
	\begin{split}
		&\overline{\bm{\Lambda}_\Phi^{23F}}(R_{xzy}^{a_2}  R_x^{b_2} R_y^{c_2} ,R_{xyz}^{a_1} R_x^{b_1} R_y^{c_1} ) = R_{xyz}^{a_2-a_1}  \overline{\bm{\Lambda}_\Phi^{222F}}(R_x^{b_2} R_y^{c_2}  ,R_x^{b_1} R_y^{c_1} ) .
	\end{split} 
\end{equation}
The coboundary  satisfying $\delta \bm{\chi} = \overline{\bm{\Lambda}_\Phi}$ is given by
\begin{equation}
	\bm{\chi}^{23F}(R_{xyz}^{a} R_x^b R_y^c) = \begin{cases}
		\bm{\chi}^{222F}(  R_x^b R_y^c), &   a=0 \\
		\begin{pmatrix}
			1 \\ 1
		\end{pmatrix}+ R_{xyz}\bm{\chi}^{222F}(R_x^b R_y^c), &   a=1 \\
		\begin{pmatrix}
			1 \\ 0
		\end{pmatrix}+ R_{xyz}^2\bm{\chi}^{222F}(R_x^b R_y^c), &   a=2 
	\end{cases}.
\end{equation}

\section{ Other nontrivial \(\overline{\Lambda_\Phi}\) cases}
Apart from the arithmetic class $mm2F$ discussed in Section~\ref{section:mm2F}, there are two other arithmetic classes, $mmmF$ and $m\bar{3}F$, for which $\overline{\Lambda_\Phi}$ belongs to a nontrivial cohomology class. We provide the explicit form of $\overline{\bm{\Lambda}_\Phi}$ for these two classes and demonstrate how it enforces nonsymmorphicity.

\subsection{$mmmF$}
For arithmetic class $mmmF$ with the flux form  
\[
\Phi = \begin{bmatrix}
	0 & \frac{1}{4} & -\frac{1}{4} \\
	-\frac{1}{4} & 0 & \frac{1}{4} \\
	\frac{1}{4} & -\frac{1}{4} & 0
\end{bmatrix},
\]  
the choice of basis is the same as in Section~\ref{section:mm2F}.  
The quotient $Z_F/L_F$ is isomorphic to $\mathbb{Z}_4 \times \mathbb{Z}_4$. A convenient basis for $\mathbb{Z}_4 \times \mathbb{Z}_4$ is given by the cosets of $\bm{Q}_1$ and $\bm{Q}_2$.  The point group of $mmmF$ is $D_{2h}$.
The elements of the $D_{2h}$  can be written as $M_x^a M_y^b M_z^c$; we have  
\begin{align}
	M_{z} \bm{Q}_1 &= \bm{Q}_3 - \bm{Q}_2, \\
	M_{z} \bm{Q}_2 &= \bm{Q}_3 - \bm{Q}_1.
\end{align}  
Using the relations $\mathbf{Q}_3 \equiv -\mathbf{Q}_1-\mathbf{Q}_2 \pmod{L_F}$ and $4\mathbf{Q}_i \equiv 0$, we obtain the induced action as matrices (mod~4):
\[
M_{z} = \begin{bmatrix}
	-1 & 2 \\
	2 & -1
\end{bmatrix},
\]  
acting by left multiplication on column vectors of $\mathbb{Z}_4 \times \mathbb{Z}_4$.

The two‑cocycle $\overline{\bm{\Lambda}_\Phi}$ is given by  
\begin{equation}
	\begin{split}
		&\overline{\bm{\Lambda}_\Phi}(M_x^{a_2}M_y^{b_2} M_z^{c_2},M_x^{a_1}M_y^{b_1} M_z^{c_1})\\
		= &\ 2 (a_2 a_1 + c_2 b_1 + b_2 c_1) \begin{pmatrix}1 \\ 0\end{pmatrix}
		+ 2( b_2 b_1 + c_2 a_1 + a_2 c_1) \begin{pmatrix}0 \\ 1\end{pmatrix} \\
		&+ 2( a_2 b_1 + b_2 a_1 + c_2 c_1) \begin{pmatrix}1 \\ 1\end{pmatrix}.
	\end{split}
\end{equation}

Notice that when we restrict $\overline{\bm{\Lambda}_\Phi}$ to the subgroup $C_{2v}$, we have $\overline{\bm{\Lambda}_\Phi^{mmmF}} = \overline{\bm{\Lambda}_\Phi^{mm2F}}$. According to the analysis in Section~\ref{section:mm2F}, the restricted cocycle already lies outside the images of the sets
$
\bigl\{ [\overline{\bm{\omega}_F}] \mid [\bm{\omega}_F] \in H^2(C_{2v},L_F) \bigr\}
$ and $\quad
\bigl\{ [\overline{\bm{\Phi}\cdot \bm{\omega}}] \mid [\bm{\omega}] \in H^2(C_{2v},L) \bigr\}.
$
Since the restriction of a cohomology class to a subgroup is well defined, the full class $[\overline{\bm{\Lambda}_\Phi}] \in H^2(G, Z_F/L_F)$ for $G= D_{2h}$ cannot belong to the corresponding sets for the full group either. Hence the same conclusion as in Section~\ref{section:mm2F} holds for $mmmF$. Because the only nonsymmorphic space group in arithmetic class $mmmF$ is $Fddd$, the flux form enforces that both the real-space and reciprocal-space groups are forced to be nonsymmorphic, i.e., $\Gamma = \Gamma_F = Fddd$.

\subsection{$m\bar{3}F$}

For arithmetic class $m\bar{3}F$ with the flux form  
\[
\Phi = \begin{bmatrix}
	0 & \frac{1}{4} & -\frac{1}{4} \\
	-\frac{1}{4} & 0 & \frac{1}{4} \\
	\frac{1}{4} & -\frac{1}{4} & 0
\end{bmatrix},
\]  
the choice of basis is the same as in Section~\ref{section:mm2F}.

The quotient $Z_F/L_F$ is isomorphic to $\mathbb{Z}_4 \times \mathbb{Z}_4$. A convenient basis for $\mathbb{Z}_4 \times \mathbb{Z}_4$ is given by the cosets of $\bm{Q}_1$ and $\bm{Q}_2$.  The point group of $m\bar{3}F$ is $T_d$.
The elements of  $T_d$ can be written as $R_{xyz}^{a} M_x^{b} M_y^{c} M_z^{d}$, where $R_{xyz}$ generates the $C_3$ subgroup. The action of $R_{xyz}$ on the $\mathbf{Q}_i$ is given by  
\begin{align}
	R_{xyz} \bm{Q}_1 = \bm{Q}_2,\quad R_{xyz} \bm{Q}_2 = \bm{Q}_3,\quad R_{xyz} \bm{Q}_3 = \bm{Q}_1.
\end{align}
Using the relations $\mathbf{Q}_3 \equiv -\mathbf{Q}_1-\mathbf{Q}_2 \pmod{L_F}$ and $4\mathbf{Q}_i \equiv 0$, we obtain the induced action as a matrix (mod~4):
\[
R_{xyz} = \begin{bmatrix}
	0 & -1 \\
	1 & -1
\end{bmatrix},
\]  
acting by left multiplication on column vectors of $\mathbb{Z}_4 \times \mathbb{Z}_4$.

The two‑cocycle $\overline{\bm{\Lambda}^{m\bar{3}F}_\Phi}$ can be obtained from $\overline{\bm{\Lambda}_\Phi^{mmmF}}$ by  
\begin{equation}
	\overline{\bm{\Lambda}_\Phi^{m\bar{3}F}}(R_{xyz}^{a_2} M_x^{b_2} M_y^{c_2} M_z^{d_2}, R_{xyz}^{a_1} M_x^{b_1} M_y^{c_1} M_z^{d_1})
	= R_{xyz}^{a_2-a_1} \overline{\bm{\Lambda}_\Phi^{mmmF}}(M_x^{b_2} M_y^{c_2} M_z^{d_2}, M_x^{b_1} M_y^{c_1} M_z^{d_1}).
\end{equation}

Notice that when we restrict $\overline{\bm{\Lambda}_\Phi}$ to the subgroup $C_{2v}$, we have $\overline{\bm{\Lambda}_\Phi^{m\bar{3}F}} = \overline{\bm{\Lambda}_\Phi^{mm2F}}$.  
According to the analysis in Section~\ref{section:mm2F}, the restricted cocycle already lies outside the images of the sets  
$
\bigl\{ [\overline{\bm{\omega}_F}] \mid [\bm{\omega}_F] \in H^2(C_{2v},L_F) \bigr\}
$ and $\quad
\bigl\{ [\overline{\bm{\Phi}\cdot \bm{\omega}}] \mid [\bm{\omega}] \in H^2(C_{2v},L) \bigr\}.
$
Since the restriction of a cohomology class to a subgroup is well defined, the full class $[\overline{\bm{\Lambda}_\Phi}] \in H^2(G, Z_F/L_F)$ for $G = T_d$ cannot belong to the corresponding sets for the full group either. Hence the same conclusion as in Section~\ref{section:mm2F} holds for $m\bar{3}F$.  
Because the only nonsymmorphic space group in arithmetic class $m\bar{3}F$ is $Fd\bar{3}$, the flux form forces both the real-space and reciprocal-space groups to be nonsymmorphic, i.e., $\Gamma = \Gamma_F = Fd\bar{3}$.
	
\end{document}